\documentclass{aa}  

\usepackage{graphicx}
\usepackage{multicol}
\usepackage{txfonts}
\usepackage{xcolor}
\usepackage{hyperref}
\usepackage{longtable}
\usepackage{pdflscape}
\usepackage{caption}
\begin{document}

   \title{Wildly Oscillating Stars}

   \subtitle{Unexplained dense ridge-like frequency agglomerations in A and F type pulsators}

   \author{V. Antoci \inst{1}
\and
        J. Labadie-Bartz \inst{1} 
\and
          M. Święch \inst{1} 
          \and
          O. Dürfeldt-Pedros \inst{1}  
          \and
          S.J. Murphy \inst{2}
          \and
          D.W. Kurtz \inst{3,4}
          \and
          T.R. Bedding\inst{5}
          \and
          G. Handler \inst{6}
          \and
          J. Fuller \inst{7}
          \and
          R.-M. Ouazzani \inst{8}
          \and
          H. Kjeldsen \inst{9}
          \and
          L. Fellay \inst{10} 
          \and
          M. Gade Pedersen \inst{5}
          \and
          E. Niemczura \inst{11}
          \and
          D.M. Bowman  \inst{12, 13}
          \and
          M. Deal\inst{14}
          \and 
          P. Mani\inst{5}
          }

   \institute{DTU Space, Technical University of Denmark, Elektrovej 327, Kgs. Lyngby, 2800, Denmark\\
              \email{antoci@space.dtu.dk}   
        \and
             Centre for Astrophysics, University of Southern Queensland, Toowoomba, QLD 4350, Australia
        \and
            Centre for Space Research, North-West University, Dr Albert Luthuli Drive, Mahikeng 2735, South Africa
        \and
            Jeremiah Horrocks Institute, University of Lancashire, Preston PR1 2HE, UK
        \and
            Sydney Institute for Astronomy (SIfA), School of Physics, University of Sydney, NSW 2006, Australia
        \and
            Nicolaus Copernicus Astronomical Center, Polish Academy of Sciences, ul. Bartycka 18, PL-00-716 Warszawa, Poland
        \and
            TAPIR, Mailcode 350-17, California Institute of Technology, Pasadena, CA 91125, USA
        \and
            LESIA, Observatoire de Paris, Université PSL, CNRS, Sorbonne Université, Université de Paris, 5 place Jules Janssen,
92195 Meudon, France
        \and
            Instutute for Physics and Astronomy, Aarhus University, Ny Munkegade 120, Aarhus C, Denamrk 
        \and
            STAR Institute, University of Liège, 19C Allée du 6 Août, B-4000 Liège, Belgium
        \and
            University of Wrocław, Astronomical Institute, Kopernika 11, 51-622 Wrocław, Poland
        \and
            School of Mathematics, Statistics and Physics, Newcastle University, Newcastle upon Tyne, NE1 7RU, United Kingdom
         \and
            Institute of Astronomy, KU Leuven, Celestijnenlaan 200D, 3001 Leuven, Belgium
        \and
           LUPM, CNRS, Université de Montpellier, Place Eugène Bataillon, 34095 Montpellier, France
        }
   \date{}

% \abstract{}{}{}{}{} 
% 5 {} token are mandatory
 
\abstract
% context heading (optional)
{}
% aims heading (mandatory)
{We investigate the origin of the dense, ridge-like frequency clusters observed in a subset of A and F type pulsating stars, which we refer to as `wildly oscillating' stars (WOS). These agglomerated frequency regions occupy a confined part of the frequency spectrum, typically below the fundamental radial mode, and are not explained by classical pulsation theory.}
% methods heading (mandatory)
{We analyse high-precision space photometry from \textit{Kepler} and TESS, construct \'echelle diagrams, and perform systematic searches for combination frequencies. We determine the expected fundamental radial mode using pulsation constants, period–luminosity relations, and stellar models in order to place the agglomerated regions in a seismic context. Rotational modulation is examined through phase-folded light curves and amplitude–phase analysis, and binarity and geometric modulation scenarios are tested. In addition, we compute non-adiabatic stability models for representative stellar parameters to assess whether standard excitation mechanisms reproduce the observed structures.}
% results heading (mandatory)
{The WOS phenomenon is confined to a narrow region of the Hertzsprung–Russell diagram near the overlap of the $\delta$~Sct and $\gamma$~Dor instability strips. The observed ridge morphology and mode density cannot be reproduced by simple asymptotic g-mode behaviour, standard low-order p modes, binarity, or typical rotational splitting. In at least two stars (KIC~5443410 and KIC~9347095), a significant fraction of peaks in the agglomerated region can be explained as nonlinear combination frequencies involving high-order g modes. However, these combinations require parent modes located within the agglomerated frequency band itself, indicating that intrinsic pulsation modes must be present there. Non-adiabatic stability calculations reproduce the classical instability domains but do not predict unstable modes with the observed density or organised ridge structure in the agglomerated region.}
% conclusions heading (optional)
{The WOS appear to represent a pulsational regime not captured by current models of mode excitation or rotational modulation. The agglomerated frequency phenomenon requires a mechanism that selects or excites a confined intermediate-frequency band and produces organised ridge structures within a narrow region of stellar parameter space.}
   \keywords{
               }

   \maketitle
%
%________________________________________________________________   %

\section{Introduction}\label{sec:intro}

Pulsating stars of intermediate-mass stars in the A and F spectral range are traditionally divided into several classes: the $\delta$~Scuti ($\delta$ Sct) stars \citep[e.g.,][]{Breger2000, Rodriguez_2001, Kurtz_2022}, $\gamma$~Doradus ($\gamma$ Dor) stars \citep{Balona_1994, Kaye_1999, Van_Reeth_2015, GangLi2020, Kurtz_2022}, the rapidly oscillating Ap (roAp) stars \citep{Kurtz_1982, Cunha_2019,Holdsworth_2021, Holdsworth_2024}, and, more recently, the ‘hump \& spike’ stars \citep{Saio_2018,Henriksen_2023_rotational_modulation, Henriksen_2023_Rossby_Modes, Antoci_2025}.

$\delta$~Sct stars pulsate in low- to intermediate-order pressure (p) modes with periods typically ranging from about 0.5 to a few hours, while $\gamma$~Dor stars oscillate in high-radial-order gravity (g) modes with periods of roughly one day. Theoretical models predict that these pulsations are driven by distinct excitation mechanisms \citep[e.g.,][]{guzik2000, dupret2004, Miglio_2008, houdek2008, antoci2014, antoci2019} and therefore their frequencies occupy different frequency regions of the amplitude spectrum. However, hybrid pulsators showing both p and g modes have been identified, providing valuable constraints on stellar interiors \citep[e.g.,][]{Handler_Shobbrook_2002, Grigahcene_2010, 2014MNRAS.444..102K, Schmid_2016}. In such stars, g modes dominate the low-frequency domain, whereas p modes appear at higher frequencies. In several cases, interactions between these two mode families have been shown to produce combination frequencies \citep{Papics_2012, Kurtz2015}.

The roAp stars are chemically peculiar and strongly magnetic, pulsating in high-radial-order p modes of the order of $5-23$\,min. Their amplitude spectra often show signatures of rotational modulation induced by chemical surface spots stabilized by the magnetic field. Unlike the $\delta$~Sct and $\gamma$~Dor stars, roAp stars are rare, with only about 130 members identified to date \citep{Holdsworth_2024}.

A more recently identified class is the `hump \& spike' stars \citep{Saio_2018}, which show a combination of unresolved Rossby modes (the hump), rotational modulation (the spike), and in approximately 40\% of cases, unresolved high-radial-order g modes \citep{Henriksen_2023_rotational_modulation, Henriksen_2023_Rossby_Modes}. The observed rotational modulation in these stars is more consistent with a dynamo-like origin, although the nature of the underlying process, as well as the mechanisms responsible for exciting both gravity and pressure modes, remain subjects of active investigation \citep{Antoci_2025}.

During a systematic visual inspection of \textit{Kepler} light curves aimed at identifying new $\gamma$~Dor stars, we encountered a group of stars whose frequency spectra defy this traditional classification. These stars show an unexpected power excess in the intermediate frequency regime, between the established g- and p-mode domains. In many cases, this region appears densely populated with peaks. So densely, in fact, that interpreting them as pulsation modes would require spherical degrees as high as $\ell \sim 15$, which is implausible due to geometric cancellation \citep{Dziembowski_1977}. Furthermore, the frequency range spanned by this region is often narrow enough to be consistent with a single radial order of p modes, raising additional doubts about a simple and purely pulsational origin.

In some stars, this agglomerated frequency region is flanked by typical g-mode pulsations at low frequencies and signatures consistent with p-mode oscillations at higher frequencies. Multiplet structures are also observed within the intermediate-frequency agglomerated frequency regions, possibly indicating rotationally induced splitting or other as yet unidentified processes. 

\begin{figure}
   \centering
\includegraphics[width=0.5\textwidth]{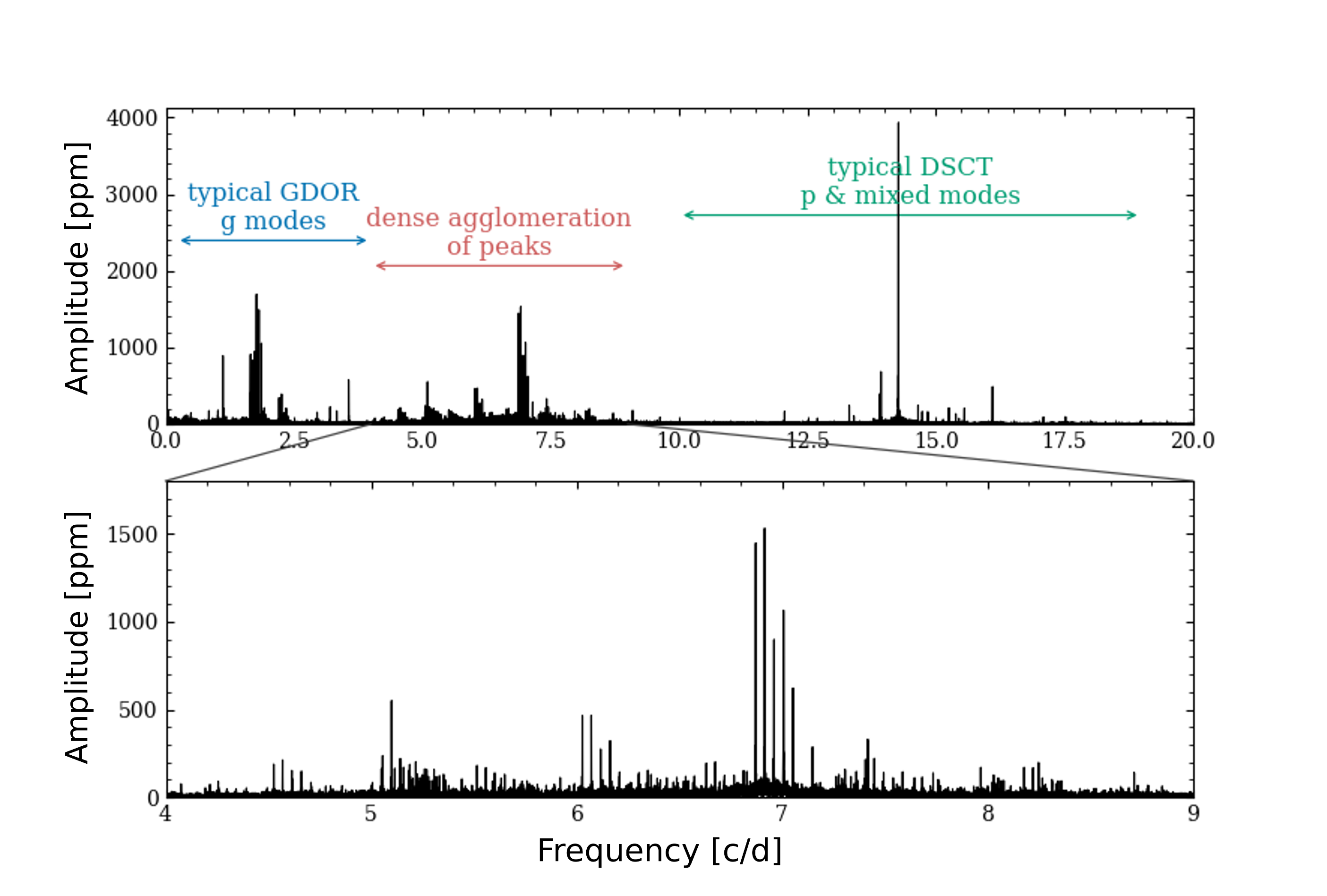}
\caption{Upper panel: Amplitude spectrum of KIC~5443410, an example of the first morphological group, Type I, in which both low-frequency $\gamma$~Dor g-mode and high-frequency $\delta$~Sct p-mode pulsations are clearly observed. Between these domains lies a densely populated agglomerated region exhibiting mutliplet structures. Lower panel: Zoom-in on the agglomerated frequency region, highlighting the forest of closely spaced peaks.}
\label{fig:WOS1_example}
\end{figure}

\begin{figure}
   \centering
\includegraphics[width=0.5\textwidth]{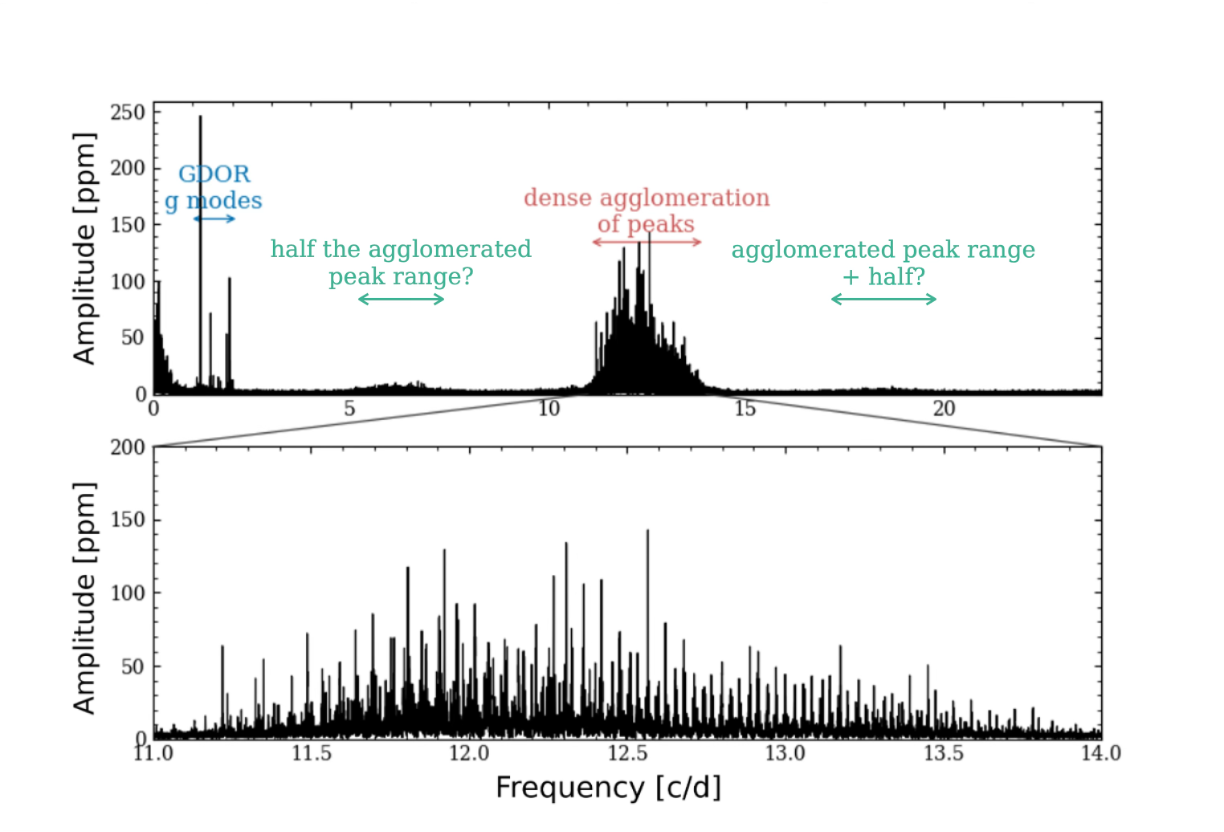}
\caption{Upper panel: Amplitude spectrum of KIC~6875337, representative of the second morphological group, Type II, where low-frequency variability is present due to either $\gamma$~Dor g-mode pulsations or rotational modulation (as seen, for example, in KIC~7430757). A prominent agglomerated region of dense and complex frequency structure is observed at higher frequencies. Lower panel: Zoom-in on the agglomerated region, showing a broad, forested hump inconsistent with typical p-mode spectra.}
\label{fig:WOS2_example}
\end{figure}

We have compiled a sample of 39 such stars, which we classify into two morphological types. Type I stars exhibit g modes, a dense intermediate-frequency hump, and conventional p-mode frequencies (see Fig.~\ref{fig:WOS1_example}). Type II stars display g modes and/or low-frequency modulation and an anomalously dense region of peaks at higher frequencies (see Fig.~\ref{fig:WOS2_example}).

This paper presents an observational description of these WOS, whose nature remains enigmatic. While we do not yet offer a definitive theoretical explanation, we explore several hypotheses and rule out potential causes such as instrumental artifacts or data processing issues. We base our analysis primarily on long-baseline \textit{Kepler} data, and complement it with TESS for some stars, plus \textit{Gaia} observations to derive stellar parameters. We examine individual frequency spectra and highlight the peculiar intermediate-frequency humps. \'Echelle diagrams are used to reveal underlying structure in these frequency groupings. For nine stars, we identify period spacing patterns among the low-frequency signals consistent with dipole and quadrupole g modes, allowing us to estimate the asymptotic period spacing $\Pi_0$ and to predict the location of the fundamental radial mode. We further apply the period--luminosity relation \citep[e.g.][]{barac2022} to independently estimate the expected radial p-mode frequencies. Notably, 25 out of the 39 stars in our sample are classified as Am stars. These chemically peculiar stars, especially those observed with TESS, are confined to a remarkably narrow region in the Hertzsprung–Russell diagram, suggesting a common mechanism related perhaps to stellar structure or evolutionary stage that may be linked to the observed spectral features. This paper thus serves to document the unusual features seen in these wildly oscillating intermediate-mass pulsators, and to provide a foundation for future observational analyses and theoretical modeling.
%__________________________________________________________________

\section{Data}

The analysis presented in this study is based primarily on photometric time-series observations from the \textit{Kepler} \citep{Koch2010_Kepler} and TESS \citep{Ricker2015} space missions, complemented by astrometric and photometric parameters from \textit{Gaia} Data Release 3 (\citet{GaiaDR3}; DR3).

\subsection{Kepler photometry}

The 23 stars in our sample observed with \textit{Kepler} were monitored in long-cadence mode, providing a time baseline of approximately 4.0~yr. This extensive coverage yields a frequency resolution better than 0.001~d$^{-1}$, which is essential for resolving densely packed frequency regions and identifying multiplet structures, particularly in the intermediate-frequency humps, described in Section~\ref{sec:intro}. The long-cadence integration time (29.4\,min) is sufficient to capture high-radial-order gravity modes and lower-order pressure modes. However, the very highest-frequency p modes may experience amplitude suppression due to temporal averaging, and some of these are above the Nyquist frequency for the long-cadence data, necessitating the use of Super-Nyquist techniques to study them \citep{murphy2013}.

We used the Simple Aperture Photometry (SAP) and applied a basic detrending procedure by adjusting each quarter to a common zero-point level. We do not observe any evidence of long-term variability in the light curves. However, we note that our detrending procedure may suppress signals with timescales longer than a Kepler quarter. This does not affect our analysis, as the features of interest occur on timescales of approximately 13\,d or shorter. We also do not detect any eclipses. This is not unexpected, as a photometric analysis of 1742 Am stars found that 4\% are eclipsing binaries \citep{Smalley_2014}. 

The long time baseline of \textit{Kepler} allows for reliable super-Nyquist analysis \citep{murphy2013}, because the slight irregularities in the sampling (due to variations in the light arrival time over the spacecraft orbit) break the strict degeneracy between true and aliased frequencies. Using this approach, we find no evidence that the intermediate-frequency humps are caused by aliasing. This conclusion is further supported by three targets with available short-cadence (60\,s) data, which independently confirm that the relevant signals originate at low frequencies. Only in one case (KIC 7352776) do we detect a small number of peaks above the long-cadence Nyquist frequency, but these are consistently recovered by the Super-Nyquist analysis.

\subsection{TESS photometry}\label{sec:tessdata}

A further 16 stars were observed by the TESS mission \citep{Ricker2015}. For these targets, we used the Simple Aperture Photometry (SAP) flux provided in the 2-minute cadence mode. However, the observational coverage for most TESS targets is significantly shorter than that of \textit{Kepler}, 
typically one to two consecutive TESS sectors (i.e. $\sim$one to two months).
Consequently, the temporal resolution is limited and not always sufficient to resolve individual frequency groupings or determine fine structure within the intermediate hump region. These stars were included based on the visual morphology of their amplitude spectra.

The TESS and \textit{Kepler} samples serve somewhat different scientific purposes in the context of this paper. 
\textit{Kepler} provides the first discovery of the phenomenon, as well as sufficient frequency resolution and precision to obtain a detailed and resolved view of the agglomerated power humps, rotational modulation, and g-mode ridges. In contrast, TESS enables expansion of the sample, improved stellar parameters for precise placement in the Hertzsprung--Russell diagram, and access to brighter stars suitable for spectroscopic and complementary follow-up observations. Because of these different roles, we do not apply the full suite of seismic and modulation analyses to the TESS stars that we perform for the \textit{Kepler} sample; the TESS data instead provide population-level context and confirmation that the phenomenon is not limited to the \textit{Kepler} field. For completeness, all available TESS observations of the WOS, including 2-minute cadence and full-frame image (FFI) sectors, are listed in Appendix~\ref{app:tesssectors}.

\subsection{Assessing photometric contamination}

As wide-field surveys with large pixels, both \textit{Kepler} and TESS can suffer from blending, where flux from neighbouring stars on the sky falls into the aperture used to extract light curves for our targets. Variations in brightness of neighbouring stars can therefore contaminate the light curves. It is thus crucial to confirm that photometric signals are intrinsic to the target star, and not due to contamination, especially when the signals in question are unusual for the class of star being considered \citep[e.g.][]{2023AJ....165..239P}. Our target list was cross-matched to the \textit{Gaia} catalogue to search for neighbouring sources with a magnitude difference of less than five (in the \textit{Gaia} $G$ band), at a separation from the target star below 18 arcsec for \textit{Kepler}, and 120 arcsec for TESS (roughly three times the typical PSF for the two instruments). About half of our targets have no neighbouring stars meeting these criteria. However, nineteen of our targets do have one or more neighbouring stars meeting these criteria. These are listed in Table~\ref{tab:neighbours}. A blending analysis was performed for these stars, discussed in Appendix~\ref{sec:blending_analysis}, and we found no indication that any of the signals of interest originate in neighbouring stars. We therefore assume that all signals identified and discussed in this work are not originating in a neighbouring \textit{Gaia} source.

\subsection{Prevalence of Am stars}

The \textit{Kepler} sample used in this study is relatively unbiased. Targets were selected from the Kepler Input Catalog (KIC; \citealt{brown2011}) based on effective temperatures in the range 6000--10\,000~K and surface gravities between $\log g = 3$ and $5$, encompassing pre-main-sequence, main-sequence, and slightly evolved A and F stars. From this broad sample of approximately 7000 stars, we identified 23 candidates exhibiting the characteristic features of WOS, of which 9 (approximately 40\%) are classified as Am stars in the literature \citep{tian2023}.

In contrast, the TESS sample analysed here is drawn from the Am-star catalogue used in \citet{duerfeldt2024} and is therefore strongly biased toward chemically peculiar stars. Within this focused set of approximately 1200 Am stars, we identified 16 WOS candidates.

In the region of the Hertzsprung–Russell diagram where the WOS are found (Sec.~\ref{sec:discussion}), the prevalence of Am stars is high. This is especially true for slow rotators (with equatorial rotation velocities $< 100$ km s$^{-1}$), where the Am fraction is nearly 100\% \citep{1973ApJ...182..809A, 1989ApJS...70..623G}. The large majority of WOS stars in our sample seem to be slow rotators (Sec.~\ref{sec:binarity}, \ref{sec:periodspacing}), although we currently lack the spectroscopic data and/or asteroseismic inferences required to determine the distribution of $v \sin i$ and/or $P_{\rm rot}$ for the full sample.
Although the current sample sizes remain small and selection biases must be considered, the relatively large fraction of Am stars among the WOS may be noteworthy. While the TESS sample cannot be used to infer occurrence rates for the general stellar population, both the substantial Am-star fraction in the unbiased \textit{Kepler} sample and the high detection rate within the Am-star sample may suggest a possible connection between chemical peculiarity and the observed frequency structures. 
Alternatively, the WOS phenomenon may also depend on other stellar properties, such as rotation rate or a specific structural configuration associated with a narrow $T_{\rm eff}$ range.

\subsection{Gaia DR3 parameters}\label{sec:gaiadata}

The stellar parameters used in this study (see Table~\ref{tab:params}) are primarily taken from the Gaia Data Release~3 (DR3; \citealt{GaiaDR3}) astrophysical parameter catalogue. Effective temperatures and surface gravities are adopted from the GSP-Phot module, which derives atmospheric parameters by fitting the Gaia BP/RP low-resolution spectrophotometry with grids of stellar atmosphere models under the assumption of single-star spectra. Stellar luminosities ($L$) are taken from the FLAME module, which combines Gaia parallaxes with the atmospheric parameters from GSP-Phot and stellar evolution constraints to infer stellar radii and luminosities \citep{creevey2023}. We assume a representative uncertainty for $T_{\rm eff}$ of approximately $110\,\mathrm{K}$ for the effective temperatures, as used in \citet{duerfeldt2024}.

For five stars (KIC~5038228, KIC~5459805, KIC~9875566, KIC~10014548, and KIC~10154966), no GSP-Phot temperature estimate is available. In these cases, we adopt the primary-component temperature from the Gaia DR3 binary Multiple Star Classifier (MSC) solution. Surface gravities are taken from the TESS Input Catalog \citep{stassun2018} for all stars except KIC~5459805, for which the value from the Kepler Input Catalog \citep{brown2011} is used. For these objects, luminosities were derived independently from Gaia parallaxes using extinction-corrected absolute magnitudes and bolometric corrections following \citet{balona1994}. The adopted parameter values and associated confidence intervals are listed in Table~\ref{tab:params}. 

For all stars we calculate the absolute visual magnitudes ($M_V$) corrected for distance- and direction-dependent extinction. Observed V-band magnitudes are taken from the literature or estimated from Gaia DR3 photometry using the transformation of \citet{jordi2010}. Since Gaia DR3 gspphot parameters are unavailable for several stars, we instead use the geometric distances ($r_{\rm geo}$) and their associated uncertainties from the Bayesian inference catalogue of \citet{bailerjones2021}. For each star, the extinction profile $A_V (d)$ is extracted from the three-dimensional dust maps of \citet{lallement2019}, which provide extinction as a function of distance $d$ along specific lines of sight. The extinction values are interpolated at the median distance as well as at the lower and upper distance bounds. The extinction-corrected absolute magnitudes were then computed as
\[
M_V = V - 5\,\log_{10}(r_{\rm geo}\,[\mathrm{pc}]) + 5 - A_V .
\]
These values of $M_V$ are used below to estimate the fundamental radial mode frequency via the period--luminosity relation given in Equation~4 of \citet{barac2022}. We refer to Section \ref{sec:radialfundamental} for further details.
It is worth noting that a large fraction of stars in our sample are, or can be expected to be, in binary systems, and we cannot currently exclude systematic effects of unresolved multiplicity on the inferred stellar parameters (see next section for a discussion of binarity).

\subsection{Binarity}\label{sec:binarity}
Spectroscopic observations have been obtained for a significant fraction of our sample, primarily for the \textsc{TESS} targets and three \textit{Kepler} stars. The data were acquired using the FIES spectrograph at the Nordic Optical Telescope (NOT) and the Southern African Large Telescope (SALT) High Resolution Spectrograph (HRS) at the SALT. The SALT data have been processed with the \texttt{PySALT}\footnote{\url{http://pysalt.salt.ac.za/}}  pipeline \citep{Crawford_2010}. For 11 stars, multiple spectra were obtained, with temporal baselines ranging from several days to several years. As shown in Table\ref{tab:specinfo}, 19 stars in our sample are identified as binaries, which is not unexpected, as previous studies report that between 60\% and 90\% of Am stars are found in binary systems \citep[e.g.,][]{Smalley_2014}. 

Given this context, we did not undertake a detailed binary characterization in the present work. Instead, we assessed whether binarity could plausibly explain the observed pulsational features. 
All candidates appear to be single-line (SB1) systems. None of the spectra show signatures of a secondary component of comparable luminosity, indicating that the observed frequency structures are unlikely to be contaminated by a luminous companion. We also note that all stars with available spectra exhibit low to intermediate projected rotational velocities ($v \sin i \lesssim 50$~kms$^{-1}$), as expected for Am stars. The absence of detectable secondary spectral components further supports the conclusion that the low $v \sin i$ values, and by extension the chemical peculiarities, are intrinsic to the observed primary stars rather than caused by line blending with a similar-mass companion. A detailed spectroscopic analysis will be presented in a follow-up paper. A summary of binarity indicators, whether drawn from the literature or identified via preliminary inspection of our spectroscopic data, is provided in Table~\ref{tab:specinfo}.

\subsection{Catalogue crossmatching and spectral classification}\label{sec:catalogues}
To assess the nature of our sample stars and to search for evidence of binarity and chemical peculiarity, we conducted a crossmatch against several major astronomical catalogues using the \texttt{astroquery} Python interface.

The following catalogues were queried:

\begin{itemize}
\item The General Catalogue of Ap and Am stars by \citet{renson2009}, used to identify chemically peculiar (CP) classifications.
\item The catalogue of LAMOST-identified Am stars by \citet{tian2023}, based on spectroscopic classifications from the LAMOST DR7 low-resolution survey.
\item The spectroscopic classification catalogue of \textit{Kepler} stars by \citet{frasca2016}, based on LAMOST observations, which includes fundamental parameters, activity indicators, and binarity flags.
\item The Ninth Catalogue of Spectroscopic Binary Orbits (SB9; \citealt{pourbaix2004}) for confirmed binary systems.
\item The binary catalogue by \citet{murphy2018}, which identifies companions via phase modulation of p-mode pulsations in \textit{Kepler} stars. This method is sensitive to binaries with orbital periods between approximately 100 and 1500\,d, constrained by the mission duration and the coherence of the pulsations.
\item The Gaia RUWE astrometric binary catalogue \citep[e.g.,][]{2021ApJ...907L..33S}.

\end{itemize}

The results are summarized in Table~\ref{tab:specinfo}, where the designation “PB” refers to pulsation binaries identified via the \citet{murphy2018} method. This catalogue complements the spectroscopic information by enabling detection of longer-period systems that may not show detectable radial velocity variations over the timescales sampled by our observations.

Information about possible binarity with longer orbital periods can also be gathered from the Gaia Renormalized Unit Weight Error (RUWE) parameter \citep[see][for a recent review, including caveats]{2024NewAR..9801694E}. In brief, a RUWE value in excess of 1.4 is indicative of astrometric binarity \citep[e.g.,][]{2021ApJ...907L..33S} although a low RUWE does not guarantee that the star is single. The corresponding values listed in Table~\ref{tab:specinfo} corroborate some of the detections of binarity from other methods, but also indicate some other binary candidates not revealed otherwise.

As noted in section \ref{sec:tessdata}, a subset of our targets are classified as chemically peculiar, particularly Am stars, based on the Renson, Tian et al., and Frasca et al. catalogues. Several targets also appear in SB9 with confirmed orbital solutions, while others show no recorded binarity.

\begin{figure*}
\centering
\includegraphics[width=1\linewidth]{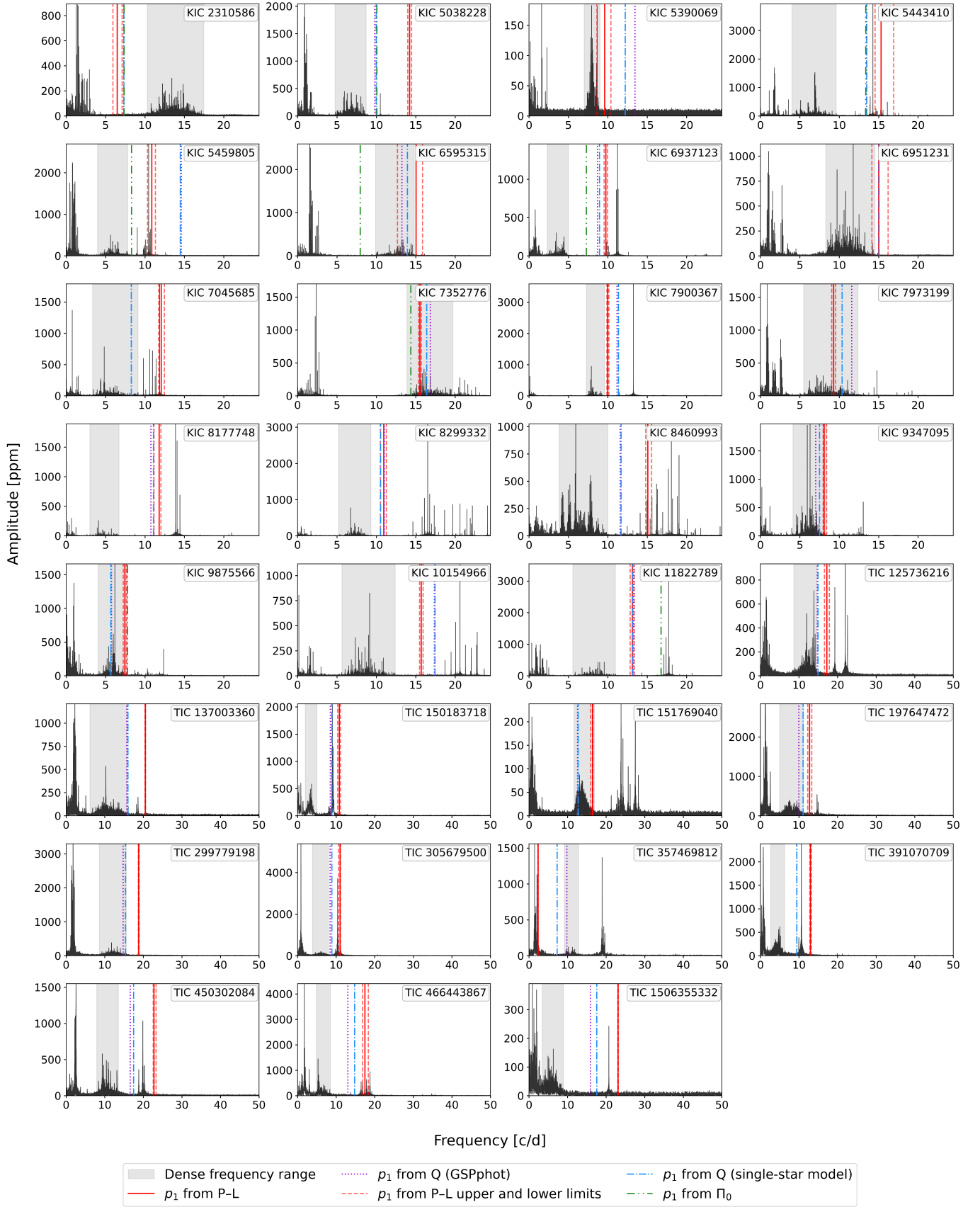}
\caption{Amplitude amplitude spectra of the type I stars.  
The region of agglomerated peaks is shaded in each panel, and predicted fundamental radial mode frequencies are overplotted based on Section~\ref{sec:radialfundamental}.  
Where available, multiple estimates are shown per star. Note that for Kepler stars we plot the data up to 24.47~d$^{-1}$ which corresponds to the Nyquist frequency.}
\label{fig:overview4a}
\end{figure*}

\begin{figure*}[ht!]
\centering
    \includegraphics[width=1\linewidth]{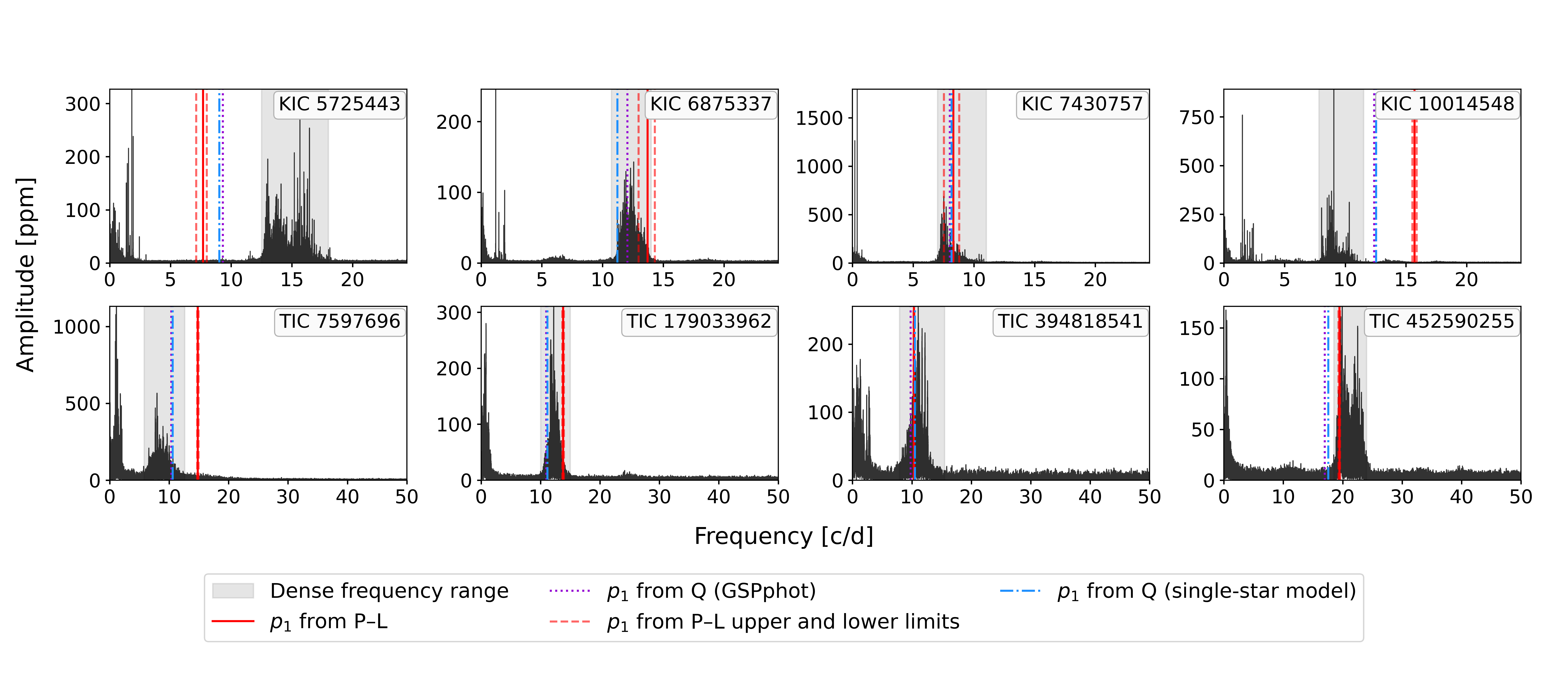}
\caption{Same as Fig.~\ref{fig:overview4a}, but for the type II stars.}
\label{fig:overview4b}
\end{figure*}

Importantly, several stars in our sample may be unresolved, relatively wide binaries, particularly those for which the fundamental mode frequencies derived from different methods show discrepancies (see Figures~\ref{fig:overview4a},\ref{fig:overview4b} and Sec.~\ref{sec:discussion}). Such mismatches may indicate unresolved companions affecting the luminosity estimate or contamination from the light curves of multiple stars. This interpretation is consistent with prior studies suggesting elevated binarity rates among chemically peculiar stars, particularly Am stars \citep[e.g.][]{murphy2018}. Stars that were identified as binaries by any method are marked with an asterisk preceding their name in Table~\ref{tab:specinfo}.

\begin{table*}[ht!]
\caption{Stellar identifiers and classification information for the sample stars, including KIC, TIC and HD designations, variability type, binarity diagnostics, the Gaia re-normalised unit weight error (RUWE), and spectral type where available. Variability types are defined as follows: Type I stars exhibit g modes, a dense intermediate-frequency hump, and conventional p-mode
frequencies; Type II stars display g modes and/or low-frequency modulation and an anomalously dense region of peaks at higher frequencies. The Binarity column indicates the method used to conclude binarity: ASTR (astrometric, based on elevated RUWE > 1.4), SPEC (spectroscopic), and PB (pulsation binary, following Murphy et al.). “uncertain” denotes inconclusive evidence, and “none” indicates no indication of binarity from the available data. Stars marked with an asterisk (*) show evidence of binarity from at least one method.}
\label{tab:specinfo}
\centering
\begin{tabular}{clcllll}
\hline\hline
 & Name & HD & var. Type & Binarity & RUWE & Spectral type \\
\hline
* & KIC 2310586     &       & Type I & ASTR & 11.48 &  \\
* & KIC 5038228     &       & Type I & SPEC (this study)    & 0.856 & Am$^{a,b}$ \\
& KIC 5390069     &       & Type I & none  &  0.991 & \\
& KIC 5443410     &       & Type I & uncertain SPEC$^{b, c}$ & 1.065 & Am$^{b}$ \\
* & KIC 5459805     &       & Type I & PB$^{d}$           & 1.196 &  \\
& KIC 5725443     &       & Type II & none & 1.039 &  \\
& KIC 6595315     &       & Type I & none            &  1.07 & \\
* & KIC 6875337     &       & Type II & ASTR, uncertain SPEC$^{c,e}$     &  8.758 & \\
& KIC 6937123     &       & Type I & none           &   0.959 & \\
& KIC 6951231     &       & Type I & none      & 1.115 & Am$^{b}$ \\
& KIC 7045685     &       & Type I & uncertain PB$^{d}$ &  1.014 & \\
& KIC 7352776     &       & Type I & none & 0.945 & Am$^{b}$ \\
* & KIC 7430757     &       & Type II & SPEC$^{f}$ & 1.009 & Am$^{b}$ \\
* & KIC 7900367     &       & Type I & SPEC (this study)    & 0.934 &  \\
* & KIC 7973199     &       & Type I & ASTR & 1.95 & Am$^{b}$ \\
& KIC 8177748     &       & Type I & none & 0.957 &  \\
* & KIC 8299332     &       & Type I & PB$^{d}$, ASTR, uncertain SPEC$^{c,e}$ & 2.342 & Am$^{b}$ \\
& KIC 8460993     &       & Type I & none     &  0.866 & \\
& KIC 9347095     &       & Type I & uncertain SPEC$^{c}$     &  0.918 & \\
& KIC 9875566     &       & Type I & none      & 0.96 &  \\
& KIC 10014548    &       & Type II & uncertain SPEC$^{c, e}$     &  0.979 & \\
& KIC 10154966    &       & Type I & none & 1.061 & Am$^{b}$ \\
& KIC 11822789    &       & Type I & none & 1.022 & Am$^{b}$ \\
& TIC 7597696     & HD 27079 & Type II & none    & 0.98 & Am$^{a}$ \\
*& TIC 125736216   & HD 23488 & Type I & SPEC$^{g}$, ASTR      & 6.652 & Am$^{a}$ \\
* & TIC 137003360   & HD 112515& Type I & SPEC (this study)  & 0.882 & Am$^{a}$ \\
* & TIC 150183718   & HD 78388 & Type I & ASTR          & 1.666 & Am$^{a}$ \\
& TIC 151769040   & HD 97160 & Type I & none          & 1.086 & Am$^{a}$ \\
* & TIC 179033962   & HD 27230 & Type II & SPEC (this study), ASTR  & 1.609 & Am$^{a}$ \\
* & TIC 197647472   & HD 208139& Type I & SPEC (this study), ASTR  & 5.24 & Am$^{a}$ \\
& TIC 299779198   & HD 16232 & Type I & none           & 0.899 & Am$^{a}$ \\
* & TIC 305679500   & HD 201032& Type I & SPEC$^{h}$          & 1.059 & Am$^{a}$ \\
* & TIC 357469812   & HD 107340& Type I & ASTR$^{i}$ & 1.799 & Am$^{a}$ \\
* & TIC 391070709   & HD 187258& Type I & SPEC$^{j}$  & 1.254 & Am$^{a}$ \\
& TIC 394818541   & HD 17784 & Type II & none           & 0.858 & Am$^{a}$ \\
* & TIC 450302084   & HD 86167 & Type I & SPEC (this study)  & 0.877 & Am$^{a}$ \\
& TIC 452590255   & HD 83094 & Type II & none           & 1.025 & Am$^{a}$ \\
* & TIC 466443867   & HD 2523A & Type I & SPEC (this study), ASTR  & 6.271 & Am$^{a}$ \\
* & TIC 1506355332  & HD 158251A & Type I & SPEC (this study) & 0.91 & Am$^{a}$ \\
\hline
\end{tabular}
\tablefoot{
$^{a}$ Renson catalogue \citep{renson2009}. \\
$^{b}$ \citet{tian2023}. \\
$^{c}$ RV shifts are reported in the literature, but uncertainties are too large to conclude binarity. \\
$^{d}$ \citet{murphy2018}\\
$^{e}$ \citet{frasca2016}\\
$^{f}$ \citet{2019RAA....19...64Q}, see Section~\ref{sec:kic7430757}.\\
$^{g}$ \citet{Torres_2021}\\
$^{h}$ \citet{Tanner_1949}\\
$^{i}$ Classified as an SB2 system in the Gaia DR3 non-single star catalogue; however, no supporting data are available for inspection.\\
$^{j}$ \citet{Ginestet_2003}\\

}

\end{table*}

\section{Data analysis}

\subsection{Frequency spectra and the agglomerated power region}\label{sec:freanalyses}

For all WOS in our sample, we computed Fourier transforms of the light curves using standard discrete Fourier techniques. The morphology of the amplitude spectra can broadly be classified into Type I and II. In the first group, Type I, we observe clear low-frequency g modes typical of $\gamma$ Doradus pulsators, alongside high-frequency p modes characteristic of $\delta$ Scuti stars. Between these two domains lies an intermediate frequency region containing a densely packed cluster of peaks, the agglomerated region, which often shows repeating or structured features (see Fig.~\ref{fig:WOS1_example}). Type II stars exhibit either low-frequency rotational modulation or low-order g modes, but no clear high-frequency p modes (see Fig.~\ref{fig:WOS2_example}). Instead, they show a dense, prominent hump of power at higher frequencies that is too rich in structure to be explained by standard p-mode pulsations alone, as, e.g., shown by \citet{Gautam2025}. Further, as highlighted in Fig.~\ref{fig:WOS2_example}, we also observe additional power excesses near frequencies corresponding to approximately half the value of the agglomerated hump, as well as at frequencies near the hump plus this half-value offset.  The amplitude spectra for all stars are displayed in Figs.~\ref{fig:overview4a} and ~\ref{fig:overview4b} with shaded regions indicating the frequency intervals interpreted as anomalously dense or “agglomerated” power clusters.

\subsection{Estimation of the radial fundamental mode frequency}\label{sec:radialfundamental}

The purpose of estimating the frequency of the fundamental radial mode is to assess the nature of the dense group of agglomerated frequencies observed in the intermediate region of the amplitude spectrum. If these frequencies lie below the fundamental radial mode, and if they are indeed due to pulsations, they must correspond either to  g modes with low-to-intermediate radial order, to mixed p-g modes in evolved stars, or perhaps to some other kind of pulsation. Conversely, frequencies above the fundamental radial mode would, under a pulsational interpretation, have to correspond to p modes, mixed modes, very low-order g modes, depending on the evolutionary stage of the star.

Establishing the frequency of the fundamental radial mode is therefore essential for constraining the physical origin of this spectral feature.

To this end, we applied three complementary approaches. First, we used the period--luminosity (P--L) relation from \citet{barac2022}, based on a fit to TESS and Gaia measurements of $\delta$~Scuti stars:
\begin{equation}
M_V = -3.01 \log(P/{\rm d}) - 1.40.
\end{equation}
Here, $P$ is the period of the fundamental radial mode in days. This empirical relation allows us to predict the frequency of the fundamental radial mode using the absolute visual magnitude ($M_V$), which we compute from \textit{Gaia} DR3 parallaxes and photometry (see Section~\ref{sec:gaiadata}).

Second, for the nine stars that exhibit coherent g-mode ridges, we determined the asymptotic period spacing parameter $\Pi_0$, as described in Section~\ref{sec:periodspacing}. Together with the stellar parameters, this value provides a strong diagnostic of the stellar evolutionary stage. We compared the measured $\Pi_0$ to a grid of MESA stellar models \citep{Paxton2011, Paxton2013, Paxton2015, Paxton2018, Paxton2019, Jermyn2023}, release r24.08.01, with pulsation properties calculated using GYRE v.7.2.1 \citep{Townsend2013, Townsend2018}, to infer the mass and internal structure of the star. Our aim was to obtain a reasonable first estimate without carrying out detailed modelling of these stars. We computed evolutionary models for stars with masses from 1.4 $\rm{M}_\odot$ to 2.0 $\rm{M_\odot}$ in steps of 0.02 $\rm{M_\odot}$ and metallicity ranging from $Z=0.01$ to $Z=0.03$ in steps of 0.0025. The GS98 solar metal mixture reference value is used, i.e. $Z_\odot = 0.02$ \citep{GS98}. We used the OPAL opacity tables \citep{Iglesias1993} and determined the convective/radiative boundaries using the Ledoux criterion. We used \citet{Cox1968}'s prescription for mixing length theory, with $\alpha_{\rm{MLT}} = 1.8$. Convective-core overshooting was activated with $f_{\rm{ov}}=0.017$, a best-fitting value for $\gamma$ Dor stars taken from \citet{Mombarg2021}. We included diffusive mixing in the envelope as $D_{\rm{ext}} = 10\,\rm{cm}^2/s$. The mesh resolution was set with $\texttt{mesh\_delta\_coeff} = 0.4$ and $\texttt{max\_dq} = 0.001$. Further details on the MESA input parameters can be found in the MESA inlist in Appendix \ref{sec:MESAinlist}. We used a custom saving routine implemented in MESA to save stellar profiles at regular intervals\footnote{available on GitHub: \url{https://github.com/Durfeldt/mesa_saving_routine.git}}. For each MESA profile in our model grid, we computed the theoretical period spacing by integrating the Brunt-Väisälä frequency over the stellar radius:
\begin{equation}
    \Pi_0 = 2\pi^2\int_R \frac{N}{r} dr,
\end{equation}
\noindent where $N$ is the Brunt-Väisälä frequency.

\begin{figure*}
\centering
\includegraphics[width=0.49\textwidth]{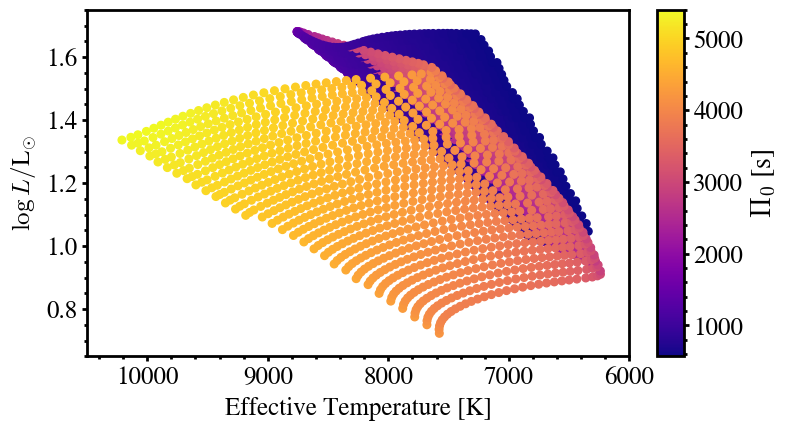}
\includegraphics[width=0.49\textwidth]{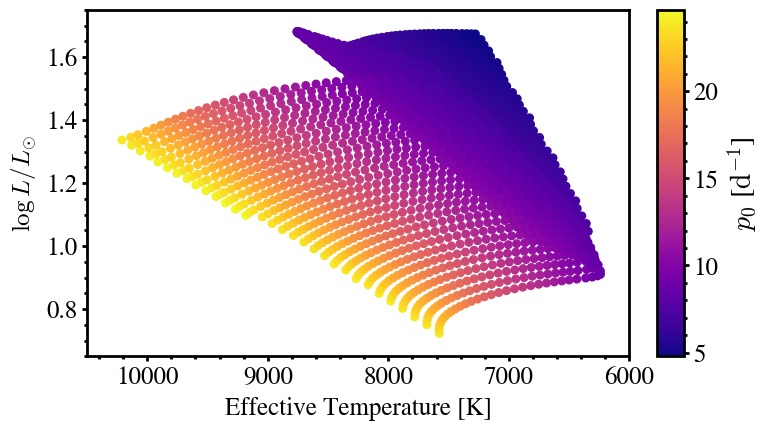}
\caption{Hertzsprung–Russell diagrams from our MESA model grid for $Z=0.01$. The evolutionary tracks cover the range 1.4 to 2.0 $M_\odot$. \textit{Left:} Color-coded $\Pi_0$ values, indicating the asymptotic period spacing for g modes as a function of stellar parameters. \textit{Right:} Corresponding fundamental radial mode frequencies. These diagrams provide theoretical guidance for interpreting the observed oscillation spectra.}
\label{fig:hrd_puls}
\end{figure*}

The values of $\Pi_0$ are shown in the left panel of Fig.~\ref{fig:hrd_puls} for a subset of our grid, corresponding to the tracks with $Z=0.01$. Each point corresponds to a profile along a MESA evolutionary track. The asymptotic period spacing decreases as the star evolves on the main sequence. 

The frequency of the fundamental radial mode, $p_1$, was computed for all MESA profiles by running GYRE. The template inlist is provided in Appendix \ref{sec:GYRE_inlist}. We used the \texttt{VACUUM} outer boundary option, ignored rotation for simplicity, and scanned the range of frequencies $1-40$\,d$^{-1}$ with a grid of 1000 points. The fundamental radial mode frequencies are shown in the right panel of Fig.~\ref{fig:hrd_puls} for the subset of our grid. It can be seen that $p_1$ shifts towards lower frequencies as the star evolves on the main sequence. 

We determined the best MESA model that reproduces the observed $T_{\rm{eff}}$, $L$, and $\Pi_0$ for each star using a custom-made interactive visualization tool that allows us to explore the grid of models and retrieve information on the values of the parameters used to find an appropriate model. Starting from the observed stellar parameters, we identified the MESA profile that lies closest in the HR diagram with an appropriate $\Pi_0$. Using this best fit, we report the frequency of $p_1$ extracted for the given profile. The outcome of our modeling is summarized in Table \ref{tab:MESAFit}. We emphasize that the aim was not to perform a detailed seismic analysis of individual stars, but rather to obtain an approximate location of the fundamental radial mode. 

Finding a model that simultaneously reproduces all three observed parameters has proven challenging. The observed $\Pi_0$ values are relatively low, indicating that the stars are near the end of the main sequence, while their effective temperatures remain comparatively high. In our modeling, we therefore prioritized achieving a good match to $\Pi_0$, as this quantity is most directly linked to the determination of $p_1$, and allowed for greater flexibility in $T_{\mathrm{eff}}$ and luminosity. This approach is motivated by the possibility that some targets are in binary systems, which may affect the reliability of the stellar parameters derived from \textit{Gaia} DR3. To reproduce the observed $\Pi_0$ values, the models generally require low metallicity for most targets. 
This is counter-intuitive, as Am stars generally have high atmospheric metal abundances. Am star surface abundances do not, however, necessarily reflect interior composition. 
Future, more detailed modeling will explore a broader parameter space, including variations in overshooting ($f_{\rm ov}$), mixing length parameter, envelope mixing,  and helium abundance, in order to better reconcile all observables.

\begin{table*}
\caption{Best fits from the MESA model grid for each star with an observed $\Pi_0$.}
\label{tab:MESAFit}
\centering
\begin{tabular}{lcrllcrc}
\hline\hline
Name & $T_{\rm{eff}}$ [K] & $L_\odot$ & $Z$ & $M$ [M$_\odot$] & $\Pi_0$ [s] & $p_1\, [\rm{d}^{-1}]$ & $X_c$ \\
\hline
KIC 2310586 & 7000 & 20.0 & 0.01 & 1.74 & 3358 & 7.346 & 0.03 \\
KIC 5038228 & 7010 & 13.5 & 0.0125 & 1.64 & 4038 & 10.064 & 0.19 \\
KIC 5443410 & 7200 & 9.8 & 0.01 & 1.5 & 3921 & 13.367 & 0.28 \\
KIC 5459805 & 7020 & 19.1 & 0.0175 & 1.86 & 4462 & 8.298 & 0.17 \\
KIC 6595315 & 6700 & 15.1 & 0.01 & 1.6 & 2945 & 7.950 & 0.01 \\
KIC 6937123 & 7000 & 21.4 & 0.01 & 1.76 & 3396 & 7.285 & 0.03 \\
KIC 7352776 & 6950 & 7.2 & 0.01 & 1.4 & 3731 & 14.373 & 0.29 \\
KIC 9875566 & 7180 & 23.4 & 0.02 & 2.0 & 4776 & 7.798 & 0.17 \\
KIC 11822789 & 7210 & 6.9 & 0.01 & 1.4 & 3825 & 16.751 & 0.38 \\
\hline
\end{tabular}
\end{table*}

Finally, we applied a third method based on the classical pulsation constant, $Q$, adopting $Q = 0.033$ for the radial fundamental mode \citep{Fitch1981}, which relates the period of radial modes to stellar surface parameters. The relation is given by:
\begin{equation}
\log Q = -6.454 + \log P + 0.5 \log g + 0.1 M_{\rm bol} + \log T_{\rm eff},
\end{equation}
where $P$ is the pulsation period (in days), $\log g$ the surface gravity, $M_{\rm bol}$ the bolometric magnitude, and $T_{\rm eff}$ the effective temperature (in kelvin). This method has been employed in a range of asteroseismic contexts \citep[e.g.][]{breger1993, zwintz2020, lovekin2017}, and most recently by \citet{duerfeldt2024} in the analysis of chemically peculiar metallic-lined Am and Fm pulsators. For stars with well-determined atmospheric parameters, this provides an additional, albeit not entirely independent, estimate of the radial mode frequency.

Together, these three methods--empirical (P--L), structural ($\Pi_0$-based modeling), and parameter-based ($Q$)---allow us to cross-validate the fundamental radial mode frequency and to assess the plausibility of interpreting the agglomerated intermediate-frequency group as g-mode or mixed-mode pulsations. The values determined can be found in Table \ref{tab:fundamental}.

Figures~\ref{fig:overview4a} and \ref{fig:overview4b} provide an overview of the observed pulsation spectra. For each star, we show the  amplitude spectrum with the agglomerated peak region highlighted. Overlaid markers indicate the predicted fundamental radial mode frequencies as derived from the three methods described above. Not all stars in the sample have complete parameter sets: in several cases, either g-mode ridges were not observed or the available data did not permit secure identification of $\ell = 1$ or $\ell = 2$ modes needed for $\Pi_0$ determination. Nonetheless, the ensemble visualization allows a direct assessment of whether the intermediate-frequency groups lie below the fundamental radial mode, as expected for g- or mixed-mode pulsations.

 For stars with multiple estimates from different methods, the agreement between approaches serves as a qualitative check on their robustness. In cases where the predicted fundamental radial mode frequencies differ significantly between methods, this may point to unresolved binarity, as a luminous companion can bias the inferred luminosity or effective temperature, particularly when derived from photometric data. Given this goal, we do not pursue a rigorous uncertainty analysis for the derived values.
Across the sample, the agglomerated power lies below the estimated fundamental radial mode in approximately 70\% of the stars. In about 18\% of cases the two overlap or fall close together, while in the remaining 12\% the fundamental radial mode is found below the agglomerated region.

%\begin{figure}
%   \centering
%\includegraphics[width=0.5\textwidth]{HRD_f_radial_PL.png}
%\caption{Hertzsprung-Russell diagram showing the location of stars in our sample, colour-coded by the estimated fundamental radial frequency from the period-luminosity relation.}
%\label{fig:HRD_f_radial_PL}
%\end{figure}

\subsubsection{Determining the period spacing for g modes}\label{sec:periodspacing}

For the \textit{Kepler} sample, all but one star exhibit signals consistent with low-frequency g modes, enabling mode identification based on their period spacing patterns. The overarching goal of the analysis described in this subsection is to determine the buoyancy radius, $\Pi_0$, from g-mode period spacing patterns. This value is then combined with stellar parameters from \textit{Gaia} ($T_{\rm eff}$ and luminosity) and a grid of stellar models computed with MESA (Sec.~\ref{sec:radialfundamental}) to estimate the frequency of the fundamental radial mode.

The frequency range covering the g-mode signals was manually selected for each star to include the low frequency signals and exclude those in the agglomerated power region. Frequencies and amplitudes were extracted using standard iterative prewhitening, applied to the first 60 peaks. While amplitude thresholds were applied at a later stage, no more than 30 peaks were used in the final analysis for any given star. An additional constraint in the prewhitening process required that any new peak be sufficiently resolved from existing ones, with a frequency separation of at least three times the inverse time baseline (i.e. $\sim$0.002 d$^{-1}$). This ensures adequate resolution while remaining well below the expected period spacing values.

\begin{figure*}
\centering
\includegraphics[width=0.95\textwidth]{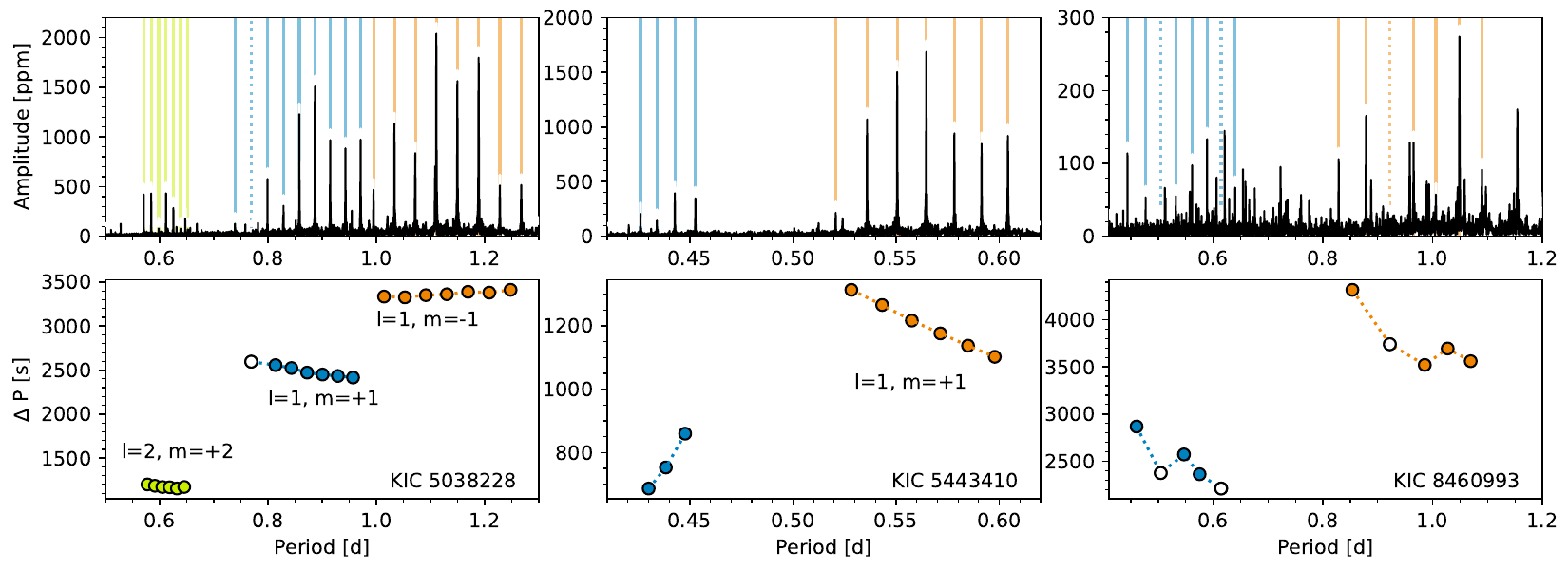}
\caption{Period spacing patterns of g modes for three stars in our sample. The upper panels show a portion of the photometric amplitude spectrum (black), with vertical lines indicating signals used to construct the period spacing patterns plotted below. The lower panels show the period spacing patterns for the identified g mode ridges, with colours corresponding to the different groups as indicated in the top panels. The signals for a given ridge are presumed to be of consecutive radial orders. Whenever a signal is missing from the photometry but is predicted by the period spacing pattern, a vertical dotted line marks the expected location. Correspondingly, an open circle in the bottom panel indicates half of the period difference between the two adjacent photometric periods. The left panel shows the ridges detected for KIC 5038228, and are labelled according to the mode identification reported in \citet{GangLi2020} which are consistent with our identification. The same is plotted in the middle panels for KIC 5443410, along with the mode identification for the more populous g mode ridge which was used to determine $\Pi_{0}$. The right panel shows the measurements for KIC 8460993 -- here the mode identification is not secure.  }
\label{fig:gmode_ridges}
\end{figure*}

Once the low-frequency signals had been identified, the \texttt{FLOSSY}\footnote{\url{https://github.com/IvS-KULeuven/FLOSSY}} package was used to manually identify period spacing patterns. Groups of signals were characterised by a common average period, an average spacing, and a period spacing slope in the $\Delta P - P$ diagram, following the approach of \citet{Garcia2022}. Depending on the star, zero to four distinct groups were identified. The largest group contained 13 peaks, but most groups comprised about five signals. Three examples are shown in Fig.~\ref{fig:gmode_ridges} for a slow rotator with three well-identified groups having very minor slopes in the $\Delta P - P$ diagram (KIC 5038228), a faster rotator as indicated by the higher slopes in the $\Delta P - P$ diagram with two groups (one of which was confidently identified; KIC 5443410), and a star with two groups, neither of which were confidently identified (KIC 8460993). 

The working assumption behind identifying such patterns is that the signals belong to g modes of common spherical degree $\ell$ and azimuthal order $m$—that is, they share the same mode geometry and direction of propagation relative to stellar rotation—while each signal corresponds to a different radial order $n$. Each identified group was then analysed using the \texttt{morse}\footnote{\url{https://github.com/schristophe/morse}} Python package \citep{Christophe2018}, which estimates $\Pi_0$ by fitting observed periods to theoretical expectations. All combinations of $\ell = 1$ ($m = -1, 0, 1$) and $\ell = 2$ ($m = -2, -1, 0, 1, 2$) were tested. Negative $m$ values correspond to prograde modes. In many cases, the most plausible ($\ell$, $m$) combinations could be inferred from the average period, spacing, and slope in the $\Delta P - P$ diagram, as discussed in \citet{GangLi2020}.

We obtained reliable $\Pi_0$ values for nine of the 23 stars in the \textit{Kepler} sample (see Table~\ref{tab:fundamental}). For four of these, $\Pi_0$ values have also been reported by \citet{GangLi2020}, and we find agreement to within 5\%. Although \texttt{morse} also provides estimates of the near-core rotation frequency, this parameter was generally poorly constrained due to the intrinsically slow rotation of most of the stars in our sample. 
 
\subsection{Combination frequency search}\label{sec:combi}

%To test whether the observed frequency excess could be explained through non-linear mode coupling, we performed an automated search for combination frequencies between the detected g-mode and p-mode regions. 

%\textbf{These 1st 2 paragraphs to be merged.}
One star in the sample (KIC~7430757) shows no low-frequency g modes, and several others lack high-frequency p modes. This immediately rules out non-linear mode coupling between \textit{observable} parent g and/or p modes as a \textit{universal} explanation for the agglomerated power clusters, even before testing whether combination frequencies match their location or structure. We note, however, that the possibility of unobserved parent modes is discussed in Section~\ref{sec:wojtek}. We performed both automated and manual searches for linear combination frequencies to assess whether the agglomerated structures in individual stars could be partially explained by nonlinear mode interaction. For the automated search, we used the low-frequency peaks extracted via the iterative pre-whitening process described in Sect.~\ref{sec:periodspacing}. Rather than restricting the search to peaks defining a clear period-spacing pattern (which is absent in some stars), we included all peaks with signal-to-noise ratio greater than 4 \citep{Breger1999}, where S/N was computed as the amplitude relative to the median in a 0.2~d$^{-1}$ window. Up to 60 peaks were considered per star. We computed simple two-term combinations of the form
\[
\nu_{\mathrm{comb}} = N_1 \nu_i \pm N_2 \nu_j,
\]
with integer coefficients $N_1, N_2 \leq 2$. Following the approach of \citet{Papics_2012}, we assumed that if simple ($\nu_i \pm \nu_j$) combination frequencies are not clearly present, then matches involving higher-order or more complex combinations are likely to arise by chance rather than reflecting genuine nonlinear mode coupling. In particular, combinations with large integer coefficients (e.g., $N_1\nu_i \pm N_2\nu_j$ with $N_1,N_2>3$) or involving a third parent mode (e.g., $N_1\nu_i \pm N_2\nu_j \pm N_3\nu_k$) were not considered if simple combination frequencies were absent, since the probability of accidental matches increases rapidly with the number of terms and coefficients, especially in such densely populated frequency regions as the agglomerated regions of our WOS.

While some combination frequencies were identified in the low-frequency domain, as commonly observed in g-mode pulsators \citep[e.g.][]{Kurtz2015}, these combinations do not systematically reproduce the location or morphology of the agglomerated clusters in most stars. For stars exhibiting high-frequency p modes, we extended the search to g/p combinations as well as p/p difference frequencies. However, these also fail to reproduce the intermediate-frequency agglomerations in a consistent manner; in particular, p/p difference frequencies generally lie well below the agglomerated region. 

In addition to the automated search, we performed a manual exploration of selected stars, allowing parent modes not only from the g- and p-mode regimes but also from within the agglomerated region itself. In this analysis, we restricted the search to low-order combinations with $N_1 = N_2 = 1$ and required that the number of identified simple combination frequencies within the agglomerated region exceed the number of assumed parent modes. This criterion was adopted to avoid over-fitting dense spectra with arbitrary high-order coefficients and to provide a minimal quantitative constraint.

Under these restrictions, most stars still do not show compelling evidence that the agglomerated region is dominated by simple combination frequencies. However, KIC~5443410 represents a notable exception (Sect.~\ref{sec:kic5443410}). In this star, one of the main multiplets in the agglomerated region appears to behave as an independent structure, while several surrounding repeating patterns can be reproduced as linear combinations between this dominant multiplet and selected g modes. Interestingly, even relatively low-amplitude g modes participate in these combinations, indicating that the nonlinear interaction does not scale trivially with the observed mode amplitudes. This behaviour further supports the interpretation that the peaks forming the high-amplitude multiplet in the agglomerated region (Figs.~\ref{fig:WOS2_example} and \ref{fig:overview4a}) correspond to genuine pulsation modes rather than purely geometric or surface-modulation effects. We identify at least one additional star (KIC~9347095) exhibiting qualitatively similar behaviour, although the correspondence is less pronounced than in KIC~5443410. Taken together, these results indicate that nonlinear mode interaction can shape parts of the agglomerated structures in certain stars, but it does not provide a general explanation for the agglomerated modes phenomenon. In particular, even in KIC~5443410, the organising multiplet itself cannot be reproduced as a simple combination of lower-frequency modes, leaving the origin of the characteristic spacing and excitation mechanism unresolved.

\subsection{Implications for mode classification}

Many, but not all, of the WOS in our sample are hybrid pulsators, exhibiting both $\gamma$~Dor-type g modes and $\delta$~Sct-type p modes. Determining the location of the fundamental radial p mode allows us to assess the nature of the agglomerated peaks seen in the intermediate frequency range. If these peaks are caused by stellar pulsations, then their position relative to the fundamental radial mode provides key information: peaks below the fundamental p-mode frequency can be interpreted as intermediate- to low-radial-order g modes or, in more evolved stars, as mixed modes.

We note that g-mode pulsations are detected in nearly all stars in our sample. In approximately 70\% of cases, the agglomerated power lies clearly below the estimated fundamental radial p mode, supporting a g mode or mixed-mode interpretation. In about 18\% of stars, the agglomerated region overlaps with or lies near the predicted fundamental p-mode frequency, but the methods used to estimate the mode yield conflicting results, some placing it just below, others just above the cluster, leading to inconclusive classification. In the remaining 12\% of stars, the fundamental p mode appears to lie below the agglomerated region. If the frequency estimates are correct, this would suggest a p-mode or mixed-mode nature, or potentially a non-pulsational origin. However, inaccuracies in the fundamental mode estimate, such as those arising from unresolved binarity, may also be responsible for this discrepancy.

Even in cases where the frequency domain strongly suggests a g-mode nature, this identification alone does not resolve the question of mode excitation. In $\gamma$~Dor stars, high-radial-order g modes are thought to be excited through the convective blocking (or convecting shunting) mechanism at the base of the envelope convection zone \citep{guzik2000, dupret2004}. In $\delta$~Scuti stars, pulsation driving is instead dominated by the classical $\kappa$ mechanism \citep[e.g.,][]{Pamyatnykh1999}, with an additional contribution from turbulent pressure in the hydrogen and helium ionisation zones \citep{houdek2008, antoci2014, antoci2019}. The frequencies of the agglomerated regions often lie between these domains, raising the possibility that neither of the classical excitation mechanisms is effective in this regime. One possible explanation is that chemical stratification, common in Am stars due to atomic diffusion and radiative levitation, may alter the opacity profile in a way that enables additional or modified excitation mechanisms. However, such effects are not included in current models. Moreover, a standard pattern of p or g modes does not account for the strikingly regular ridge structures seen in the \'echelle diagrams (Section~\ref{sec:echelle_ridges}), which suggest that additional modulation or structural effects may be at play. We refer to Section~\ref{sec:MAD} for more details on the excitation of pulsations. 
In general, mode identification in A- and F stars remains challenging, not only because of their often complex pulsation spectra, but also because the mode-selection mechanism is still poorly understood for both g and p modes \citep[e.g., ][]{Smolec2014, Bowman2018}. Effects such as rotation \citep[e.g.,][]{GangLi2020}, mixing \citep[][]{antoci2019, duerfeldt2024, Berry2025}, magnetic fields \citep[][]{Lecoanet2022, duerfeldt2026} may contribute to this diversity, with some stars showing rich spectra and others exciting only a small subset of the modes predicted to be unstable, possibly because additional modes remain below the detection threshold.

\subsection{\'Echelle analysis}\label{sec:echelle_ridges}

To further investigate the structure of the agglomerated frequency region, we employed autocorrelation analysis and constructed frequency \'echelle diagrams. The autocorrelation functions of the power spectra often revealed recurring splittings with structures consistent with quasi-regular spacing of peaks. These spacings, which we refer to as $\delta f$, were used as trial modulo values in the \'echelle diagrams. An \'echelle diagram is constructed by folding the pulsation frequencies modulo a chosen frequency spacing, allowing regularly spaced modes to align in near-vertical ridges and thereby making frequency patterns easier to identify \citep[e.g., ][]{Suarez2010, bedding2023}.

\begin{figure}
   \centering
\includegraphics[width=0.5\textwidth]{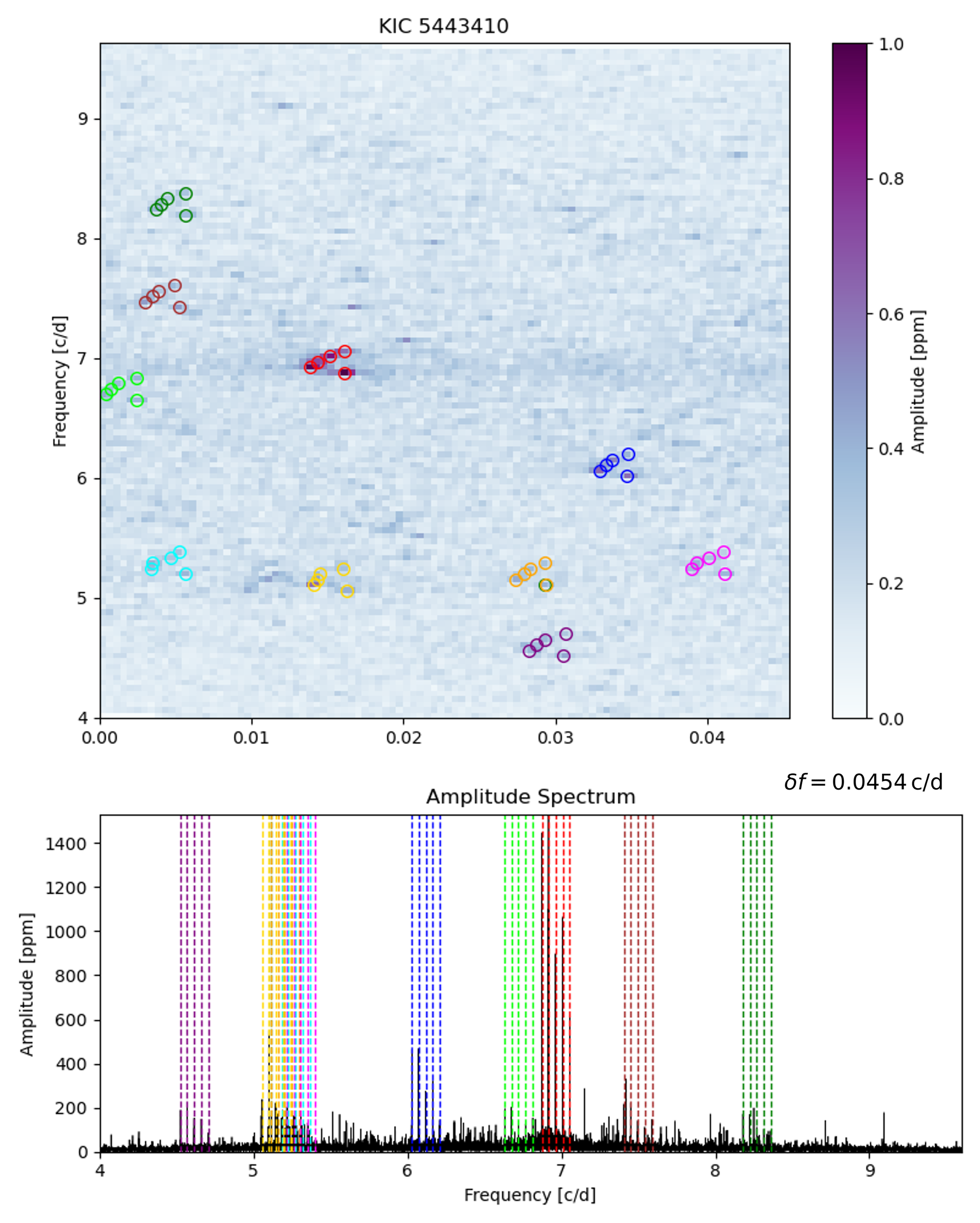}
\caption{Upper panel: \'Echelle diagram for KIC~5443410 with identified ridges or recurring patterns. Lower panel: Corresponding amplitude spectrum, where ridge components are highlighted.}
\label{fig:echelle_KIC5443410}
\end{figure}

\begin{figure}
   \centering
\includegraphics[width=0.5\textwidth]{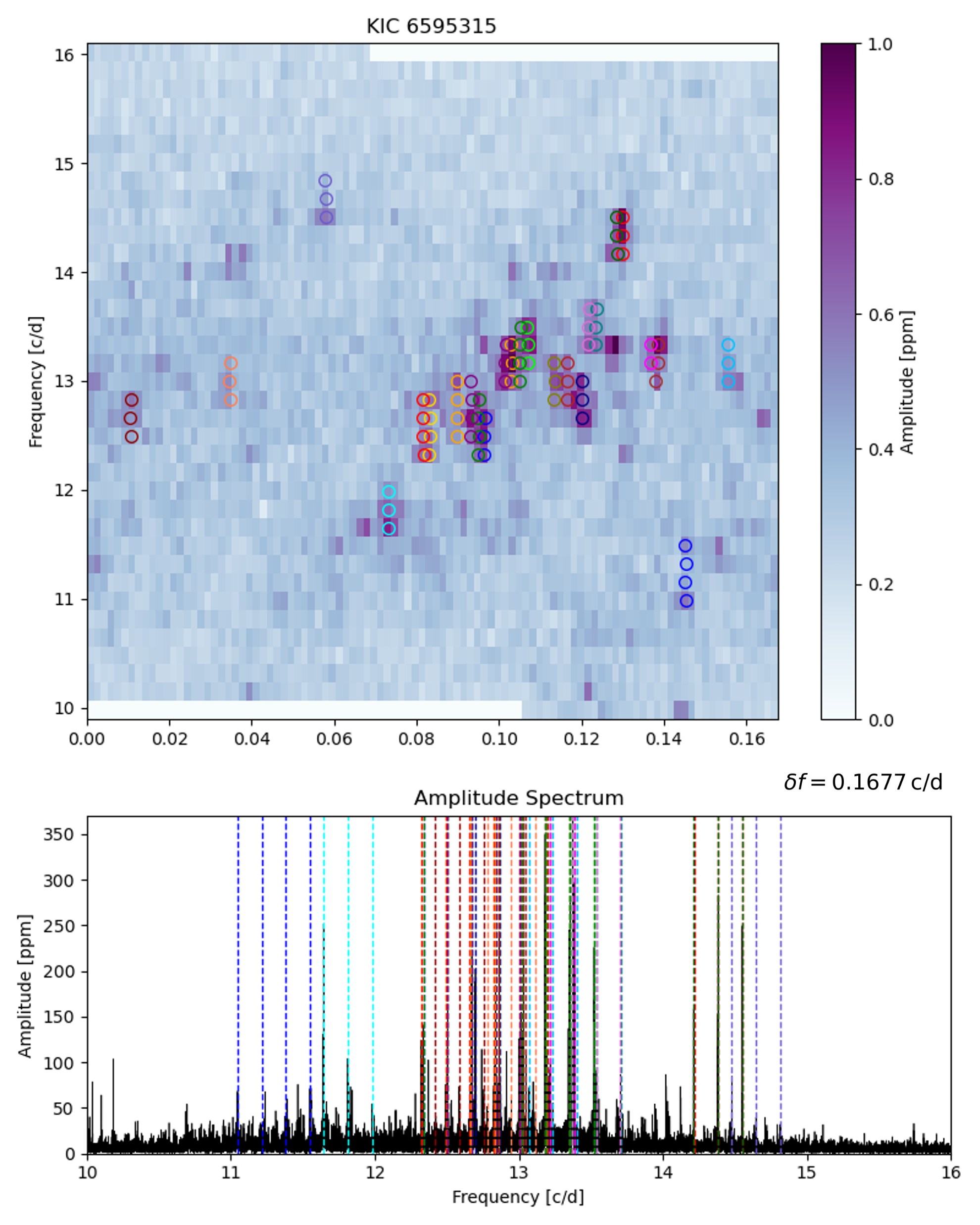}
\caption{Upper panel: \'Echelle diagram for KIC~6595315 with identified ridges or recurring patterns. Lower panel: Corresponding amplitude spectrum, where ridge components are highlighted.}
\label{fig:echelle_KIC6595315}
\end{figure}

\begin{figure}
   \centering
\includegraphics[width=0.5\textwidth]{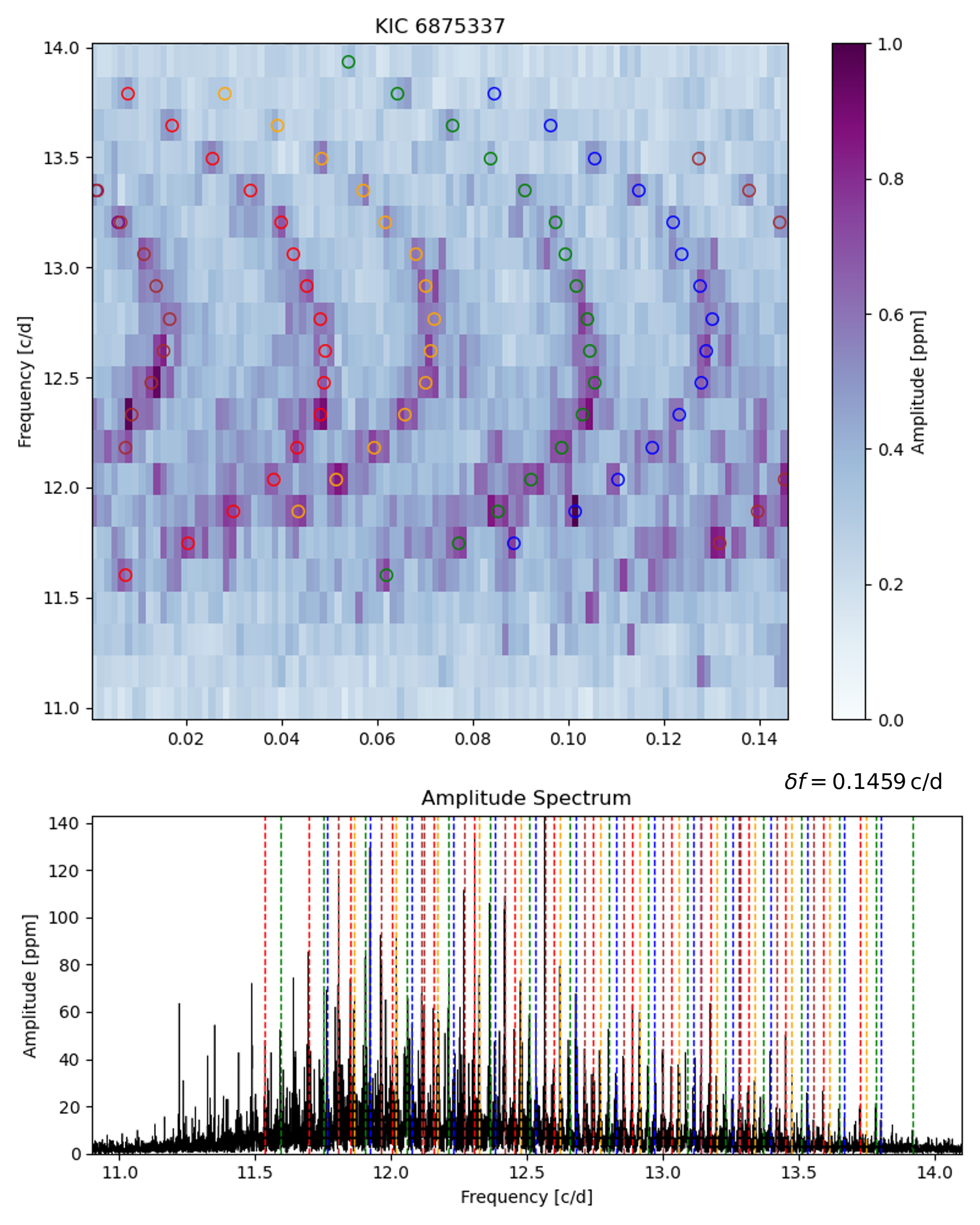}
\caption{Upper panel: \'Echelle diagram for KIC~6875337 with identified ridges or recurring patterns. Lower panel: Corresponding amplitude spectrum, where ridge components are highlighted.}
\label{fig:echelle_KIC6875337}
\end{figure}

\begin{figure}
   \centering
\includegraphics[width=0.5\textwidth]{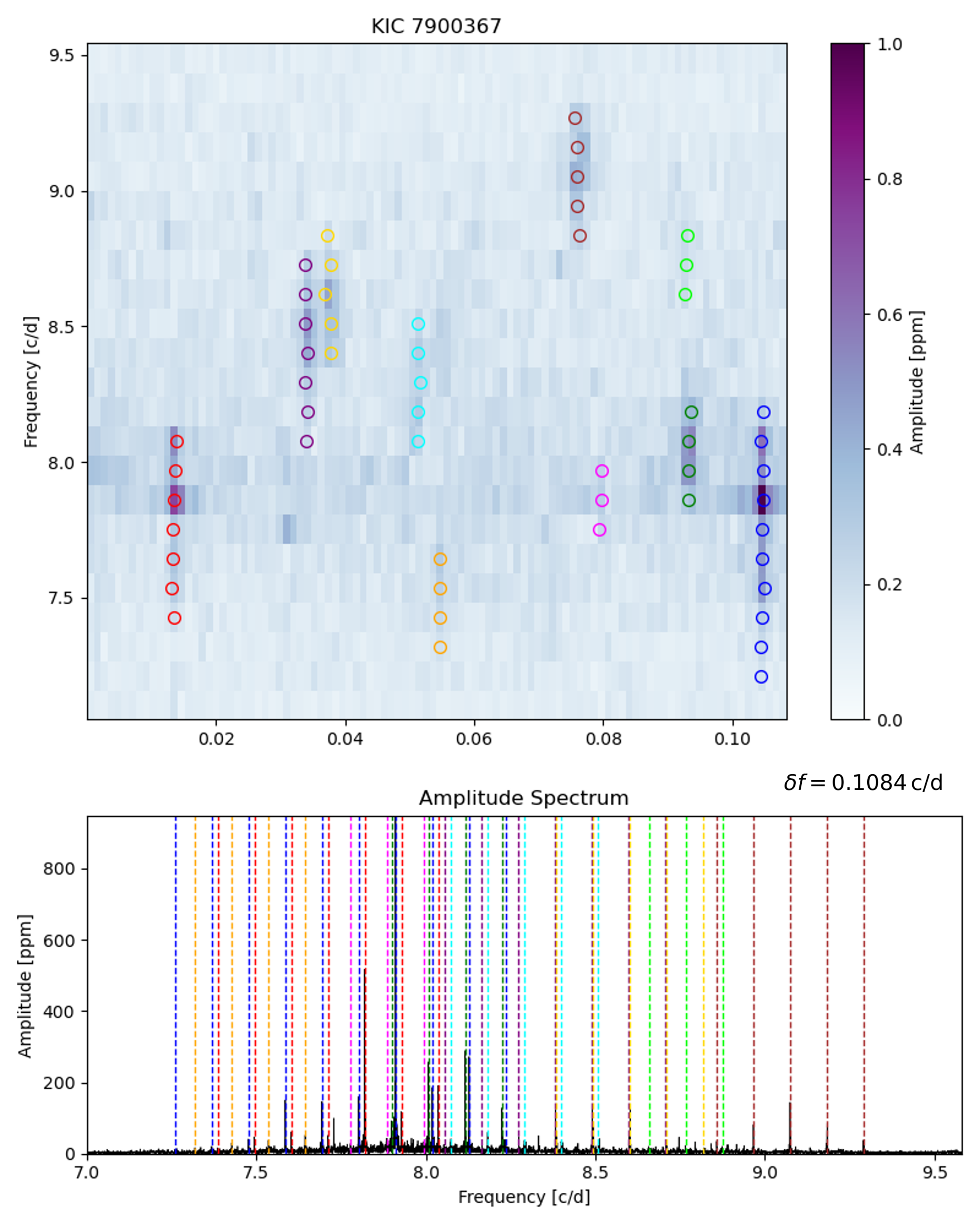}
\caption{Upper panel: \'Echelle diagram for KIC~7900367 with identified ridges or recurring patterns. Lower panel: Corresponding amplitude spectrum, where ridge components are highlighted.}
\label{fig:echelle_KIC7900367}
\end{figure}

\begin{figure}
   \centering
\includegraphics[width=0.5\textwidth]{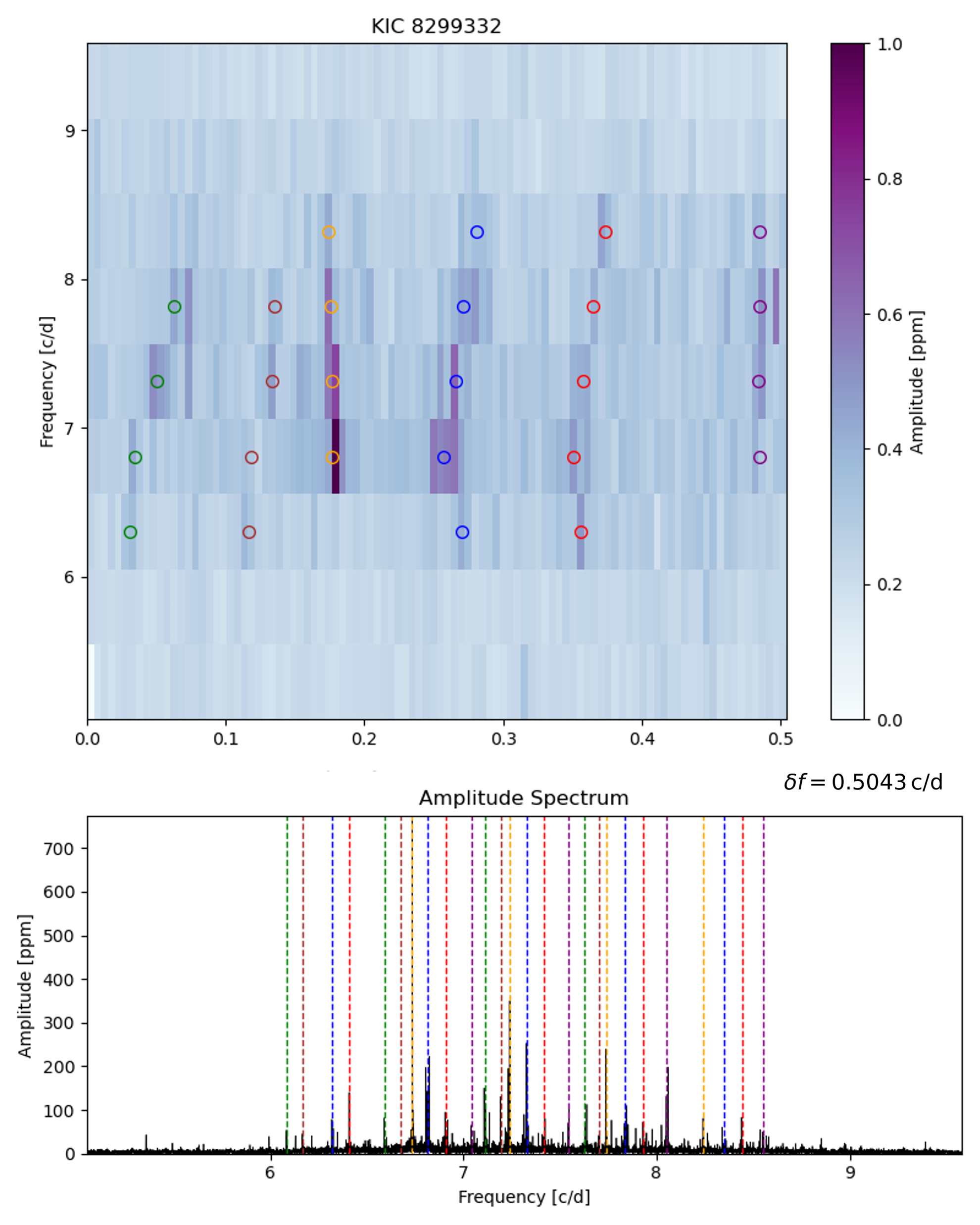}
\caption{Upper panel: \'Echelle diagram for KIC~8299332 with identified ridges or recurring patterns. Lower panel: Corresponding amplitude spectrum, where ridge components are highlighted.}
\label{fig:echelle_KIC8299332}
\end{figure}

\begin{figure}
   \centering
\includegraphics[width=0.5\textwidth]{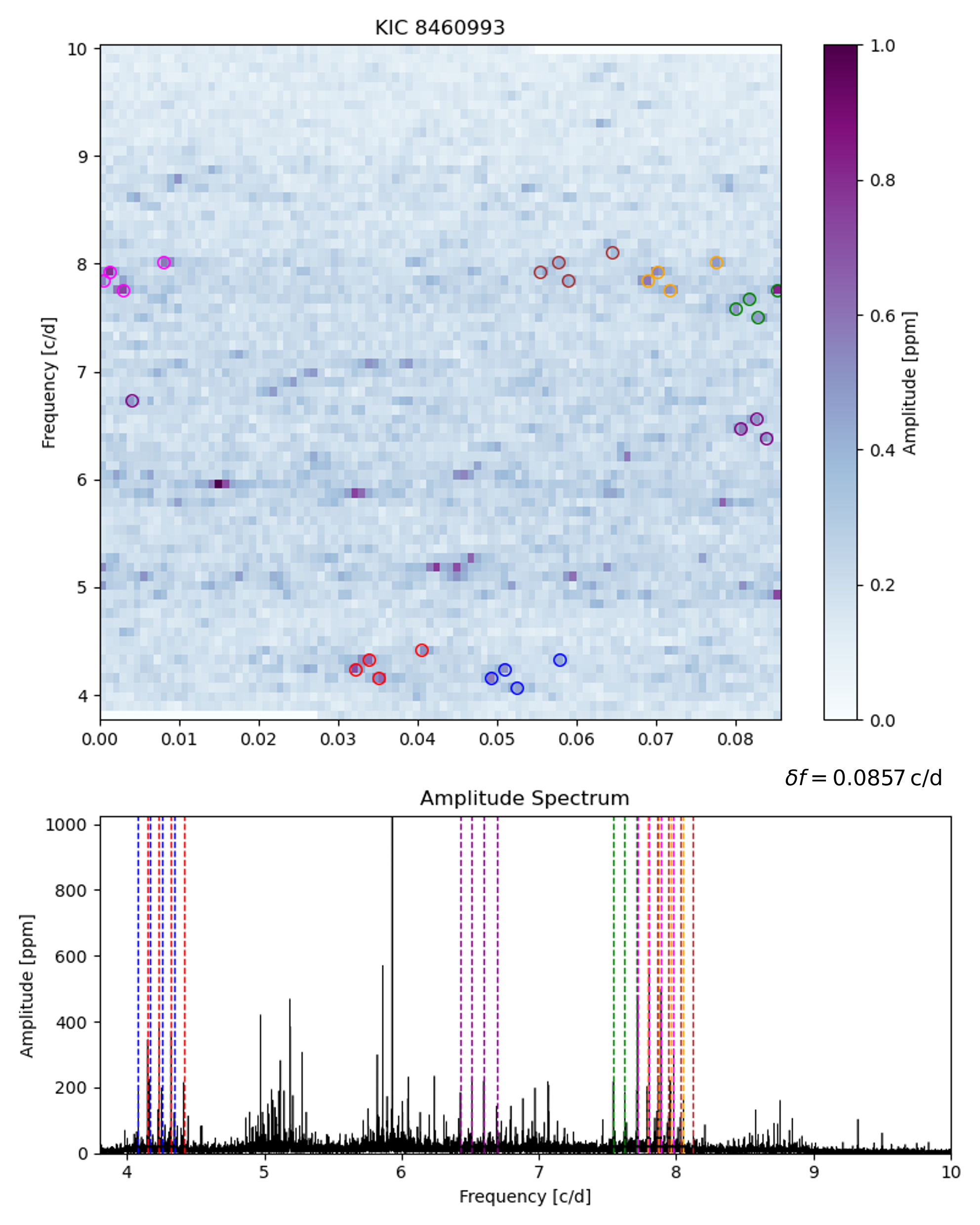}
\caption{Upper panel: \'Echelle diagram for KIC~8460993 with identified ridges or recurring patterns. Lower panel: Corresponding amplitude spectrum, where ridge components are highlighted.}
\label{fig:echelle_KIC8460993}
\end{figure}

\begin{figure}
   \centering
\includegraphics[width=0.5\textwidth]{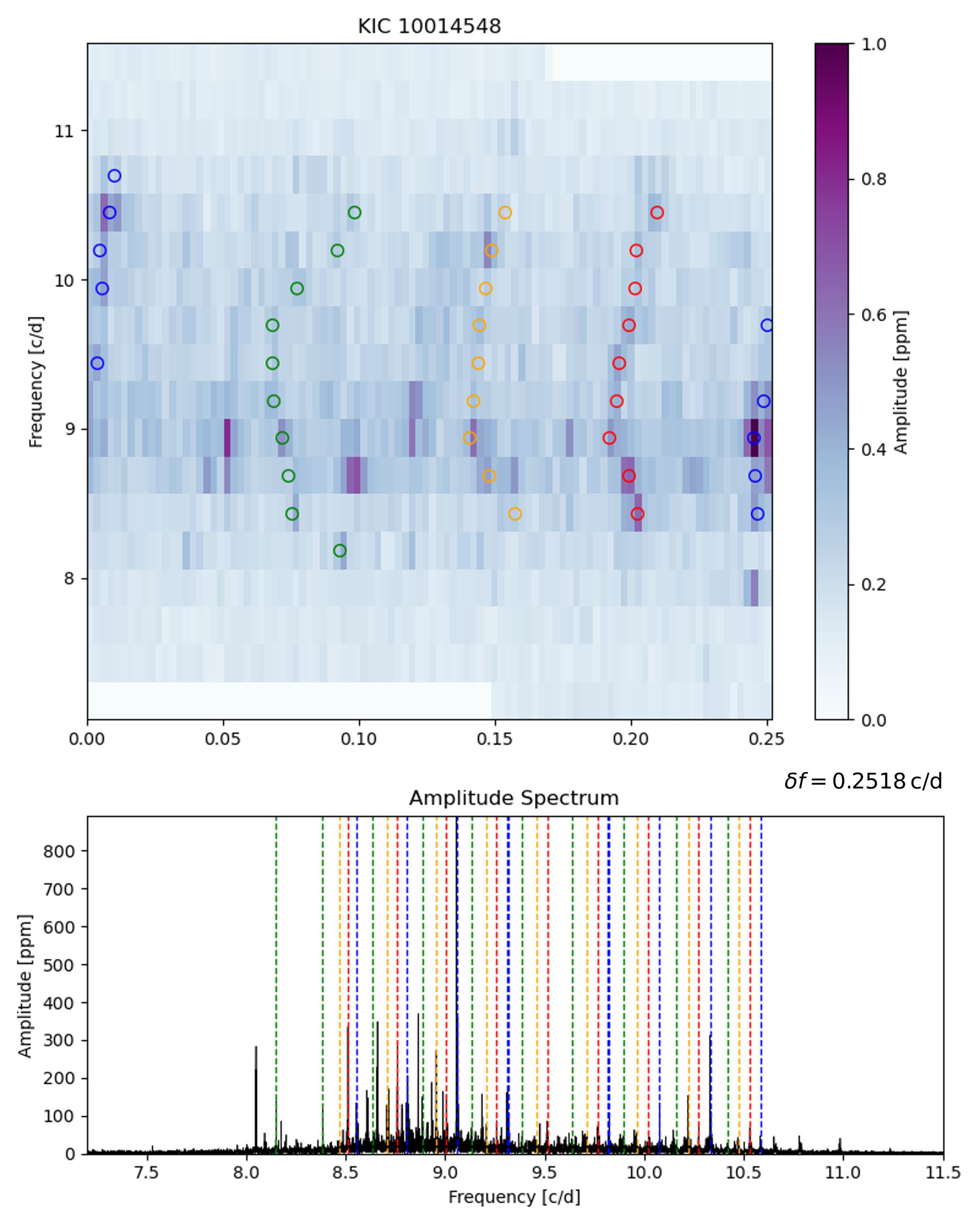}
\caption{Upper panel: \'Echelle diagram for KIC~10014548 with identified ridges or recurring patterns. Lower panel: Corresponding amplitude spectrum, where ridge components are highlighted.}
\label{fig:echelle_KIC10014548}
\end{figure}

In Figs~\ref{fig:echelle_KIC5443410}--\ref{fig:echelle_KIC10014548}, we present representative \'echelle diagrams for a subset of stars that display the clearest patterns. Each figure shows the \'echelle diagram (upper panel), where the identified ridges are colour-coded, together with the corresponding amplitude spectrum (lower panel), in which the same ridges are highlighted. In some cases, additional low signal-to-noise multiplets may be present but are not marked. Several stars, KIC~5443410, KIC~6595315, KIC~6875337, KIC~7900367, KIC~8299332, KIC~8460993, and KIC~10014548, exhibit coherent ridge structures that repeat throughout the agglomerated region. However, the spacing between peaks is generally \textit{not} constant across the full frequency range. In other words, although the ridge morphology is clearly visible and recurrent, it is not equidistant in a manner that would be expected from simple rotational splitting.

The presence of repeating, but non-equidistant, ridges poses a challenge for conventional interpretations. In known oblique pulsators, such as the rapidly oscillating Ap (roAp) stars \citep[e.g.][]{Kurtz_1982}, pulsation multiplets arise due to geometric modulation with a spacing exactly equal to the stellar rotation frequency. In our case, the spacings vary within the ridges, and yet the morphology is preserved, i.e., the same ridge structure appears in several multiplets.
This observation argues against a purely geometric origin. If the ridges were due to `traditional' rotational splitting of pulsation modes, one would expect small but measurable deviations in splitting patterns between modes of different radial orders due to Coriolis and structural effects. Moreover, explaining the observed multiplets through rotational splitting would imply high values of the spherical degree $\ell$, which is difficult to reconcile with photometric observations due to geometric cancellation \citep{Jagoda2002}. For instance, while in KIC~8460993 (Fig.~\ref{fig:echelle_KIC8460993}) we observe clear quadruplets and in KIC~5443410 (Fig.~\ref{fig:echelle_KIC5443410}) we identify quintuplets, possibly consistent with an $\ell=2$ mode where all $2l+1$ components are visible, the case of KIC~6875337 (Fig.~\ref{fig:echelle_KIC6875337}) is more extreme, showing multiplets with up to 15 components. This would require $\ell \geq 7$, which is highly improbable in photometry.

The persistence of identical ridge structures across the agglomerated region, despite these inconsistencies, suggests a different, yet unidentified, modulation mechanism beyond simple rotational splitting. Furthermore, in most cases the region of agglomerated modes is relatively narrow (see Table~\ref{tab:fundamental}), which allows us to exclude the possibility that these peaks are rotationally split p modes or mixed modes. The observed frequency span is smaller than would be expected for modes of one or two radial orders. This interpretation is further supported by the fact that, for the majority of stars, the predicted fundamental radial modes lie at higher frequencies than the agglomerated region. In the case of g or mixed modes, the spacing between consecutive radial orders becomes very small for high radial orders, yet the observed structures do not match the expected morphology for pure asymptotic sequences. We refer to Section \ref{sec:discussion} for a detailed discussion.

\subsection{Amplitude and phase modulation analyses}

For stars exhibiting clear ridge structures in the \'echelle diagrams, we analyse the phase-dependent amplitude and phase variability of selected peaks. We use the same rotation or related modulation frequency identified from the \'echelle diagrams ($\delta f$) to phase-fold the light curves. For visualisation, the light curves are binned into 200 phase bins, while for the analysis the data are subdivided into ten equally spaced phase bins. In each bin, we determine the amplitude and phase of selected central or near-central peaks within the ridges using \textsc{Period04} \citep{lenz2005}, yielding amplitude and phase as a function of rotation phase.

This approach is motivated by rotationally split pulsation spectra in oblique pulsators, such as rapidly oscillating Ap stars, where equidistant multiplets allow amplitude and phase variability to be interpreted in terms of a common modulation. We test whether a similar interpretation applies to the ridge structures observed here. However, in contrast to the strictly equidistant splittings in such systems, the ridges in our stars are often tilted or curved in the \'echelle diagrams and therefore not exactly commensurate with a single modulation frequency.

In practice, the frequency is held fixed, while only the amplitude and phase are fitted independently in each rotation-phase bin. In the presence of nearby unresolved modes, the fitted signal can therefore represent the combined contribution of multiple frequencies. As a result, the recovered amplitude and phase may vary with rotation phase even when the underlying oscillations are intrinsically stable. The magnitude and structure of this effect depend on the frequency spacing, with non-commensurate frequencies leading to quasi-periodic variations in the recovered parameters.

While interference can produce such variability, it is not clear whether it can reproduce the observed behaviour, in particular coherent variations across multiple peaks. We therefore use Monte Carlo simulations that reproduce the observed ridge-like frequency structures and Kepler time sampling, and analyse them in the same way as the observations.

We find that correlations between two peaks can be reproduced under this interference-only scenario, indicating that pairwise coherence is not a sufficient diagnostic. We therefore extend the analysis to three peaks and require that all pairwise combinations simultaneously satisfy strict coherence criteria. The simulations show that such triplet coherence is rarely produced by interference alone, implying that the probability of a chance occurrence is low. When three peaks exhibit coherent correlation or anticorrelation in the observations, this provides strong evidence that their variability reflects a shared underlying mechanism. Once this is established, the shape of the phase-dependent variability can be interpreted as a physically meaningful observable, tracing a common modulation pattern.

Details of the Monte Carlo simulations are provided in Appendix~\ref{sec:montecarlo}. The resulting amplitude and phase variability for individual stars are shown and discussed in Sect.~\ref{sec:singlestars}.

\section{Comments on individual stars}\label{sec:singlestars}

In this section, we summarise the main observational properties of the selected stars on an individual basis. The data analysis has been described in the preceding sections, so here we present the relevant results for each object and highlight the diversity of behaviour across the sample.

\subsection{KIC 5443410}\label{sec:kic5443410}

KIC~5443410 is a hybrid pulsator, exhibiting high-radial-order g modes characteristic of $\gamma$~Dor stars together with several isolated peaks in the p-mode regime consistent with $\delta$~Sct pulsations. The g-mode period spacing $\Pi_0$, determined as described in section \ref{sec:periodspacing},  is  3984~s, which is consistent with high-order g-mode behaviour in intermediate-mass stars. The parameters of the best-fitting MESA model used to determine the fundamental radial mode are listed in Table~\ref{tab:MESAFit}.

KIC~5443410 displays one of the clearest examples of recurring, regular frequency multiplets among our sample. As shown in Fig.~\ref{fig:echelle_KIC5443410}, the star exhibits several well-defined quintuplets with consistent morphology and spacing across the agglomerated region. These structures are easier to identify than in other stars due to both the relatively high amplitudes of the peaks and the relative sparsity of surrounding peaks. The characteristic frequency spacing organising the multiplets in the \'echelle diagram is $\delta f=0.04541$~d$^{-1}$.

Within the dense agglomerated region, the multiplet structures are not all identical. One multiplet appears consistent with a set of independent modes, while most of the other repeating patterns can be reproduced as linear combination frequencies involving the dominant multiplet (the highest-amplitude structure shown in Fig.~\ref{fig:WOS1_example}) and several g modes. Notably, even one of the lower-amplitude g or Rossby ($r$) modes participates in such combinations. This indicates efficient nonlinear coupling between the g-mode cavity and the frequency region producing the dense multiplet structures. However, the dominant multiplet cannot be reproduced as a combination of the observed g and/or p modes, and its physical origin therefore remains unexplained.

To test whether sets of closely spaced peaks in the dense agglomerated region could form part of a g-mode sequence, we compared their frequency and period spacings. For representative sets of closely spaced peaks at 7.0082 and 6.9620~d$^{-1}$, the frequency spacing is $\Delta f \approx 0.046$~d$^{-1}$, close to $\delta f$. For representative high-radial-order g modes at 1.7711 and 1.7294~d$^{-1}$, the difference is $ 0.04533$~d$^{-1}$, which is numerically similar in frequency space. However, high-order g modes tend to be evenly spaced in period rather than in frequency. Converting to period spacings yields $\Delta P_{\rm dense} \approx 80$~s, whereas the asymptotic $\ell=1$ g-mode spacing is $\Delta P_{\ell=1} = 2817$~s ($\Pi_0=3984$ s). Thus, despite the apparent similarity in frequency separations, the closely spaced peaks in the dense agglomerated region cannot belong to the same asymptotic g-mode sequence and, if interpreted as g modes, would imply unrealistically high spherical degree.

Interestingly, the two highest-amplitude g modes are very close to integer multiples of the organising frequency $\delta f=0.04541$~d$^{-1}$. Specifically, the frequencies at 1.81638 and 1.771050~d$^{-1}$ are consistent with $40\delta f$ and $39\delta f$, respectively, within the frequency resolution of $1/T \approx 6.8\times10^{-4}$~d$^{-1}$. Whether this near-integer commensurability is coincidental or reflects a physical connection between the g-mode spectrum and the frequency scale organising the agglomerated multiplets remains unclear.

\begin{figure}
   \centering
\includegraphics[width=0.5\textwidth]{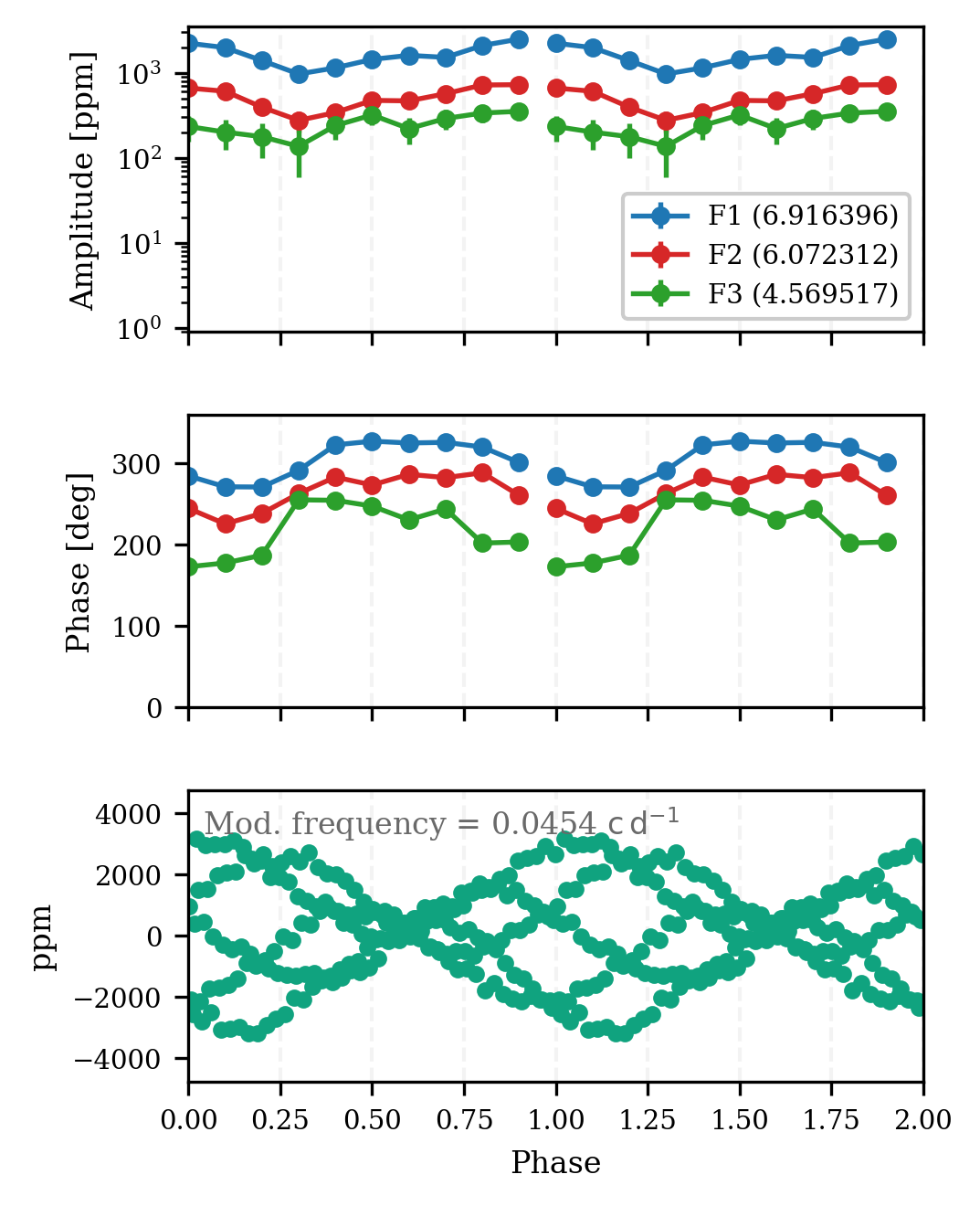}
\caption{Amplitude and phase variability for KIC~5443410. \textit{Upper panel:} Amplitude variation of selected peaks. \textit{Middle panel:} Corresponding phase variations. \textit{Lower panel:} Light curve phase-folded with $\delta f$, which also corresponds to the spacing derived from the \'echelle diagram. Uncertainties are the analytical errors from Period04.}
\label{fig:kic5443410_amplivar}
\end{figure}

We examined the amplitude and phase modulation of the central peaks of the highest-frequency quintuplet, and of a second and third quintuplet that appears to originate from combination frequencies. Both exhibit the same modulation behaviour (Fig.~\ref{fig:kic5443410_amplivar}). The lower panel shows the phase curve folded with $\delta f = 0.04541$~d$^{-1}$ which, unlike in several other stars, does not resemble rotational modulation. The two dominant g modes are separated by $\Delta f \approx 0.04533$~d$^{-1}$, corresponding to a beating timescale of $\sim 22$~d, consistent with $1/\Delta f$. The observed modulation pattern is therefore best explained by interference between these closely spaced modes rather than by rotational modulation.

Although such integer-like commensurabilities can resemble tidally excited oscillations in close binaries, KIC~5443410 is not known to be a binary based on current spectroscopic evidence (see Table~\ref{tab:specinfo}), and the physical origin of the commensurability remains unclear. Since most Am stars are in close binaries, further observations are required to test if binarity plays any role in the organization of the photometric frequency patterns.

From the g-mode pattern, we infer a near-core rotation frequency of $0.78 \pm 0.20$~d$^{-1}$. This rotation rate is nearly twenty times larger than the characteristic spacing $\delta f$ organising the multiplet structures in the \'echelle diagram (Fig.~\ref{fig:echelle_KIC5443410}). If the multiplet spacing were interpreted as a direct rotational signature, this discrepancy would require strong radial differential rotation between the near-core region probed by the g modes and the layer responsible for the observed multiplet spacing.

%KIC5443410 (together with KIC 8460993) differs from the remainder of our sample in that its near-core rotation rate is not consistent with slow internal rotation. The other stars exhibit near-core rotation frequencies compatible with relatively slow rotation, making KIC5443410 a potential outlier within the current sample.

The fundamental radial mode lies well above the agglomerated frequency region, irrespective of the method used to determine it (see Fig.~\ref{fig:overview4a} and Table~\ref{tab:fundamental}). If the peaks in the agglomerated region are pulsational in origin, they must correspond to modes distinct from the classical high-order asymptotic g-mode sequence, with their regularity hinting at an as-yet unidentified excitation or modulation mechanism.

\subsection{KIC 6595315}\label{sec:kic6595315}

\begin{figure}
   \centering
\includegraphics[width=0.5\textwidth]{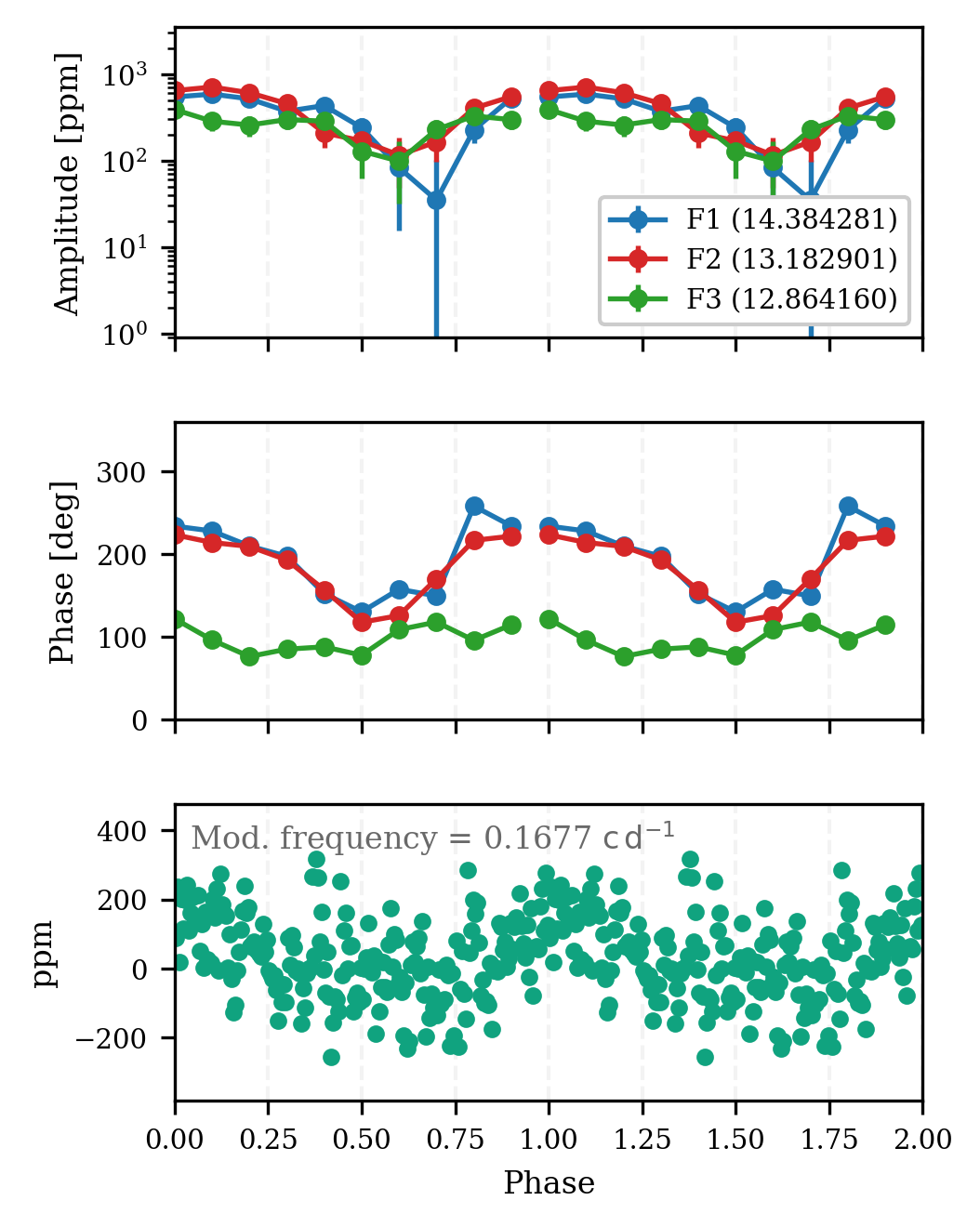}
\caption{Amplitude and phase variability for KIC~6595315. \textit{Upper panel:} Amplitude variation of selected peaks. \textit{Middle panel:} Corresponding phase variations. \textit{Lower panel:} Light curve phase-folded with the rotation frequency, which also corresponds to the spacing derived from the \'echelle diagram. Uncertainties are the analytical errors from Period04.}
\label{fig:kic6595315_amplivar}
\end{figure}

KIC~6595315 displays high-radial-order g modes characteristic of $\gamma$~Dor stars, with a spacing of $\Pi_0 = 2967$~s. This star exhibits clear ridge structures in the \'echelle diagram (Fig.~\ref{fig:echelle_KIC6595315}), where several repeating multiplets are evident across the agglomerated frequency region. Using a characteristic spacing of 0.1677~d$^{-1}$, determined from the ridge morphology in the \'echelle diagram, we applied the same phase-binned amplitude and phase analysis as described above, and phase-folded the light curve with the corresponding period.
As shown in Fig.~\ref{fig:kic6595315_amplivar}, the resulting variations reveal a coherent modulation pattern in both amplitude and phase. In light of the tests described above, such behaviour is unlikely to arise from interference alone and instead indicates that the peaks respond to a common underlying mechanism, with rotation providing a natural interpretation.

Regarding the nature of the agglomerated peaks, the estimated fundamental radial mode for this star (see Fig.~\ref{fig:overview4a} and Table\ref{tab:fundamental}) lies either just below or just above the dense frequency hump, depending on the method used. This placement suggests that the observed peaks may correspond to low-radial-order g modes, mixed modes, or possibly even low-order p modes, if they indeed originate from pulsations.

\subsection{KIC 6875337}\label{sec:kic6875337}
KIC~6875337 was the first WOS to be identified in our sample. It belongs to the Type II group, characterised by low-frequency variability consistent with rotational modulation or low-order g modes, but lacking clear high-frequency p-mode pulsations (see Fig.~\ref{fig:WOS2_example}). No regular g-mode patterns are detected. Instead, the star exhibits a pronounced agglomerated hump at intermediate-to-high frequencies that cannot be explained by standard p-mode pulsations. In addition, we identify peaks at approximately half the agglomerated frequency and at higher frequencies corresponding to the agglomerated region plus this half-frequency offset (Fig.~\ref{fig:WOS2_example}). These features are neither consistent with combination frequencies (see Section \ref{sec:combi}) nor harmonics, and they are not attributable to instrumental effects. Their presence in other stars, including KIC~6875337, KIC~7430757, and TIC~452590255, strongly suggests an astrophysical origin.

If the agglomerated structure in KIC~6875337 is due to pulsations, the frequencies are compatible with low-radial-order g or p modes, as they lie close to the predicted fundamental radial mode (Fig.~\ref{fig:overview4b}; Table\ref{tab:fundamental}). \'Echelle diagrams reveal several coherent ridges within this region, each consisting of 13–17 peaks with similar morphology. Interpreting these as rotationally split multiplets would require spherical degrees of at least ($\ell \gtrsim 6$). Such high-l modes are expected to suffer strong geometric cancellation in disk-integrated photometry, making their detection highly unlikely \citep[e.g.][]{Jagoda2002, Rappaport2026}. Moreover, the absence of uniform frequency spacings further challenges a simple interpretation in terms of rotational splitting or the oblique pulsator model (see Sections~\ref{sec:intro} and \ref{sec:rotmod}).

\begin{figure}
   \centering
\includegraphics[width=0.5\textwidth]{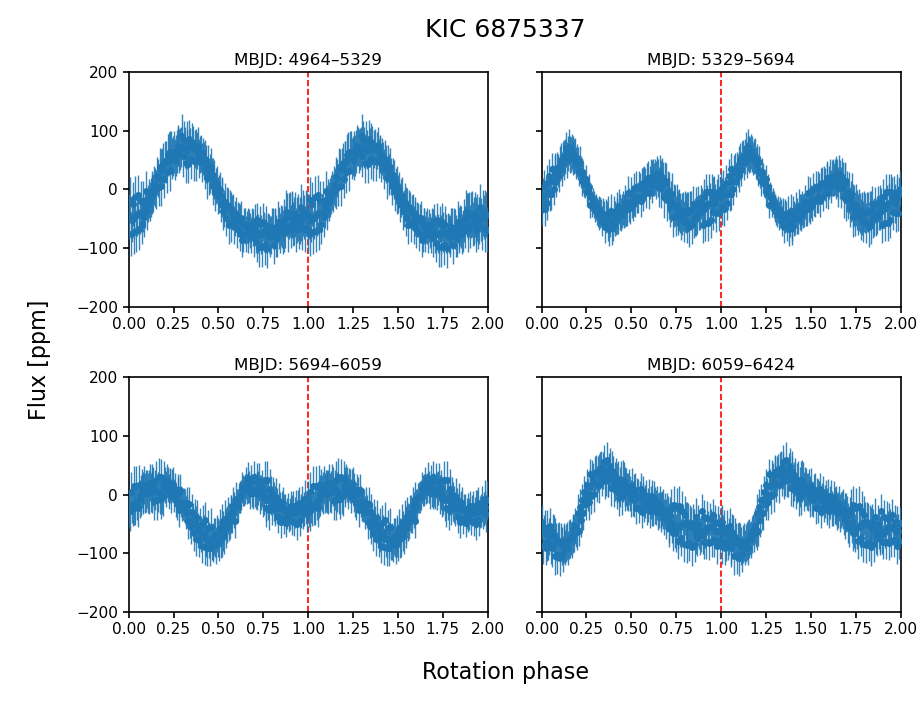}
\caption{Band-pass filtered, phase-folded Kepler light curves of KIC~6875337, using the $\delta f$ derived from the \'echelle diagram. The four panels show consecutive time segments (MBJD ranges indicated), each displaying clear rotational modulation with persistent maxima and slowly evolving surface features.}
\label{fig:kic6875337_rotvar}
\end{figure}

Using the characteristic spacing derived from the \'echelle diagram ($\delta f = 0.14594$~d$^{-1}$), we examined the light curve in the same rotational framework as for the previous stars. However, unlike in those cases, we do not identify three peaks with correlated amplitude and phase variability, and therefore do not present the amplitude-phase-$\delta f$ analysis for this star. This is perhaps not surprising given the extremely high density of peaks. Instead, in Fig.~\ref{fig:kic6875337_rotvar} we show only the phase-folded light curve, using $\delta f = 0.14594$~d$^{-1}$ and binning into 200 bins. The data were additionally filtered with a passband below $0.6$~d$^{-1}$ to isolate the low-frequency signal. As in the case of KIC~7900367, the resulting variation resembles that expected from rotational modulation, while exhibiting significant temporal variability.

\subsection{KIC 7900367}\label{sec:kic7900367}

\begin{figure}
    \centering
    \includegraphics[width=1\linewidth]{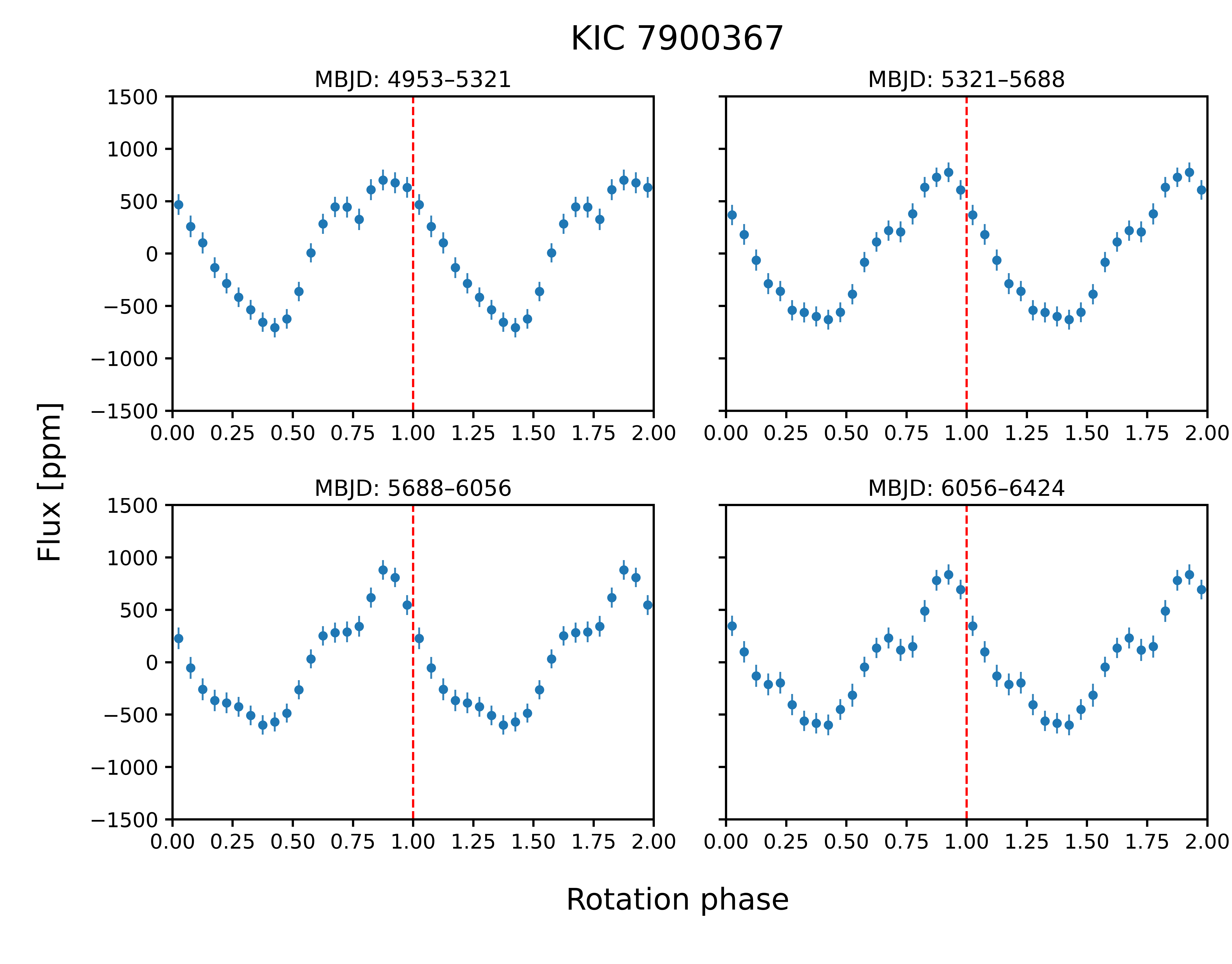}
    \caption{Phase-folded Kepler light curves of KIC~7900367 in four consecutive time segments (MBJD ranges indicated in each panel). Each panel reveals clear rotational modulation, with persistent maxima and slowly evolving surface features. The data are binned into 20 phase bins.}
    \label{fig:kic7900367_rotmod}
\end{figure}

KIC~7900367 exhibits a low-frequency, high-amplitude signal at 0.10841~d$^{-1}$ (Fig.~\ref{fig:overview4a}), consistent with rotational modulation from surface inhomogeneities such as stellar spots, and shows only a few peaks that may be attributed to g modes. The phase-folded light curve at 0.10841~d$^{-1}$ maintains a persistent maximum across the full \textit{Kepler} baseline, while its evolving shape indicates slowly changing spots. In Fig.~\ref{fig:kic7900367_rotmod}, we present the light curve divided into four time segments, each binned and phase-folded using the inferred rotation period. The resulting curves highlight both the long-lived nature of the modulation and its subtle temporal evolution, supporting a rotational origin and consistent with the sharp peak in the amplitude spectrum.

What makes this star particularly noteworthy is the structure of its agglomerated frequency region. The \'echelle diagram (Fig.~\ref{fig:echelle_KIC7900367}) reveals several ridges split by exactly the same value as the rotational signal. These ridges comprise up to ten components and form highly regular multiplets. The agreement between the splitting and the photometric rotation frequency suggests a direct connection between the surface modulation and the frequency structure.

\begin{figure}
   \centering
\includegraphics[width=0.5\textwidth]{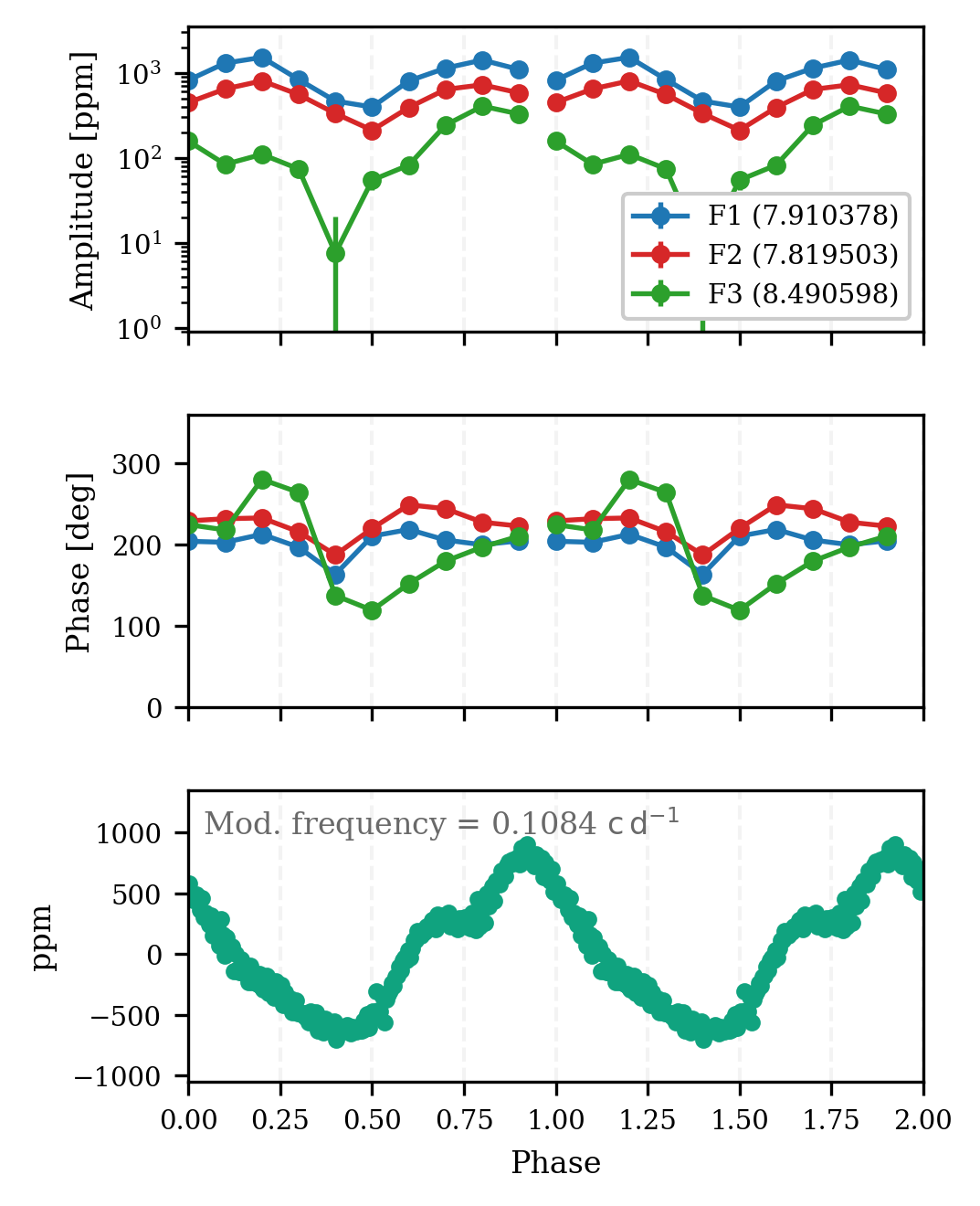}
\caption{Amplitude and phase variability for KIC~7900367. \textit{Upper panel:} Amplitude variation of selected peaks. \textit{Middle panel:} Corresponding phase variations. \textit{Lower panel:} Light curve phase-folded with the rotation frequency, which also corresponds to the spacing derived from the \'echelle diagram. Uncertainties are the analytical errors from Period04.}
\label{fig:kic7900367_amplivar}
\end{figure}

For KIC~7900367, the predicted fundamental radial mode lies at slightly higher frequency than the agglomerated region, independent of the method used (see Fig.~\ref{fig:overview4a} and Table~\ref{tab:fundamental}). This places the multiplets in a regime consistent with mixed modes or low-radial-order g modes, if interpreted as pulsational in origin. The results (Fig.~\ref{fig:kic7900367_amplivar}) show pronounced amplitude and phase modulation that varies coherently with rotational phase. This strongly suggests that the observed ridges are modulated by surface features co-rotating with the star.

\subsection{KIC 7430757}\label{sec:kic7430757}

KIC~7430757 is classified spectroscopically as an Am star (see Table~\ref{tab:specinfo}), and stands out within our sample as the only star that does not show evidence for g-mode pulsations (see Fig.~\ref{fig:overview4b}). 
Although there does not yet seem to be a spectroscopic orbit published, four low-resolution spectra from the LAMOST survey have a $max - min$ radial velocity variance of 77 km s$^{-1}$, with typical per-point errors of $\sim$10 km s$^{-1}$ \citep{2019RAA....19...64Q}. Visual inspection of these four heliocentric RV corrected spectra corroborate this variability -- RV shifts are clearly evident.

\begin{figure}
    \centering
    \includegraphics[width=1\linewidth]{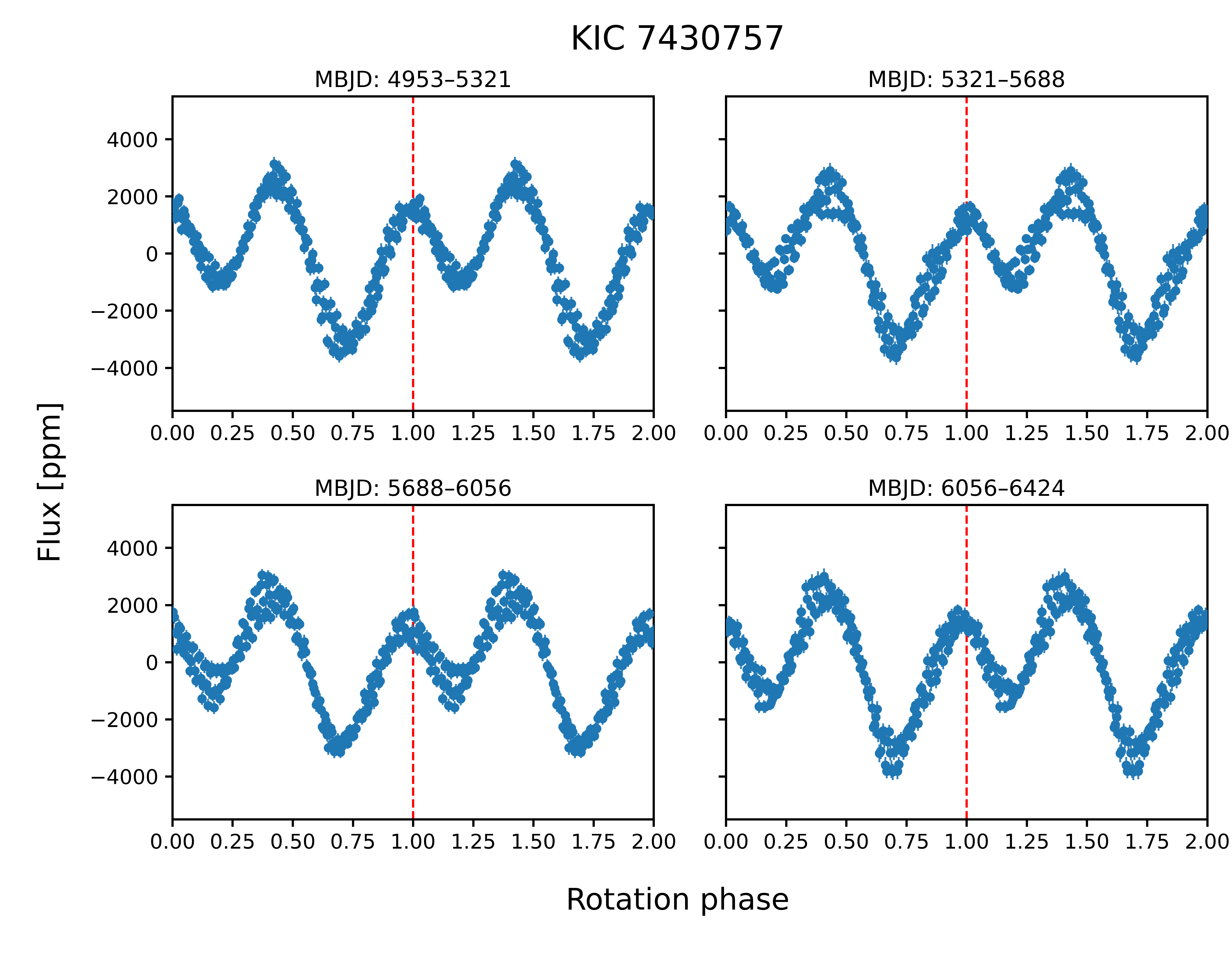}
\caption{
%Phase-folded 
\textit{Kepler} light curves of KIC~7430757 phase-folded to a period of 5.1680 d in four consecutive time segments (MBJD ranges indicated in each panel), each binned into 200 phase bins. Each panel reveals clear, stable orbital or rotational modulation, indicating that some of the peaks are multiplets of the rotation frequency.}
\label{fig:kic7430757_rotmod}
\end{figure}

The light curve for KIC~7430757 displays very clear and stable double-waved modulation at 0.19350~d$^{-1}$ (5.17 d). 
As shown in Fig.~\ref{fig:kic7430757_rotmod}, the light curve, divided into multiple segments, reveals a persistent 
modulation pattern. The shape of the phase-folded light curve remains remarkably stable across the entire \textit{Kepler} time span, strongly suggesting the underlying physical cause of the photometric variation is unchanging during the 4 years of \textit{Kepler}. Considering the large radial velocity variance, the relatively high photometric amplitude, the stability of the photometric signal, and the relatively short period, KIC~7430757 is possibly a close binary with ellipsoidal variation. If this is true, it is also likely that the rotation is phase locked to the same 5.17 d period -- hereafter `rotation frequency' and `orbital frequency' are used interchangeably for KIC~7430757. The unequal maxima in the phased photometry \citep[i.e. the O'Connell effect][]{1951PRCO....2...85O} is sometimes seen in ellipsoidal variables, but lacks a universal explanation. We phase-folded the available RV measurements with the photometric period of 5.17~d, and they are consistent with a binary interpretation. However, the data consist of only four RV measurements, two of which were obtained at the same epoch, and span only 0.3 in phase. It is therefore not possible to determine whether the orbital period is indeed equal to the photometric period. If an orbital period is longer, it remains possible that the photometric modulation is purely rotational in nature.

Within the agglomerated frequency region of KIC~7430757, several peaks coincide with integer multiples of the rotational modulation frequency. These give rise to oscillatory signals that are phase-aligned with the rotation, yet their amplitudes and detailed profiles vary over time. This behaviour suggests a connection between the agglomerated peaks and the stellar rotation period. In addition to the prominent high-frequency agglomerated hump, there is also a notable excess of power near half the central frequency of this hump, as well as at the sum of the two. This pattern is similarly observed in other Type~II stars. While we do not include the \'echelle diagram for this star, visual inspection reveals the presence of ridge-like structures; however, the identification of coherent multiplet patterns is ambiguous, and the connection between individual peaks remains unclear.

\subsection{KIC 8299332}\label{sec:kic8299332}
KIC~8299332 shows both p and g modes. We identify several coherent ridges in the agglomerated region of the amplitude spectrum, as shown in Fig.~\ref{fig:echelle_KIC8299332}. These ridges exhibit a spacing consistent with a frequency of 0.504320~d$^{-1}$, suggesting a link between surface modulation and the observed frequency structure. The fundamental radial mode for this star is located at significantly higher frequencies, with all methods yielding consistent results (see Fig.~\ref{fig:overview4a} and Table~\ref{tab:fundamental}). This implies that the peaks forming the ridges, if indeed caused by stellar pulsations, correspond to low- to intermediate-radial-order g modes or possibly mixed modes.

\begin{figure}
   \centering
\includegraphics[width=0.5\textwidth]{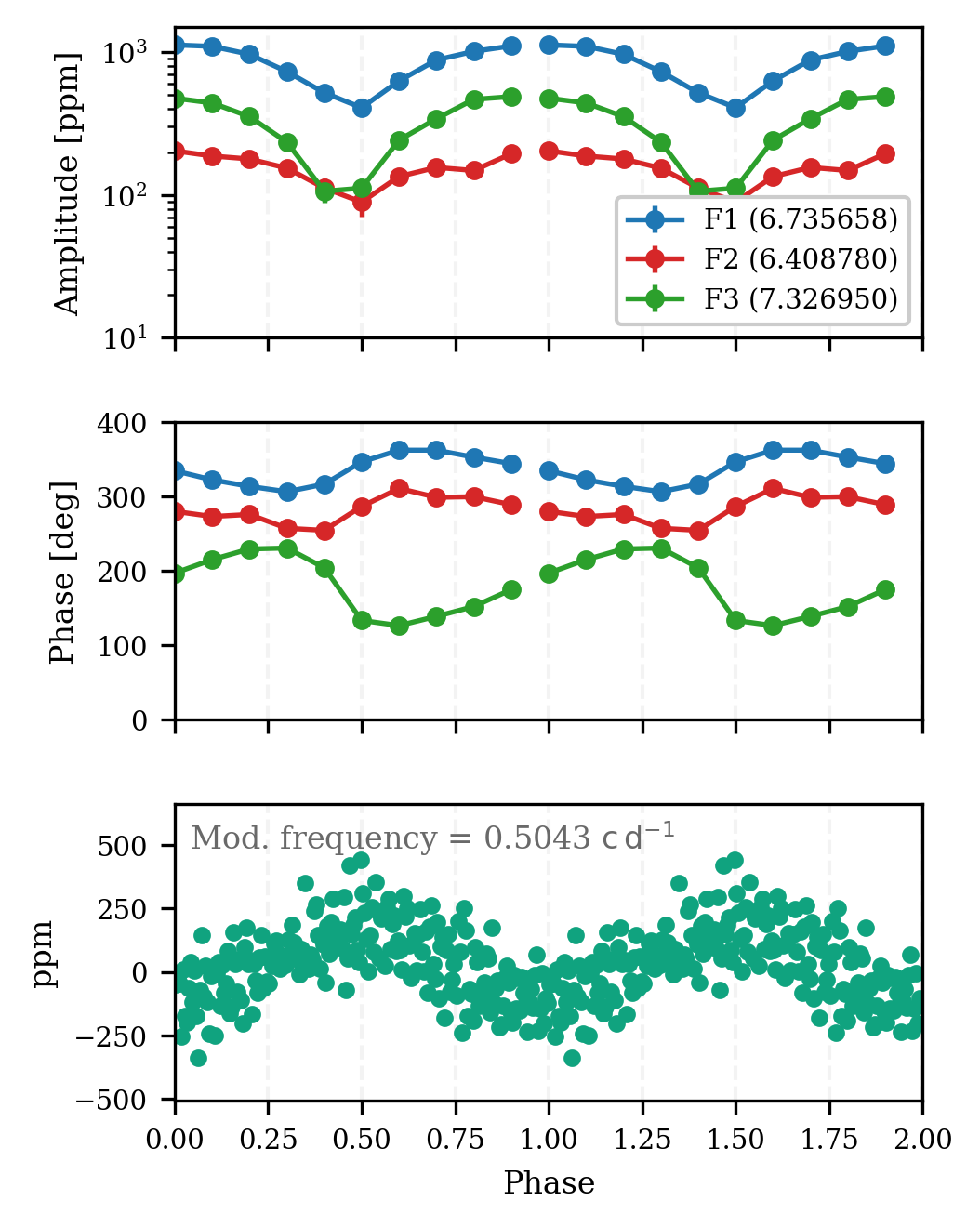}
\caption{Amplitude and phase variability for KIC~8299332. \textit{Upper panel:} Amplitude variation of selected peaks. \textit{Middle panel:} Corresponding phase variations. \textit{Lower panel:} Light curve phase-folded with the rotation frequency, which also corresponds to the spacing derived from the \'echelle diagram. Uncertainties are the analytical errors from Period04.}
\label{fig:kic8299332_amplivar}
\end{figure}

To investigate the nature of the modulation, we applied the same phase-binned amplitude and phase analysis as for KIC~6875337 and KIC~7900367, using the spacing derived from the \'echelle diagram and phase-folding the light curve with the corresponding period. The results, shown in Fig.~\ref{fig:kic8299332_amplivar}, reveal a clear modulation pattern in both amplitude and phase as a function of rotation phase, reinforcing the hypothesis that the ridge structure may be linked to rotational modulation or spot-related surface features.

\subsection{KIC 8460993}\label{sec:kic8460993}
For the hybrid p- and g-mode pulsator KIC~8460993, we applied the same analysis as for the previous stars. The \'echelle diagram (Fig.~\ref{fig:echelle_KIC8460993}) reveals frequency groups that are not as sharply defined as in KIC~5443410, yet still exhibit repeating patterns indicative of an underlying regularity. In particular, we observe recurrent quadruplets across the agglomerated region, where one of the higher-frequency components often appears slightly offset. Despite this asymmetry, the recurrence of the same morphological structure in multiple frequency groups makes the pattern readily recognisable.

The amplitude spectrum of this star is especially dense, with a large number of peaks filling the agglomerated region. Nevertheless, when we phase-fold the band-pass filtered light curve using the spacing of $0.08574$~d$^{-1}$ and apply the same phase-binned amplitude and phase analysis as for the other stars, we detect no significant amplitude or phase modulation. This lack of variability may be related to the unusually dense amplitude spectrum. From the g-mode series (Section~\ref{sec:periodspacing}), we estimate a near-core rotational frequency of $0.55 \pm 0.2$~d$^{-1}$. As in the case of KIC~5443410 (see Section~\ref{sec:kic5443410}), if the spacing of $0.08574$~d$^{-1}$ is associated with rotation, this would imply that the stellar core rotates approximately 4--9 times faster than the envelope.

The predicted fundamental radial mode lies well above the agglomerated region (see Fig.~\ref{fig:overview4a} and Table~\ref{tab:fundamental}), implying that the peaks, if they are due to pulsations, likely correspond to low-radial-order g modes or mixed modes.

\subsection{KIC 10014548}\label{sec:kic10014548}

\begin{figure}
   \centering
\includegraphics[width=0.5\textwidth]{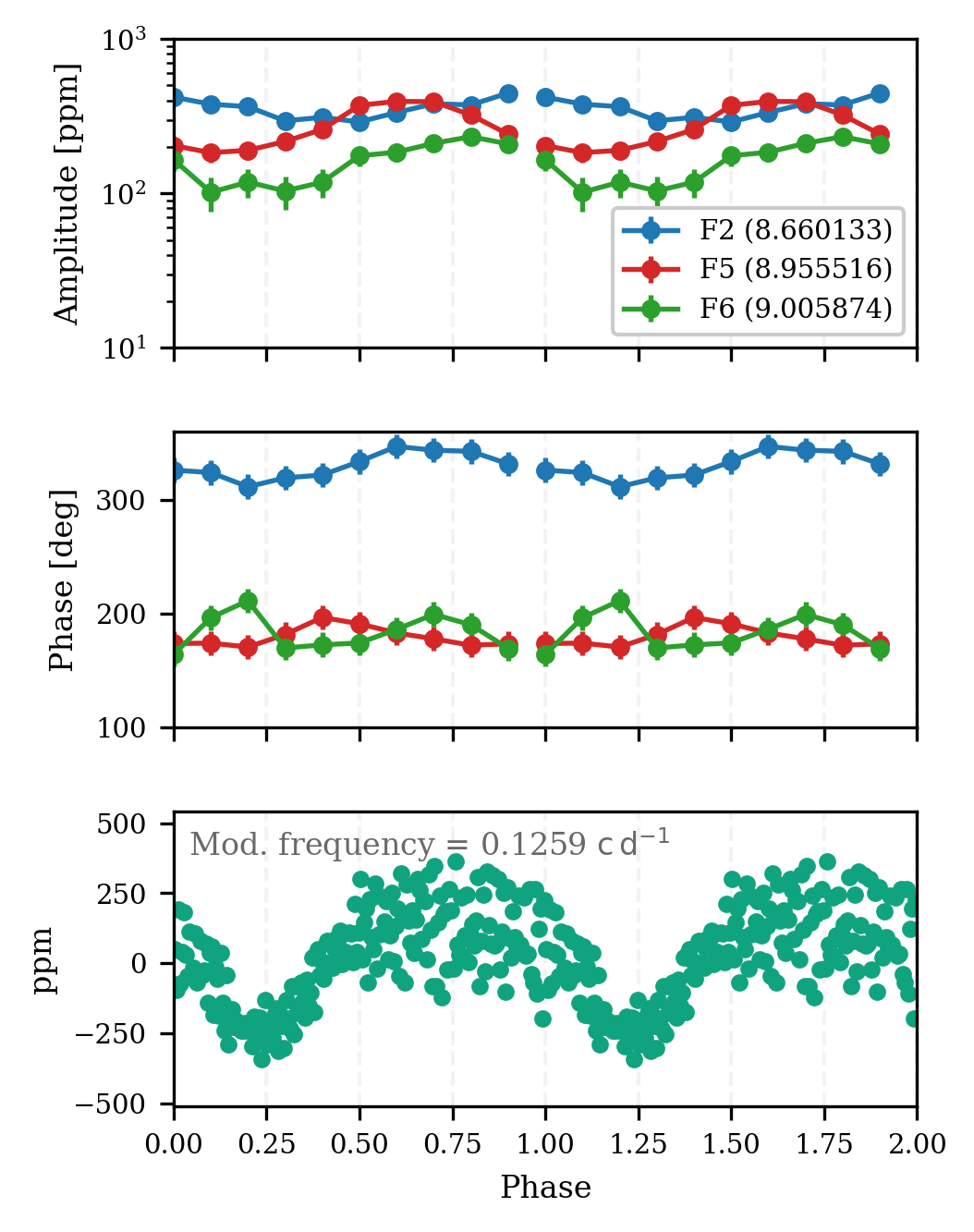}
\caption{Amplitude and phase variability for KIC~10014548. \textit{Upper panel:} Amplitude variation of selected peaks. \textit{Middle panel:} Corresponding phase variations. \textit{Lower panel:} Light curve phase-folded with the rotation frequency, which also corresponds to the spacing derived from the \'echelle diagram. Uncertainties are the analytical errors from Period04.}
\label{fig:kic10014548_amplivar}
\end{figure}

For the type~II $\gamma$~Dor star KIC~10014548, we performed the same analysis as for the previous targets. The \'echelle diagram (Fig.~\ref{fig:echelle_KIC10014548}) reveals decuplet\footnote{Multiplet consisting of ten peaks.} ridge structures with a characteristic spacing of $\delta f = 0.2518$~d$^{-1}$. However, the clearest amplitude and phase modulation occurs at half this value, $\delta f/2 = 0.1259$~d$^{-1}$ (Fig.~\ref{fig:kic10014548_amplivar}), where phase-folding shows coherent variability in both amplitude and phase, suggesting a possible link between the multiplets and the surface modulation timescale.

The estimated fundamental radial mode lies well above the agglomerated region (Fig.~\ref{fig:overview4b}; Table~\ref{tab:fundamental}), suggesting that the peaks, if pulsational, correspond to low-radial-order g modes or mixed modes. The amplitude spectrum also shows additional power at half the agglomerated frequency, at twice its value, and at the agglomerated region plus its half-frequency. The origin of these features remains unclear and may point to nonlinear coupling or a more complex modulation mechanism.

\section{Discussion}\label{sec:discussion}

\begin{figure}
   \centering
\includegraphics[width=1\linewidth]{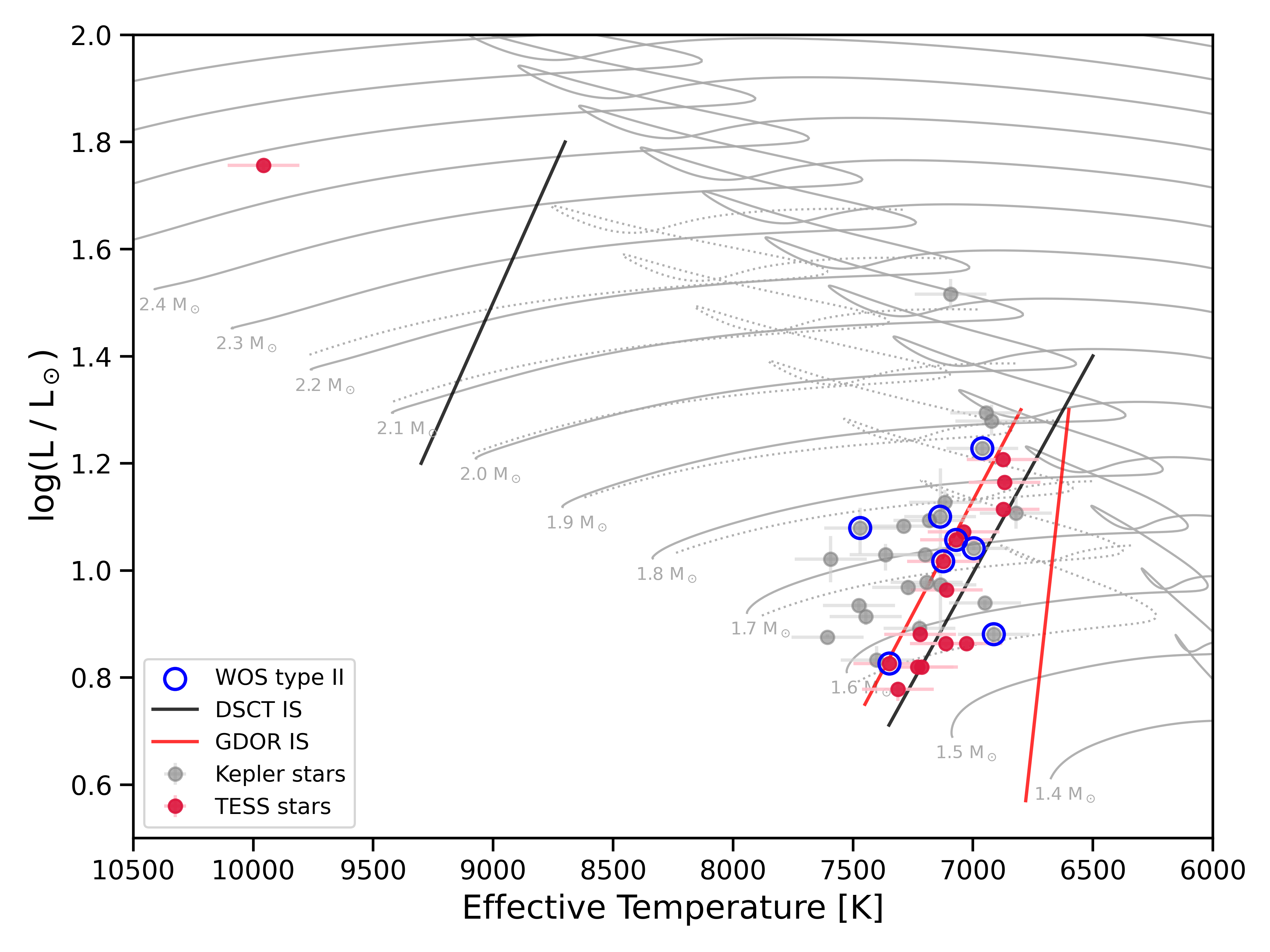}
\caption{Hertzsprung--Russell diagram showing the distribution of our \textit{Kepler} and \textit{TESS} targets. Continuous lines indicate MESA evolutionary tracks computed with solar metallicity, while dotted lines correspond to models with $Z=0.01$, which provide the best match to the stars with measured $\Pi_0$. The low-metallicity tracks span masses starting at 1.4 M$_\odot$ and follow the same 0.1 M$_\odot$ spacing as the solar-metallicity models.}
\label{fig:HRD_Kep_TESS}
\end{figure}

In Fig.~\ref{fig:HRD_Kep_TESS}, we present the location of our sample in the Hertzsprung–Russell diagram using the \textit{Gaia}-derived parameters described in Section~\ref{sec:gaiadata}. The TESS targets occupy a remarkably narrow region of the HR diagram. Because these stars are significantly brighter than the \textit{Kepler} targets, their stellar parameters are more precisely constrained. Note that the outlier is an astrometric and perhaps a spectroscopic binary (TIC~357469812, Table~\ref{tab:specinfo}). It is therefore plausible that the \textit{Kepler} stars would follow a similarly narrow sequence if their parameters were known with comparable precision. For reference, we show the classical $\delta$~Sct instability strip \citep{murphy2019} and the $\gamma$~Dor instability region \citep{dupret2004}. We also include MESA evolutionary tracks, as described in Section~\ref{sec:radialfundamental}. The WOS cluster near the overlap of these instability domains where hybrid p and g modes pulsators are expected.

This confinement already places strong constraints on possible explanations. If the observed frequency structures were caused by binary companions, a much broader distribution across the HR diagram would be expected; instead, the phenomenon is restricted to a well-defined region of parameter space. At the same time, the stars span a range of evolutionary states within this region, indicating that the effect is neither limited to a narrow evolutionary phase nor governed by a simple monotonic mass dependence. In addition, we find no evidence that the observed splittings originate from binarity. Neither orbital light-curve modulation nor spectroscopic signatures support tidally excited oscillations or frequency multiplets produced by orbital phase modulation. Moreover, orbital or triaxial modulation would produce strictly regular frequency spacings commensurate with the orbital period \citep[e.g.,][]{Rappaport2026}, whereas the observed ridge spacings are not perfectly equidistant and vary across the agglomerated region. While binarity alone is therefore insufficient to explain the observations, we cannot exclude that close binarity (or past binary interaction) acting together with specific stellar structural properties may contribute; assessing this possibility will require more comprehensive modelling and observational constraints in future work.

A further structural constraint arises from the position of the fundamental radial mode. In most stars, the agglomerated frequency region lies below the predicted fundamental radial frequency, independent of whether this frequency is estimated via pulsation constants, period–luminosity relations, or stellar models (Sect.~\ref{sec:radialfundamental}). If the agglomerated peaks are pulsational, they therefore cannot correspond to ordinary low-order p modes. Moreover, the frequency span of the agglomerated region is typically too narrow to represent one or two full radial order of p modes and too dense to match a simple asymptotic g-mode sequence. These morphological constraints already exclude several standard interpretations.

Stars that exhibit clear g-mode ridges and for which $\Pi_0$ can be determined appear to require low metallicity in the models. This is intriguing, as many of these objects are classified as Am stars, which typically display chemically peculiar surface abundances, while their bulk metallicity is not known to be systematically higher or lower than solar. However, the presence of TIC~125736216 as a confirmed Pleiades member \citep{bedding2023} indicates that the situation is likely more complex (the Pleiades have a near-Solar metallicity), and highlights the need for improved constraints on binarity as well as more reliable determinations of $T_{\rm{eff}}$ and luminosity.

The analysis of KIC~5443410 provides additional insight. In this star, a substantial fraction of the peaks in the agglomerated region can be reproduced as linear combination frequencies between the dominant quintuplet in the agglomerated frequency region and high-order g modes. Interestingly, even relatively low-amplitude g modes participate in these combinations, indicating that the nonlinear interaction does not scale trivially with observed amplitude. Because nonlinear coupling requires participating frequencies to correspond to genuine pulsation modes, this strongly implies that the dominant multiplet peaks are intrinsic oscillations rather than purely geometric surface modulation. However, the dominant multiplet itself cannot be reproduced as a combination of lower-frequency modes, leaving the origin of the organising spacing unresolved.

In several other stars, individual peaks in the agglomerated region can also be approximated by linear combinations of lower-frequency modes. Yet such identifications typically require higher-order coefficients rather than simple low-order sum or difference relations. In dense spectra, the probability of accidental numerical matches increases rapidly with coefficient order. While nonlinear mode coupling can in principle generate higher-order combinations, it remains uncertain to what extent the apparent matches represent genuine physical interaction or coincidental alignments. A dedicated statistical assessment is required to quantify the significance of these identifications.

\begin{table*}
\caption{Status of proposed explanations for the widely oscillating stars phenomenon. "Tension" indicates that a given scenario can account for some of the observed features, but does not provide a complete explanation or requires further verification.}
\label{tab:wos_status}
\centering
\small
\setlength{\tabcolsep}{5pt}
\renewcommand{\arraystretch}{1.35}
\begin{tabular}{p{4.4cm} p{2.1cm} p{8.0cm} p{1.3cm}}
\hline
\textbf{Hypothesis} & \textbf{Status} & \textbf{Keywords / constraint} & \textbf{Sect.} \\
\hline

Rotational modulation (surface structures)
& Tension
& harmonics at $f_{\rm rot}$; rotation-linked timescale; KIC~5443410: spacing $\sim$10--20$\times$ smaller than near-surface $f_{\rm rot}$; strong radial differential rotation required; rotational modulation from one spot only would require equidistant splitting.
& \ref{sec:kic5443410}, \ref{sec:kic6875337}, \ref{sec:kic7430757}, \ref{sec:kic7900367}, \ref{sec:rotmod} \\

Extra excitation of standard p/g modes
& Tension
& hump: narrow \& very dense; no clear $\Delta P$ sequence; standard $\gamma$~Dor/$\delta$~Sct driving not reproducing agglomerated peaks
& \ref{sec:MAD} \\

Close binarity (tidal excitation)
& Excluded
& no robust orbital signatures; no strictly orbital-commensurate multiplets
&  \\

Close binarity (single-sided pulsator)
& Excluded
& same: no stable orbital modulation; no orbital-locked frequency structure
& \ref{sec:discussion} \\

High-$\ell$ nonradial f/g modes
& Tension
& high-degree parent modes + harmonic visibility; Echelle structure not reproduced for KIC 6875337
& \ref{sec:wojtek}\\

Mass dependence
& Excluded
& no monotonic mass trend in HRD; clustering rather than continuous sequence
&  \ref{sec:discussion} \\

Evolutionary stage dependence
& Excluded
& no narrow evolutionary locus in HRD; not a short-lived phase signature
&  \ref{sec:discussion} \\

Combination frequencies (nonlinear coupling)
& Tension
& KIC~5443410: one multiplet independent; others consistent with g-mode interaction; not accounting for primary multiplet; in other stars: matches only with high-order coefficients; chance-alignment risk in dense spectra
&   \ref{sec:combi} , \ref{sec:discussion} \\

Oblique pulsator model
& Excluded
& predicts strictly symmetric, regularly split multiplets; observed ridges non-equidistant
& \ref{sec:rotmod} \\

Normal p/g/mixed modes with rotational splitting
& Excluded
& rotational splitting alone insufficient; implied mode density too high; hump width too small for single p radial order; not matching asymptotic g-mode morphology
& \ref{sec:singlestars}, \ref{sec:discussion} \\

\hline
\end{tabular}
\end{table*}

\subsection{Rotational modulation}\label{sec:rotmod}

Rotational effects provide several possible mechanisms for producing structured frequency patterns. 
In oblique pulsators such as the roAp stars \citep{Kurtz_1982, Shibahashi_1985, Bigot2002}, 
a pulsation mode whose symmetry axis is inclined relative to the rotation axis produces 
multiplets spaced exactly by the stellar rotation frequency. Such multiplets are strictly 
equidistant and symmetric. In our stars, however, the ridge spacings are not perfectly 
equidistant and their detailed morphology is inconsistent with simple oblique geometries. 
Explaining the observed structures via conventional rotational splitting would furthermore 
require relatively high spherical degree modes in several cases. Because high-$\ell$ modes 
suffer strong geometric cancellation in disk-integrated photometry, their visibility is 
expected to be low. Although KIC~7900367 exhibits equidistant spacing compatible with 
rotation, this behaviour is not representative of the class as a whole.

In addition to oblique pulsation, rotational modulation by surface inhomogeneities such as 
temperature or chemical spots must be considered. In its simplest form, spot-induced 
variability produces a fundamental rotation frequency and its harmonics. If intrinsic 
pulsation modes are present, their amplitudes may be modulated at the rotation frequency. 
Such amplitude modulation generates symmetric sidelobes at 
$f_{\rm puls} \pm n f_{\rm rot}$, potentially forming multiplet-like structures around 
a central pulsation frequency. If spots occur at different latitudes in the presence of 
differential rotation, different modes might be modulated with slightly different 
rotation rates, leading to multiple sets of closely spaced sidelobes. Over long time 
baselines, the superposition of these signals could in principle produce ridge-like 
features in the amplitude spectrum or \'echelle diagram.
However, the amplitude of this effect may be to small to be detectable. 

In particular, KIC~7900367 shows clear low-frequency rotational modulation with distinct rotational harmonics, and the 
amplitudes and phases of peaks in the intermediate-frequency region appear modulated on 
the same timescale. This behaviour is qualitatively similar to the rotation-linked amplitude and phase modulation discussed by \citet{Mathys1985} in the context of roAp stars, and may therefore be relevant here as a possible analogy, though not necessarily in the same physical setting. This suggests that surface rotation influences at least part of the 
observed frequency structure. KIC~6875337 also shows evidence of rotationally linked 
variability, although its agglomerated region displays a more complex morphology.
KIC~7430757 also exhibits a photometric signal consistent with rotational modulation, but this signal is perhaps better explained as being induced by a close binary companion (Sec.~\ref{sec:kic7430757}). 

Despite this, purely spot-induced variability would generally produce harmonic series 
of the rotation frequency and symmetric amplitude-modulation sidelobes, rather than 
the highly structured, persistent, and densely populated multiplets observed in many 
of our target stars. In addition, spot modulation does not naturally account for the presence of 
combination frequencies involving high-order g modes, as observed in KIC~5443410. 
While rotational modulation by surface inhomogeneities clearly contributes to the 
variability in some objects, particularly 
%KIC~7430757 and 
KIC~7900367, it does not 
straightforwardly reproduce the full morphology, density, and stability of the 
agglomerated frequency regions across the sample. Furthermore, spot evolution combined with differential rotation would also be expected to introduce measurable phase drifts and distortions in phase-folded light curves over long time baselines. 
%However, at least in KIC~7430757, the phase behaviour remains remarkably stable over the entire observing period, suggesting that the dominant surface structures do not exhibit the rapid evolution or latitudinal shear that would be required to generate the observed ridge morphology through spot modulation alone.

A hybrid scenario in which surface rotation modulates intrinsically excited pulsation 
modes cannot be excluded. However, assessing whether differential rotation, spot 
evolution, and pulsation–rotation coupling can collectively reproduce the observed 
ridge structures will require detailed forward modelling, which we plan for the future.

\subsection{High-degree nonradial modes and harmonic visibility}\label{sec:wojtek}

An alternative structural explanation invokes high-degree nonradial modes whose harmonics become visible in integrated photometry, inspired by the interpretation of \citet{Dziembowski2016} for RR~Lyrae stars and Cepheids. In this framework, high-$\ell$ f- (or g) modes with intrinsic frequencies near half the observed values are preferentially excited in the outer envelope. Their first harmonics, suffering less geometric cancellation, would then appear as the observed peaks in the agglomerated region. 

\begin{figure}
   \centering
\includegraphics[width=0.5\textwidth]{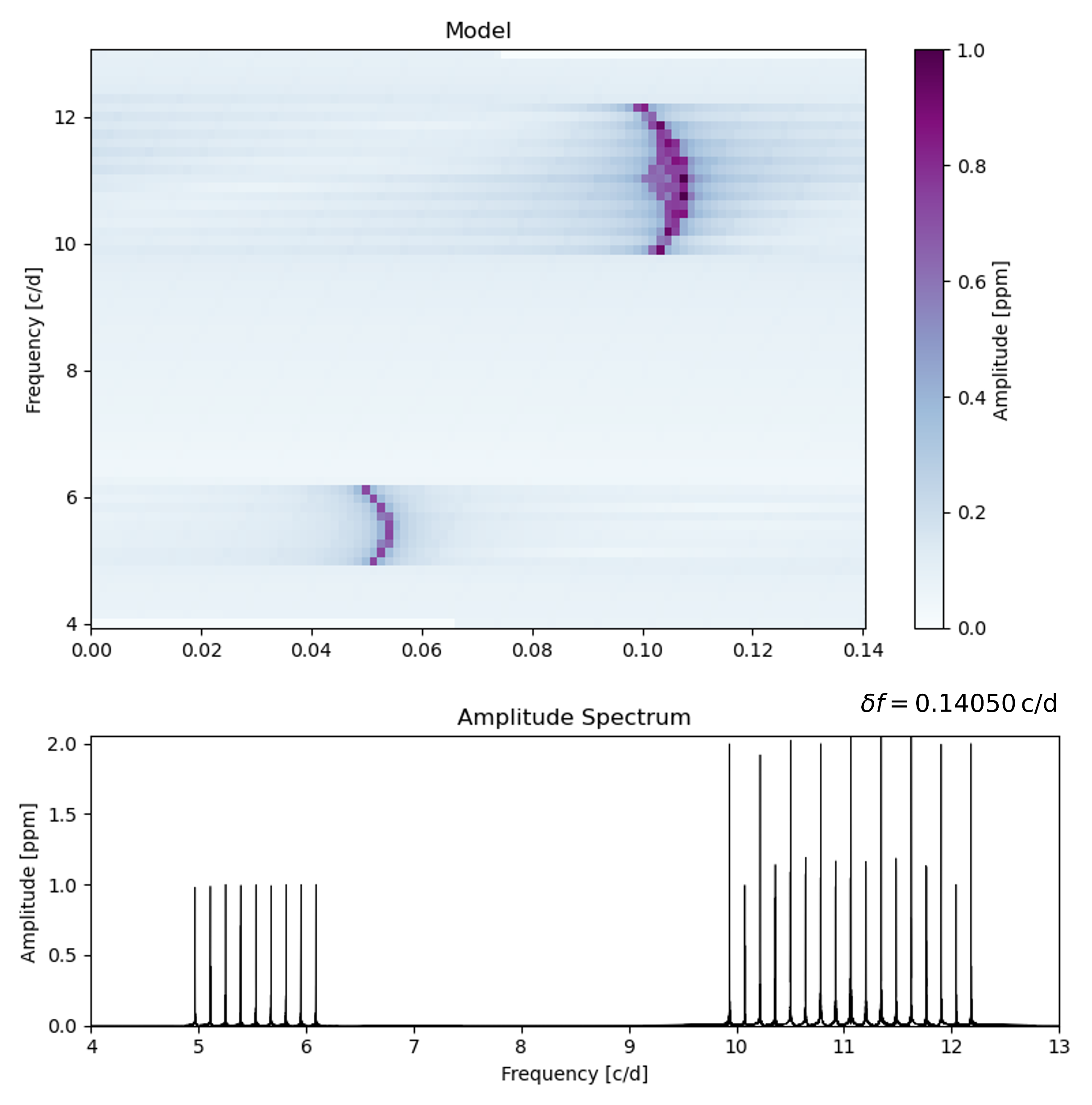}
\caption{Upper panel: \'echelle diagram for $\ell=4$ multiplet split by rotation. Lower panel: Corresponding amplitude spectrum with the $\ell=4$ multiplet and the corresponding harmonics and combination frequencies.}
\label{fig:Echelle_model}
\end{figure}

Even in this scenario, however, the underlying parent modes must be pulsational in nature. The harmonic interpretation therefore does not eliminate pulsation physics but instead shifts the observable signal from the parent mode to its harmonic. We explored this possibility qualitatively for KIC~6875337 by constructing a synthetic $\ell=4$ multiplet including rotational splitting, harmonics, and combination terms (see Fig.~\ref{fig:Echelle_model}). While the resulting model produces curved ridges in the \'echelle diagram (see Fig.~\ref{fig:echelle_KIC6875337}), the curvature does not reproduce the observed morphology for the adopted rotation rate. Although this mechanism cannot be excluded, it requires a more detailed study of rotational effects, and photometric visibility than is feasible here.

\subsection{Excitation mechanisms}\label{sec:MAD}

For a representative stellar mass of 1.55 M$_\odot$, whose evolutionary track spans the effective temperature and luminosity range occupied by the WOS, we computed a full evolutionary sequence using the Code Liégeois d’Evolution Stellaire (\textsc{CLES}; \citealt{Scuflaire2008a}). Along this sequence, we performed a non-adiabatic stability analysis of the oscillation modes with the \textsc{MAD} pulsation code \citep{Dupret2001,Dupret2002,Dupret2005} in order to assess whether the observed agglomerated frequency regions in our stars can be reproduced theoretically. The \textsc{MAD} code includes a time-dependent treatment of convection–oscillation coupling following \citet{Grigahcene2005}, which consistently models convective luminosity blocking (a.k.a. convective shunting; see discussion in \citep{Houdek2015}), an accepted driving mechanism for $\delta$~Scuti and $\gamma$~Dor pulsations in this temperature regime \citep{Dupret2005,Houdek2015}.

\begin{figure}[h]
\centering
\includegraphics[width=\hsize]{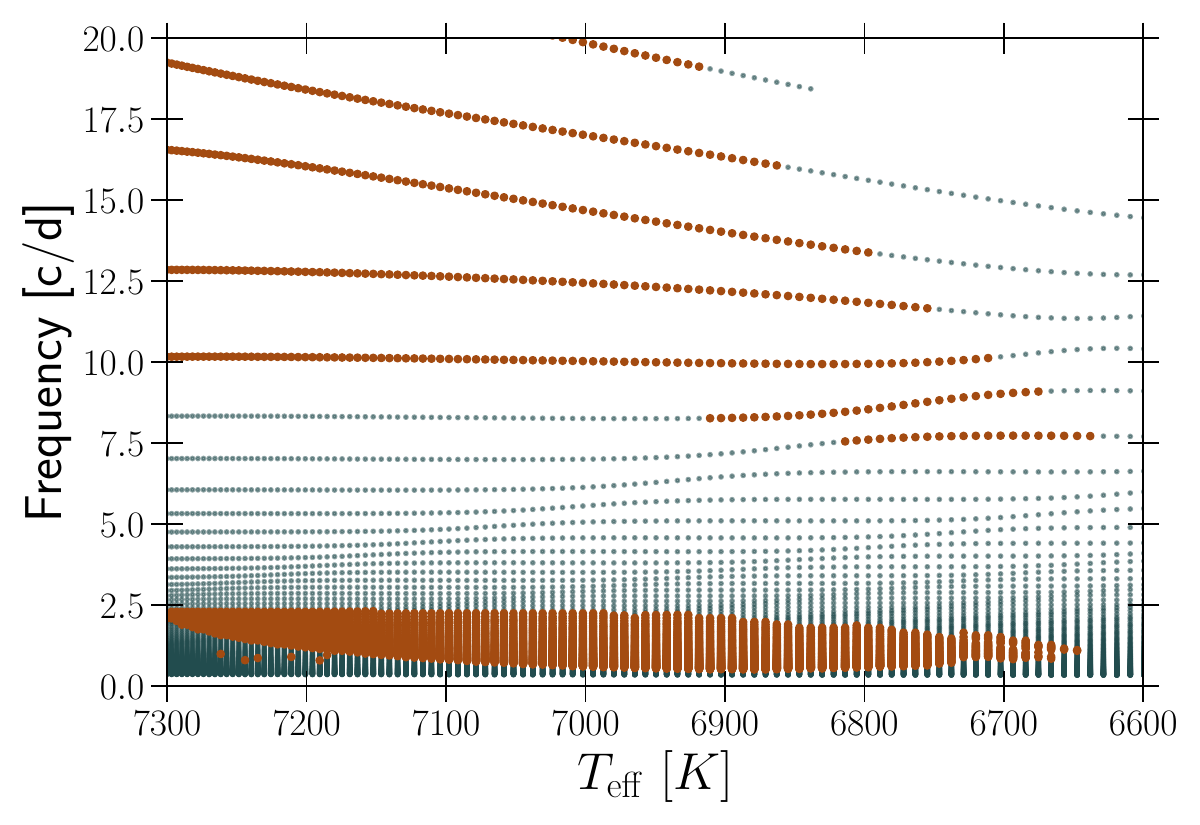}
\caption{Evolution of the frequencies of the $\ell=2$ modes as a function of the effective temperature for the 1.55 M$_\odot$ star along its evolution. Orange dots indicate unstable modes, while grey dots correspond to stable modes.}
\label{fig.excitation_l1_MAD}
\end{figure}

The evolutionary models were computed assuming an initial hydrogen mass fraction $X = 0.72$, solar metallicity $Z = 0.015$ (AGSS09 chemical mixture; \citet{Asplund2009}), and a mixing-length parameter $\alpha_{\mathrm{MLT}} = 2.0$ \citep{Cox1968}, consistent with values required to reproduce the observed $\gamma$~Dor instability strip with these tools \citep{Dupret2005}. Step overshooting with $\alpha_{\mathrm{OV}} = 0.2$ was included at the convective core boundary. We adopted the FreeEOS equation of state \citep{Irwin2012}, OPAL opacity tables \citep{Iglesias1996}, the $T(\tau)$ atmospheric relation from model C of \citet{Atmosphere1981}, and nuclear reaction rates from \citet{Reaction2011}. Figure~\ref{fig.excitation_l1_MAD} shows the predicted unstable-mode frequency range as a function of effective temperature along the 1.55 M$_\odot$ sequence.

Rotation was not included in the present computations. Although many of the stars are Am objects and therefore expected to rotate moderately or slowly, rotation modifies gravity-mode frequencies through the Coriolis force and alters the mode density in the inertial frame \citep[e.g.][]{Bouabid2013}. Nevertheless, rotation does not remove the fundamental separation between the low-frequency g-mode domain and the higher-frequency mixed/p-mode domain predicted by the models.

As shown in Fig.~\ref{fig.excitation_l1_MAD}, the calculations reproduce the characteristic hybrid behaviour expected in this effective-temperature regime: high-radial-order g modes are excited at low frequencies ($\lesssim 2.5\,\mathrm{d^{-1}}$), while mixed modes and low-order p modes are excited at higher frequencies, consistent with the classical $\delta$~Scuti instability strip \citep{Dupret2005,Houdek2015,antoci2019}. In some models, the predicted mixed and low-order p modes partially overlap in frequency with the observed agglomerated region. However, the unstable modes in this regime remain too sparsely distributed to reproduce the observed dense and structured hump. Moreover, the models exhibit a clear separation between the high-order g-mode domain and the mixed/p-mode domain, leaving a frequency interval in which no modes are predicted to be unstable. The overall morphology and mode density of the agglomerated region are therefore not reproduced by the standard non-adiabatic calculations.

The detection of combination frequencies in several stars strongly indicates that the agglomerated frequency regions are pulsational in origin, yet their excitation is not reproduced by standard convective-blocking (convective shunting) or classical $\kappa$-mechanism driving (e.g. \citealt{Dupret2005,Grigahcene2005}). Nearly all spectroscopically classified stars in our sample are Am stars, characterised by slow to moderate rotation and vertical chemical gradients produced by atomic diffusion \citep[e.g.,][]{theado2009,deal2016}. Such chemical stratification modifies opacity profiles and can alter mode driving \citep[e.g.,][]{antoci2019, duerfeldt2024}. It is therefore conceivable that diffusion-induced opacity variations or altered envelope structure modify the excitation conditions in a way not captured by standard models. However, whether such effects can produce the observed narrow and densely populated agglomerated frequency regions remains uncertain.

To summarise, the combined observational and theoretical constraints indicate that the agglomerated frequency phenomenon requires a mechanism that (i) operates within a narrow region of stellar parameter space, (ii) excites or selects a confined frequency band below the fundamental radial mode, and (iii) produces the organised ridge-like structures observed in the amplitude spectrum. In several stars, the ridges appear linked to rotational timescales, yet they are not consistently explained by classical rotational splitting, oblique pulsation, or simple spot modulation. None of the mechanisms explored here simultaneously accounts for the confinement in the HR diagram, the dense and structured ridge morphology, and the partial association with rotation. The WOS therefore appear to occupy a pulsational regime not reproduced by standard models of excitation or rotational modulation. The relative viability of the proposed scenarios and the principal constraints discussed above are summarised in Table~\ref{tab:wos_status}.

\section{Conclusions}

We have investigated the dense, ridge-like frequency structures observed in a sample of intermediate-mass pulsating stars, which we refer to as WOS. These objects occupy a narrow region of the Hertzsprung–Russell diagram near the overlap of the $\delta$~Sct and $\gamma$~Dor instability strips and exhibit agglomerated frequency regions typically located below the fundamental radial mode.

The observed ridge morphology and mode density cannot be reproduced by simple asymptotic g mode behaviour, standard low-order p modes, binarity, or classical rotational splitting. Non-adiabatic stability calculations reproduce the expected classical instability domains but do not predict unstable modes with the observed density or organised ridge structure in the agglomerated region.

In at least two stars (KIC~5443410 and KIC~9347095), a significant fraction of peaks in the agglomerated region can be reproduced as nonlinear combination frequencies involving high-order g modes. However, these combinations require parent modes that themselves reside within the agglomerated frequency band. Linear combinations between classical g and p modes alone are insufficient to reproduce the structure. This implies that, for the combinations based on observable parents, intrinsic pulsation modes must be present in the agglomerated region, even though their excitation is not predicted by current models. We note, however, that combinations involving geometrically suppressed, unobserved parent modes may also be present, as discussed in Section~\ref{sec:wojtek}.

Taken together, the WOS appear to represent a pulsational regime not yet captured by standard models of mode excitation or rotational modulation. The relative viability of the considered scenarios and the principal observational constraints are summarised in Table~\ref{tab:wos_status}. Future work will include detailed spectroscopic analyses of targets lacking high-quality data, refined spectral classification and binarity assessments, in-depth forward modelling, and expansion of the sample to determine the prevalence and parameter dependence of the phenomenon. In this context, TIC~125736216 is particularly valuable, as it is a confirmed member of the Pleiades cluster and therefore benefits from the well-constrained age and metallicity of the cluster.

\begin{acknowledgements}
   Co-funded by the European Union (ERC, MAGNIFY, Project 101126182 ). Views and opinions expressed are however those of the author(s) only and do not necessarily reflect those of the European Union or the European Research Council. Neither the European Union nor the granting authority can be held responsible for them.
    The Australian Research Council has supported SJM through Future Fellowship FT210100485, and TRB and PM through Laureate Fellowship FL220100117. GH thanks the Polish National Center for Science (NCN) for financial support through grant 2021/43/B/ST9/02972. L.F was supported by the Fonds de la Recherche Scientifique F.R.S-FNRS as a Research Fellow. MGP is the recipient of an Australian Research Council Australian Discovery Early Career Award (project number DE250100146) funded by the Australian Government.
    DMB gratefully acknowledges UK Research and Innovation (UKRI) in the form of a Frontier Research grant under the UK government's ERC Horizon Europe funding guarantee (SYMPHONY; PI Bowman; grant number: EP/Y031059/1), and a Royal Society University Research Fellowship (PI Bowman; grant number: URF{\textbackslash}R1{\textbackslash}231631).
    This research made use of Lightkurve, a Python package for Kepler and TESS data analysis (Lightkurve Collaboration, 2018). This work used the software {\sc morse} developed by S. Christophe and distributed under the GNU General Public License v3.0 (GPL-3.0). Some of the observations reported in this paper were obtained with the Southern African Large Telescope (SALT) under program 2023-1-SCI-006 (PI: V. Antoci, E.Niemczura). We acknowledge the use of ChatGPT (OpenAI) to assist with language editing, improve the clarity of the manuscript, and support the development of Python scripts..

\end{acknowledgements}

%-------------------------------------------------------------------

\bibliographystyle{aa} 
\bibliography{references}

@article{duerfeldt2026,
       author = {{D{\"u}rfeldt-Pedros}, Oliver and {Antoci}, Victoria and {Lecoanet}, Daniel and {Guo}, Zao and {Labadie-Bartz}, Johnathan},
        title = {Near-core magnetic field strengths inferred from gravity modes in intermediate-mass stars},
      journal = {\aap},
         year = {2026},
       volume = {arXiv:2606.12148},
          doi = {arXiv:2606.12148 },
       adsurl = {},
}

@article{Ricker2015,
       author = {{Ricker}, George R. and {Winn}, Joshua N. and {Vanderspek}, Roland and {Latham}, David W. and {Bakos}, G{\'a}sp{\'a}r {\'A}. and {Bean}, Jacob L. and {Berta-Thompson}, Zachory K. and {Brown}, Timothy M. and {Buchhave}, Lars and {Butler}, Nathaniel R. and {Butler}, R. Paul and {Chaplin}, William J. and {Charbonneau}, David and {Christensen-Dalsgaard}, J{\o}rgen and {Clampin}, Mark and {Deming}, Drake and {Doty}, John and {De Lee}, Nathan and {Dressing}, Courtney and {Dunham}, Edward W. and {Endl}, Michael and {Fressin}, Francois and {Ge}, Jian and {Henning}, Thomas and {Holman}, Matthew J. and {Howard}, Andrew W. and {Ida}, Shigeru and {Jenkins}, Jon M. and {Jernigan}, Garrett and {Johnson}, John Asher and {Kaltenegger}, Lisa and {Kawai}, Nobuyuki and {Kjeldsen}, Hans and {Laughlin}, Gregory and {Levine}, Alan M. and {Lin}, Douglas and {Lissauer}, Jack J. and {MacQueen}, Phillip and {Marcy}, Geoffrey and {McCullough}, Peter R. and {Morton}, Timothy D. and {Narita}, Norio and {Paegert}, Martin and {Palle}, Enric and {Pepe}, Francesco and {Pepper}, Joshua and {Quirrenbach}, Andreas and {Rinehart}, Stephen A. and {Sasselov}, Dimitar and {Sato}, Bun'ei and {Seager}, Sara and {Sozzetti}, Alessandro and {Stassun}, Keivan G. and {Sullivan}, Peter and {Szentgyorgyi}, Andrew and {Torres}, Guillermo and {Udry}, Stephane and {Villasenor}, Joel},
        title = {Transiting Exoplanet Survey Satellite (TESS)},
      journal = {JATIS},
         year = {2015},
       volume = {1},
          doi = {10.1117/1.JATIS.1.1.014003},
       adsurl = {https://ui.adsabs.harvard.edu/abs/2015JATIS...1a4003R},
}

@ARTICLE{Koch2010_Kepler,
       author = {{Koch}, David G. and {Borucki}, William J. and {Basri}, Gibor and {Batalha}, Natalie M. and {Brown}, Timothy M. and {Caldwell}, Douglas and {Christensen-Dalsgaard}, J{\o}rgen and {Cochran}, William D. and {DeVore}, Edna and {Dunham}, Edward W. and {Gautier}, Thomas N., III and {Geary}, John C. and {Gilliland}, Ronald L. and {Gould}, Alan and {Jenkins}, Jon and {Kondo}, Yoji and {Latham}, David W. and {Lissauer}, Jack J. and {Marcy}, Geoffrey and {Monet}, David and {Sasselov}, Dimitar and {Boss}, Alan and {Brownlee}, Donald and {Caldwell}, John and {Dupree}, Andrea K. and {Howell}, Steve B. and {Kjeldsen}, Hans and {Meibom}, S{\o}ren and {Morrison}, David and {Owen}, Tobias and {Reitsema}, Harold and {Tarter}, Jill and {Bryson}, Stephen T. and {Dotson}, Jessie L. and {Gazis}, Paul and {Haas}, Michael R. and {Kolodziejczak}, Jeffrey and {Rowe}, Jason F. and {Van Cleve}, Jeffrey E. and {Allen}, Christopher and {Chandrasekaran}, Hema and {Clarke}, Bruce D. and {Li}, Jie and {Quintana}, Elisa V. and {Tenenbaum}, Peter and {Twicken}, Joseph D. and {Wu}, Hayley},
        title = "{Kepler Mission Design, Realized Photometric Performance, and Early Science}",
      journal = {\apjl},
         year = 2010,
        month = apr,
       volume = {713},
       number = {2},
        pages = {L79-L86},
          doi = {10.1088/2041-8205/713/2/L79},
archivePrefix = {arXiv},
       eprint = {1001.0268},
 primaryClass = {astro-ph.EP},
       adsurl = {https://ui.adsabs.harvard.edu/abs/2010ApJ...713L..79K}
}

@ARTICLE{Suarez2010,
       author = {{Su{\'a}rez}, J.~C. and {Goupil}, M.~J. and {Reese}, D.~R. and {Samadi}, R. and {Ligni{\`e}res}, F. and {Rieutord}, M. and {Lochard}, J.},
        title = "{On the Interpretation of Echelle Diagrams for Solar-like Oscillations Effect of Centrifugal Distortion}",
      journal = {\apj},
         year = 2010,
        month = sep,
       volume = {721},
       number = {1},
        pages = {537-546},
          doi = {10.1088/0004-637X/721/1/537},
archivePrefix = {arXiv},
       eprint = {1009.0123},
 primaryClass = {astro-ph.SR},
       adsurl = {https://ui.adsabs.harvard.edu/abs/2010ApJ...721..537S}
}

@ARTICLE{Breger1999,
       author = {{Breger}, M. and {Handler}, G. and {Garrido}, R. and {Audard}, N. and {Zima}, W. and {Papar{\'o}}, M. and {Beichbuchner}, F. and {Li}, Zhi-Ping and {Jiang}, Shi-Yang and {Liu}, Zong-Li and {Zhou}, Ai-Ying and {Pikall}, H. and {Stankov}, A. and {Guzik}, J.~A. and {Sperl}, M. and {Krzesinski}, J. and {Ogloza}, W. and {Pajdosz}, G. and {Zola}, S. and {Thomassen}, T. and {Solheim}, J.-E. and {Serkowitsch}, E. and {Reegen}, P. and {Rumpf}, T. and {Schmalwieser}, A. and {Montgomery}, M.~H.},
        title = "{30+ frequencies for the delta Scuti variable 4 Canum Venaticorum: results of the 1996 multisite campaign}",
      journal = {\aap},
         year = 1999,
        month = sep,
       volume = {349},
        pages = {225-235},
       adsurl = {https://ui.adsabs.harvard.edu/abs/1999A&A...349..225B}
}

@ARTICLE{Bowman2018,
       author = {{Bowman}, Dominic M. and {Kurtz}, Donald W.},
        title = "{Characterizing the observational properties of {\ensuremath{\delta}} Sct stars in the era of space photometry from the Kepler mission}",
      journal = {\mnras},
         year = 2018,
        month = may,
       volume = {476},
       number = {3},
        pages = {3169-3184},
          doi = {10.1093/mnras/sty449},
archivePrefix = {arXiv},
       eprint = {1802.05433},
 primaryClass = {astro-ph.SR},
       adsurl = {https://ui.adsabs.harvard.edu/abs/2018MNRAS.476.3169B}
}

@ARTICLE{Lecoanet2022,
       author = {{Lecoanet}, Daniel and {Bowman}, Dominic M. and {Van Reeth}, Timothy},
        title = "{Asteroseismic inference of the near-core magnetic field strength in the main-sequence B star HD 43317}",
      journal = {\mnras},
         year = 2022,
        month = may,
       volume = {512},
       number = {1},
        pages = {L16-L20},
          doi = {10.1093/mnrasl/slac013},
archivePrefix = {arXiv},
       eprint = {2202.03440},
 primaryClass = {astro-ph.SR},
       adsurl = {https://ui.adsabs.harvard.edu/abs/2022MNRAS.512L..16L}
}

@ARTICLE{Berry2025,
       author = {{Berry}, Ian and {Huber}, Daniel and {Li}, Yaguang and {Hey}, Daniel and {Bedding}, Timothy R. and {Murphy}, Simon J.},
        title = "{Discovery of 79 {\ensuremath{\delta}} Scuti Stars in NGC 3532 Suggests a Decrease of Pulsator Occurrence with Age}",
      journal = {\apj},
         year = 2025,
        month = dec,
       volume = {995},
       number = {1},
          eid = {128},
        pages = {128},
          doi = {10.3847/1538-4357/ae18c6},
archivePrefix = {arXiv},
       eprint = {2510.20048},
 primaryClass = {astro-ph.SR},
       adsurl = {https://ui.adsabs.harvard.edu/abs/2025ApJ...995..128B}
}

@INPROCEEDINGS{Smolec2014,
       author = {{Smolec}, Rados{\l}aw},
        title = "{Mode selection in pulsating stars}",
    booktitle = {Precision Asteroseismology},
         year = 2014,
       editor = {{Guzik}, Joyce A. and {Chaplin}, William J. and {Handler}, Gerald and {Pigulski}, Andrzej},
       series = {IAU Symposium},
       volume = {301},
        month = feb,
        pages = {265-272},
          doi = {10.1017/S1743921313014439},
archivePrefix = {arXiv},
       eprint = {1309.5959},
 primaryClass = {astro-ph.SR},
       adsurl = {https://ui.adsabs.harvard.edu/abs/2014IAUS..301..265S}
}

@ARTICLE{Rappaport2026,
       author = {{Rappaport}, S.~A. and {Jayaraman}, R. and {Handler}, G. and {Kurtz}, D. and {Zhang}, V. and {Gagliano}, R. and {Powell}, B. and {Fuller}, J. and {Borkovits}, T. and {Kostov}, V. and {Daszy{\'n}ska-Daszkiewicz}, J.},
        title = "{Discovery of the First Octupole Pulsation Mode in a delta Scuti Star: A Stationary l = 3 Sectoral Mode}",
      journal = {arXiv e-prints},
         year = 2026,
        month = apr,
          eid = {arXiv:2604.18836},
        pages = {arXiv:2604.18836},
          doi = {10.48550/arXiv.2604.18836},
archivePrefix = {arXiv},
       eprint = {2604.18836},
 primaryClass = {astro-ph.SR},
       adsurl = {https://ui.adsabs.harvard.edu/abs/2026arXiv260418836R}
}

@ARTICLE{Mathys1985,
       author = {{Mathys}, G.},
        title = "{The influence of surface inhomogeneities on the rapid light variations of AP stars.}",
      journal = {\aap},
         year = 1985,
        month = oct,
       volume = {151},
        pages = {315-321},
       adsurl = {https://ui.adsabs.harvard.edu/abs/1985A&A...151..315M}
}

@ARTICLE{Jagoda2002,
       author = {{Daszy{\'n}ska-Daszkiewicz}, J. and {Dziembowski}, W.~A. and {Pamyatnykh}, A.~A. and {Goupil}, M.-J.},
        title = "{Photometric amplitudes and phases of nonradial oscillation in rotating stars}",
      journal = {\aap},
         year = 2002,
        month = sep,
       volume = {392},
        pages = {151-159},
          doi = {10.1051/0004-6361:20020911},
archivePrefix = {arXiv},
       eprint = {astro-ph/0206109},
 primaryClass = {astro-ph},
       adsurl = {https://ui.adsabs.harvard.edu/abs/2002A&A...392..151D}
}

@ARTICLE{Fitch1981,
       author = {{Fitch}, W.~S.},
        title = "{L=0, 1, 2, and 3 pulsation constants for evolutionary models of del SCT stars.}",
      journal = {\apj},
         year = 1981,
        month = oct,
       volume = {249},
        pages = {218-227},
          doi = {10.1086/159278},
       adsurl = {https://ui.adsabs.harvard.edu/abs/1981ApJ...249..218F}
}

@INPROCEEDINGS{Breger2000,
       author = {{Breger}, M.},
        title = "{{\ensuremath{\delta}} Scuti stars (Review)}",
    booktitle = {Delta Scuti and Related Stars},
         year = 2000,
       editor = {{Breger}, Michel and {Montgomery}, Michael},
       series = {Astronomical Society of the Pacific Conference Series},
       volume = {210},
        month = jan,
        pages = {3},
       adsurl = {https://ui.adsabs.harvard.edu/abs/2000ASPC..210....3B}
}

@ARTICLE{lenz2005,
       author = {{Lenz}, P. and {Breger}, M.},
        title = "{Period04 User Guide}",
      journal = {Communications in Asteroseismology},
         year = 2005,
        month = jun,
       volume = {146},
        pages = {53-136},
          doi = {10.1553/cia146s53},
       adsurl = {https://ui.adsabs.harvard.edu/abs/2005CoAst.146...53L}
}

@ARTICLE{bedding2023,
       author = {{Bedding}, Timothy R. and {Murphy}, Simon J. and {Crawford}, Courtney and {Hey}, Daniel R. and {Huber}, Daniel and {Kjeldsen}, Hans and {Li}, Yaguang and {Mann}, Andrew W. and {Torres}, Guillermo and {White}, Timothy R. and {Zhou}, George},
        title = "{TESS Observations of the Pleiades Cluster: A Nursery for {\ensuremath{\delta}} Scuti Stars}",
      journal = {\apjl},
         year = 2023,
        month = mar,
       volume = {946},
       number = {1},
          eid = {L10},
        pages = {L10},
          doi = {10.3847/2041-8213/acc17a},
archivePrefix = {arXiv},
       eprint = {2212.12087},
 primaryClass = {astro-ph.SR},
       adsurl = {https://ui.adsabs.harvard.edu/abs/2023ApJ...946L..10B}
}

@ARTICLE{brown2011,
       author = {{Brown}, Timothy M. and {Latham}, David W. and {Everett}, Mark E. and {Esquerdo}, Gilbert A.},
        title = "{Kepler Input Catalog: Photometric Calibration and Stellar Classification}",
      journal = {\aj},
         year = 2011,
        month = oct,
       volume = {142},
       number = {4},
          eid = {112},
        pages = {112},
          doi = {10.1088/0004-6256/142/4/112},
archivePrefix = {arXiv},
       eprint = {1102.0342},
 primaryClass = {astro-ph.SR},
       adsurl = {https://ui.adsabs.harvard.edu/abs/2011AJ....142..112B}
}

@ARTICLE{stassun2018,
       author = {{Stassun}, Keivan G. and {Oelkers}, Ryan J. and {Pepper}, Joshua and {Paegert}, Martin and {De Lee}, Nathan and {Torres}, Guillermo and {Latham}, David W. and {Charpinet}, St{\'e}phane and {Dressing}, Courtney D. and {Huber}, Daniel and {Kane}, Stephen R. and {L{\'e}pine}, S{\'e}bastien and {Mann}, Andrew and {Muirhead}, Philip S. and {Rojas-Ayala}, B{\'a}rbara and {Silvotti}, Roberto and {Fleming}, Scott W. and {Levine}, Al and {Plavchan}, Peter},
        title = "{The TESS Input Catalog and Candidate Target List}",
      journal = {\aj},
         year = 2018,
        month = sep,
       volume = {156},
       number = {3},
          eid = {102},
        pages = {102},
          doi = {10.3847/1538-3881/aad050},
archivePrefix = {arXiv},
       eprint = {1706.00495},
 primaryClass = {astro-ph.EP},
       adsurl = {https://ui.adsabs.harvard.edu/abs/2018AJ....156..102S}
}

@ARTICLE{creevey2023,
       author = {{Creevey}, O.~L. and {Sordo}, R. and {Pailler}, F. and {Fr{\'e}mat}, Y. and {Heiter}, U. and {Th{\'e}venin}, F. and {Andrae}, R. and {Fouesneau}, M. and {Lobel}, A. and {Bailer-Jones}, C.~A.~L. and {Garabato}, D. and {Bellas-Velidis}, I. and {Brugaletta}, E. and {Lorca}, A. and {Ordenovic}, C. and {Palicio}, P.~A. and {Sarro}, L.~M. and {Delchambre}, L. and {Drimmel}, R. and {Rybizki}, J. and {Torralba Elipe}, G. and {Korn}, A.~J. and {Recio-Blanco}, A. and {Schultheis}, M.~S. and {De Angeli}, F. and {Montegriffo}, P. and {Abreu Aramburu}, A. and {Accart}, S. and {{\'A}lvarez}, M.~A. and {Bakker}, J. and {Brouillet}, N. and {Burlacu}, A. and {Carballo}, R. and {Casamiquela}, L. and {Chiavassa}, A. and {Contursi}, G. and {Cooper}, W.~J. and {Dafonte}, C. and {Dapergolas}, A. and {de Laverny}, P. and {Dharmawardena}, T.~E. and {Edvardsson}, B. and {Le Fustec}, Y. and {Garc{\'\i}a-Lario}, P. and {Garc{\'\i}a-Torres}, M. and {Gomez}, A. and {Gonz{\'a}lez-Santamar{\'\i}a}, I. and {Hatzidimitriou}, D. and {Jean-Antoine Piccolo}, A. and {Kontiza}, M. and {Kordopatis}, G. and {Lanzafame}, A.~C. and {Lebreton}, Y. and {Licata}, E.~L. and {Lindstr{\o}m}, H.~E.~P. and {Livanou}, E. and {Magdaleno Romeo}, A. and {Manteiga}, M. and {Marocco}, F. and {Marshall}, D.~J. and {Mary}, N. and {Nicolas}, C. and {Pallas-Quintela}, L. and {Panem}, C. and {Pichon}, B. and {Poggio}, E. and {Riclet}, F. and {Robin}, C. and {Santove{\~n}a}, R. and {Silvelo}, A. and {Slezak}, I. and {Smart}, R.~L. and {Soubiran}, C. and {S{\"u}veges}, M. and {Ulla}, A. and {Utrilla}, E. and {Vallenari}, A. and {Zhao}, H. and {Zorec}, J. and {Barrado}, D. and {Bijaoui}, A. and {Bouret}, J.-C. and {Blomme}, R. and {Brott}, I. and {Cassisi}, S. and {Kochukhov}, O. and {Martayan}, C. and {Shulyak}, D. and {Silvester}, J.},
        title = "{Gaia Data Release 3. Astrophysical parameters inference system (Apsis). I. Methods and content overview}",
      journal = {\aap},
         year = 2023,
        month = jun,
       volume = {674},
          eid = {A26},
        pages = {A26},
          doi = {10.1051/0004-6361/202243688},
archivePrefix = {arXiv},
       eprint = {2206.05864},
 primaryClass = {astro-ph.GA},
       adsurl = {https://ui.adsabs.harvard.edu/abs/2023A&A...674A..26C}
}

@ARTICLE{balona1994,
       author = {{Balona}, L.~A.},
        title = "{Effective Temperature Bolometric Correction and Mass Calibration of O-F}",
      journal = {\mnras},
         year = 1994,
        month = may,
       volume = {268},
        pages = {119},
          doi = {10.1093/mnras/268.1.119},
       adsurl = {https://ui.adsabs.harvard.edu/abs/1994MNRAS.268..119B}
}

@ARTICLE{theado2009,
       author = {{Th{\'e}ado}, S. and {Vauclair}, S. and {Alecian}, G. and {LeBlanc}, F.},
        title = "{Influence of Thermohaline Convection on Diffusion-Induced Iron Accumulation in a Stars}",
      journal = {\apj},
         year = 2009,
        month = oct,
       volume = {704},
       number = {2},
        pages = {1262-1273},
          doi = {10.1088/0004-637X/704/2/1262},
archivePrefix = {arXiv},
       eprint = {0908.1534},
 primaryClass = {astro-ph.SR},
       adsurl = {https://ui.adsabs.harvard.edu/abs/2009ApJ...704.1262T}
}

@ARTICLE{deal2016,
       author = {{Deal}, Morgan and {Richard}, Olivier and {Vauclair}, Sylvie},
        title = "{Hydrodynamical instabilities induced by atomic diffusion in A stars and their consequences}",
      journal = {\aap},
         year = 2016,
        month = may,
       volume = {589},
          eid = {A140},
        pages = {A140},
          doi = {10.1051/0004-6361/201628180},
archivePrefix = {arXiv},
       eprint = {1604.01241},
 primaryClass = {astro-ph.SR},
       adsurl = {https://ui.adsabs.harvard.edu/abs/2016A&A...589A.140D}
}

@ARTICLE{bigot2002,
       author = {{Bigot}, L. and {Dziembowski}, W.~A.},
        title = "{The oblique pulsator model revisited}",
      journal = {\aap},
         year = 2002,
        month = aug,
       volume = {391},
        pages = {235-245},
          doi = {10.1051/0004-6361:20020824},
       adsurl = {https://ui.adsabs.harvard.edu/abs/2002A&A...391..235B}
}

@ARTICLE{murphy2019,
       author = {{Murphy}, Simon J. and {Hey}, Daniel and {Van Reeth}, Timothy and {Bedding}, Timothy R.},
        title = "{Gaia-derived luminosities of Kepler A/F stars and the pulsator fraction across the {\ensuremath{\delta}} Scuti instability strip}",
      journal = {\mnras},
         year = 2019,
        month = may,
       volume = {485},
       number = {2},
        pages = {2380-2400},
          doi = {10.1093/mnras/stz590},
archivePrefix = {arXiv},
       eprint = {1903.00015},
 primaryClass = {astro-ph.SR},
       adsurl = {https://ui.adsabs.harvard.edu/abs/2019MNRAS.485.2380M}
}

@ARTICLE{Bouabid2013,
       author = {{Bouabid}, M.-P. and {Dupret}, M.-A. and {Salmon}, S. and {Montalb{\'a}n}, J. and {Miglio}, A. and {Noels}, A.},
        title = "{Effects of the Coriolis force on high-order g modes in {\ensuremath{\gamma}} Doradus stars}",
      journal = {\mnras},
         year = 2013,
        month = mar,
       volume = {429},
       number = {3},
        pages = {2500-2514},
          doi = {10.1093/mnras/sts517},
       adsurl = {https://ui.adsabs.harvard.edu/abs/2013MNRAS.429.2500B}
}

@ARTICLE{Houdek2015,
       author = {{Houdek}, G{\"u}nter and {Dupret}, Marc-Antoine},
        title = "{Interaction Between Convection and Pulsation}",
      journal = {Living Reviews in Solar Physics},
         year = 2015,
        month = dec,
       volume = {12},
       number = {1},
          eid = {8},
        pages = {8},
          doi = {10.1007/lrsp-2015-8},
archivePrefix = {arXiv},
       eprint = {1601.03913},
 primaryClass = {astro-ph.SR},
       adsurl = {https://ui.adsabs.harvard.edu/abs/2015LRSP...12....8H}
}

@ARTICLE{Scuflaire2008a,
   author = {{Scuflaire}, R. and {Th{\'e}ado}, S. and {Montalb{\'a}n}, J. and 
	{Miglio}, A. and {Bourge}, P.-O. and {Godart}, M. and {Thoul}, A. and 
	{Noels}, A.},
    title = "{CL{\'E}S, Code Li{\'e}geois d'{\'E}volution Stellaire}",
  journal = {ApSS},
     year = 2008,
    month = aug,
   volume = 316,
    pages = {83-91}
}

@ARTICLE{Dupret2001,
       author = {{Dupret}, M.~A.},
        title = "{Nonradial nonadiabatic stellar pulsations: A numerical method and its application to a beta Cephei model}",
      journal = {\aap},
         year = 2001,
        month = jan,
       volume = {366},
        pages = {166-173},
          doi = {10.1051/0004-6361:20000219},
       adsurl = {https://ui.adsabs.harvard.edu/abs/2001A&A...366..166D}
}

@ARTICLE{Dupret2002,
       author = {{Dupret}, M. -A. and {De Ridder}, J. and {Neuforge}, C. and {Aerts}, C. and {Scuflaire}, R.},
        title = "{Influence of non-adiabatic temperature variations on line profile variations of slowly rotating beta Cep stars and SPBs. I. Non-adiabatic eigenfunctions in the atmosphere of a pulsating star}",
      journal = {\aap},
         year = 2002,
        month = apr,
       volume = {385},
        pages = {563-571},
          doi = {10.1051/0004-6361:20020193},
       adsurl = {https://ui.adsabs.harvard.edu/abs/2002A&A...385..563D}
}

@ARTICLE{Dupret2005,
       author = {{Dupret}, M. -A. and {Grigahc{\`e}ne}, A. and {Garrido}, R. and {Gabriel}, M. and {Scuflaire}, R.},
        title = "{Convection-pulsation coupling. II. Excitation and stabilization mechanisms in {\ensuremath{\delta}} Sct and {\ensuremath{\gamma}} Dor stars}",
      journal = {\aap},
         year = 2005,
        month = jun,
       volume = {435},
       number = {3},
        pages = {927-939},
          doi = {10.1051/0004-6361:20041817},
       adsurl = {https://ui.adsabs.harvard.edu/abs/2005A&A...435..927D}
}

@ARTICLE{Grigahcene2005,
       author = {{Grigahc{\`e}ne}, A. and {Dupret}, M. -A. and {Gabriel}, M. and {Garrido}, R. and {Scuflaire}, R.},
        title = "{Convection-pulsation coupling. I. A mixing-length perturbative theory}",
      journal = {\aap},
         year = 2005,
        month = may,
       volume = {434},
       number = {3},
        pages = {1055-1062},
          doi = {10.1051/0004-6361:20041816},
       adsurl = {https://ui.adsabs.harvard.edu/abs/2005A&A...434.1055G}
}

@BOOK{Cox1968,
   author = {{Cox}, J.~P. and {Giuli}, R.~T.},
    title = "{Principles of stellar structure }",
booktitle = {Principles of stellar structure, by J.P.~Cox and R.~T.~ Giuli.~ New York: Gordon and Breach, 1968},
     year = 1968,
   adsurl = {http://adsabs.harvard.edu/abs/1968pss..book.....C}
}

@ARTICLE{Asplund2009,
       author = {{Asplund}, Martin and {Grevesse}, Nicolas and {Sauval}, A. Jacques and
         {Scott}, Pat},
        title = "{The Chemical Composition of the Sun}",
      journal = {ARA\&A},
         year = 2009,
        month = sep,
       volume = {47},
       number = {1},
        pages = {481-522},
          doi = {10.1146/annurev.astro.46.060407.145222},
archivePrefix = {arXiv},
       eprint = {0909.0948},
 primaryClass = {astro-ph.SR},
       adsurl = {https://ui.adsabs.harvard.edu/abs/2009ARA&A..47..481A}
}

@MISC{Irwin2012,
       author = {{Irwin}, Alan W.},
        title = "{FreeEOS: Equation of State for stellar interiors calculations}",
         year = 2012,
        month = nov,
          eid = {ascl:1211.002},
        pages = {ascl:1211.002},
archivePrefix = {ascl},
       eprint = {1211.002},
       adsurl = {https://ui.adsabs.harvard.edu/abs/2012ascl.soft11002I}
}

@ARTICLE{Iglesias1996,
   author = {{Iglesias}, C.~A. and {Rogers}, F.~J.},
    title = "{Updated Opal Opacities}",
  journal = {ApJ},
     year = 1996,
    month = jun,
   volume = 464,
    pages = {943}
}

@ARTICLE{Atmosphere1981,
   author = {{Vernazza}, J.~E. and {Avrett}, E.~H. and {Loeser}, R.},
    title = "{Structure of the solar chromosphere. III - Models of the EUV brightness components of the quiet-sun}",
  journal = {\apjs},
     year = 1981,
    month = apr,
   volume = 45,
    pages = {635-725},
      doi = {10.1086/190731},
   adsurl = {http://adsabs.harvard.edu/abs/1981ApJS...45..635V}
}

@ARTICLE{Reaction2011,
       author = {{Adelberger}, E.~G. and {Garc{\'\i}a}, A. and {Robertson}, R.~G. Hamish and
         {Snover}, K.~A. and {Balantekin}, A.~B. and {Heeger}, K. and
         {Ramsey-Musolf}, M.~J. and {Bemmerer}, D. and {Junghans}, A. and
         {Bertulani}, C.~A. and {Chen}, J. -W. and {Costantini}, H. and
         {Prati}, P. and {Couder}, M. and {Uberseder}, E. and {Wiescher}, M. and
         {Cyburt}, R. and {Davids}, B. and {Freedman}, S.~J. and {Gai}, M. and
         {Gazit}, D. and {Gialanella}, L. and {Imbriani}, G. and {Greife}, U. and
         {Hass}, M. and {Haxton}, W.~C. and {Itahashi}, T. and {Kubodera}, K. and
         {Langanke}, K. and {Leitner}, D. and {Leitner}, M. and {Vetter}, P. and
         {Winslow}, L. and {Marcucci}, L.~E. and {Motobayashi}, T. and
         {Mukhamedzhanov}, A. and {Tribble}, R.~E. and {Nollett}, Kenneth M. and
         {Nunes}, F.~M. and {Park}, T. -S. and {Parker}, P.~D. and
         {Schiavilla}, R. and {Simpson}, E.~C. and {Spitaleri}, C. and
         {Strieder}, F. and {Trautvetter}, H. -P. and {Suemmerer}, K. and
         {Typel}, S.},
        title = "{Solar fusion cross sections. II. The pp chain and CNO cycles}",
      journal = {Reviews of Modern Physics},
         year = 2011,
        month = jan,
       volume = {83},
       number = {1},
        pages = {195-246},
          doi = {10.1103/RevModPhys.83.195},
archivePrefix = {arXiv},
       eprint = {1004.2318},
 primaryClass = {nucl-ex},
       adsurl = {https://ui.adsabs.harvard.edu/abs/2011RvMP...83..195A}
}

@ARTICLE{Dziembowski2016,
       author = {{Dziembowski}, W.~A.},
        title = "{Nonradial oscillations in classical pulsating stars. Predictions and discoveries}",
      journal = {Commmunications of the Konkoly Observatory Hungary},
         year = 2016,
        month = may,
       volume = {105},
        pages = {23-30},
          doi = {10.48550/arXiv.1512.03708},
archivePrefix = {arXiv},
       eprint = {1512.03708},
 primaryClass = {astro-ph.SR},
       adsurl = {https://ui.adsabs.harvard.edu/abs/2016CoKon.105...23D}
}

@ARTICLE{2021ApJ...907L..33S,
       author = {{Stassun}, Keivan G. and {Torres}, Guillermo},
        title = "{Parallax Systematics and Photocenter Motions of Benchmark Eclipsing Binaries in Gaia EDR3}",
      journal = {\apjl},
         year = 2021,
        month = feb,
       volume = {907},
       number = {2},
          eid = {L33},
        pages = {L33},
          doi = {10.3847/2041-8213/abdaad},
archivePrefix = {arXiv},
       eprint = {2101.03425},
 primaryClass = {astro-ph.SR},
       adsurl = {https://ui.adsabs.harvard.edu/abs/2021ApJ...907L..33S}
}

@ARTICLE{2024NewAR..9801694E,
       author = {{El-Badry}, Kareem},
        title = "{Gaia's binary star renaissance}",
      journal = {\nar},
         year = 2024,
        month = jun,
       volume = {98},
          eid = {101694},
        pages = {101694},
          doi = {10.1016/j.newar.2024.101694},
archivePrefix = {arXiv},
       eprint = {2403.12146},
 primaryClass = {astro-ph.SR},
       adsurl = {https://ui.adsabs.harvard.edu/abs/2024NewAR..9801694E}
}

@article{Gaia,
       author = {{Gaia Collaboration} and {Prusti}, T. and {de Bruijne}, J.~H.~J. and {Brown}, A.~G.~A. and {Vallenari}, A. and {Babusiaux}, C. and {Bailer-Jones}, C.~A.~L. and {Bastian}, U. and {Biermann}, M. and {Evans}, D.~W. and {Eyer}, L. and {Jansen}, F. and {Jordi}, C. and {Klioner}, S.~A. and {Lammers}, U. and {Lindegren}, L. and {Luri}, X. and {Mignard}, F. and {Milligan}, D.~J. and {Panem}, C. and {Poinsignon}, V. and {Pourbaix}, D. and {Randich}, S. and {Sarri}, G. and {Sartoretti}, P. and {Siddiqui}, H.~I. and {Soubiran}, C. and {Valette}, V. and {van Leeuwen}, F. and {Walton}, N.~A. and {Aerts}, C. and {Arenou}, F. and {Cropper}, M. and {Drimmel}, R. and {H{\o}g}, E. and {Katz}, D. and {Lattanzi}, M.~G. and {O'Mullane}, W. and {Grebel}, E.~K. and {Holland}, A.~D. and {Huc}, C. and {Passot}, X. and {Bramante}, L. and {Cacciari}, C. and {Casta{\~n}eda}, J. and {Chaoul}, L. and {Cheek}, N. and {De Angeli}, F. and {Fabricius}, C. and {Guerra}, R. and {Hern{\'a}ndez}, J. and {Jean-Antoine-Piccolo}, A. and {Masana}, E. and {Messineo}, R. and {Mowlavi}, N. and {Nienartowicz}, K. and {Ord{\'o}{\~n}ez-Blanco}, D. and {Panuzzo}, P. and {Portell}, J. and {Richards}, P.~J. and {Riello}, M. and {Seabroke}, G.~M. and {Tanga}, P. and {Th{\'e}venin}, F. and {Torra}, J. and {Els}, S.~G. and {Gracia-Abril}, G. and {Comoretto}, G. and {Garcia-Reinaldos}, M. and {Lock}, T. and {Mercier}, E. and {Altmann}, M. and {Andrae}, R. and {Astraatmadja}, T.~L. and {Bellas-Velidis}, I. and {Benson}, K. and {Berthier}, J. and {Blomme}, R. and {Busso}, G. and {Carry}, B. and {Cellino}, A. and {Clementini}, G. and {Cowell}, S. and {Creevey}, O. and {Cuypers}, J. and {Davidson}, M. and {De Ridder}, J. and {de Torres}, A. and {Delchambre}, L. and {Dell'Oro}, A. and {Ducourant}, C. and {Fr{\'e}mat}, Y. and {Garc{\'\i}a-Torres}, M. and {Gosset}, E. and {Halbwachs}, J. -L. and {Hambly}, N.~C. and {Harrison}, D.~L. and {Hauser}, M. and {Hestroffer}, D. and {Hodgkin}, S.~T. and {Huckle}, H.~E. and {Hutton}, A. and {Jasniewicz}, G. and {Jordan}, S. and {Kontizas}, M. and {Korn}, A.~J. and {Lanzafame}, A.~C. and {Manteiga}, M. and {Moitinho}, A. and {Muinonen}, K. and {Osinde}, J. and {Pancino}, E. and {Pauwels}, T. and {Petit}, J. -M. and {Recio-Blanco}, A. and {Robin}, A.~C. and {Sarro}, L.~M. and {Siopis}, C. and {Smith}, M. and {Smith}, K.~W. and {Sozzetti}, A. and {Thuillot}, W. and {van Reeven}, W. and {Viala}, Y. and {Abbas}, U. and {Abreu Aramburu}, A. and {Accart}, S. and {Aguado}, J.~J. and {Allan}, P.~M. and {Allasia}, W. and {Altavilla}, G. and {{\'A}lvarez}, M.~A. and {Alves}, J. and {Anderson}, R.~I. and {Andrei}, A.~H. and {Anglada Varela}, E. and {Antiche}, E. and {Antoja}, T. and {Ant{\'o}n}, S. and {Arcay}, B. and {Atzei}, A. and {Ayache}, L. and {Bach}, N. and {Baker}, S.~G. and {Balaguer-N{\'u}{\~n}ez}, L. and {Barache}, C. and {Barata}, C. and {Barbier}, A. and {Barblan}, F. and {Baroni}, M. and {Barrado y Navascu{\'e}s}, D. and {Barros}, M. and {Barstow}, M.~A. and {Becciani}, U. and {Bellazzini}, M. and {Bellei}, G. and {Bello Garc{\'\i}a}, A. and {Belokurov}, V. and {Bendjoya}, P. and {Berihuete}, A. and {Bianchi}, L. and {Bienaym{\'e}}, O. and {Billebaud}, F. and {Blagorodnova}, N. and {Blanco-Cuaresma}, S. and {Boch}, T. and {Bombrun}, A. and {Borrachero}, R. and {Bouquillon}, S. and {Bourda}, G. and {Bouy}, H. and {Bragaglia}, A. and {Breddels}, M.~A. and {Brouillet}, N. and {Br{\"u}semeister}, T. and {Bucciarelli}, B. and {Budnik}, F. and {Burgess}, P. and {Burgon}, R. and {Burlacu}, A. and {Busonero}, D. and {Buzzi}, R. and {Caffau}, E. and {Cambras}, J. and {Campbell}, H. and {Cancelliere}, R. and {Cantat-Gaudin}, T. and {Carlucci}, T. and {Carrasco}, J.~M. and {Castellani}, M. and {Charlot}, P. and {Charnas}, J. and {Charvet}, P. and {Chassat}, F. and {Chiavassa}, A. and {Clotet}, M. and {Cocozza}, G. and {Collins}, R.~S. and {Collins}, P. and {Costigan}, G. and {Crifo}, F. and {Cross}, N.~J.~G. and {Crosta}, M. and {Crowley}, C. and {Dafonte}, C. and {Damerdji}, Y. and {Dapergolas}, A. and {David}, P. and {David}, M. and {De Cat}, P. and {de Felice}, F. and {de Laverny}, P. and {De Luise}, F. and {De March}, R. and {de Martino}, D. and {de Souza}, R. and {Debosscher}, J. and {del Pozo}, E. and {Delbo}, M. and {Delgado}, A. and {Delgado}, H.~E. and {di Marco}, F. and {Di Matteo}, P. and {Diakite}, S. and {Distefano}, E. and {Dolding}, C. and {Dos Anjos}, S. and {Drazinos}, P. and {Dur{\'a}n}, J. and {Dzigan}, Y. and {Ecale}, E. and {Edvardsson}, B. and {Enke}, H. and {Erdmann}, M. and {Escolar}, D. and {Espina}, M. and {Evans}, N.~W. and {Eynard Bontemps}, G. and {Fabre}, C. and {Fabrizio}, M. and {Faigler}, S. and {Falc{\~a}o}, A.~J. and {Farr{\`a}s Casas}, M. and {Faye}, F. and {Federici}, L. and {Fedorets}, G. and {Fern{\'a}ndez-Hern{\'a}ndez}, J. and {Fernique}, P. and {Fienga}, A. and {Figueras}, F. and {Filippi}, F. and {Findeisen}, K. and {Fonti}, A. and {Fouesneau}, M. and {Fraile}, E. and {Fraser}, M. and {Fuchs}, J. and {Furnell}, R. and {Gai}, M. and {Galleti}, S. and {Galluccio}, L. and {Garabato}, D. and {Garc{\'\i}a-Sedano}, F. and {Gar{\'e}}, P. and {Garofalo}, A. and {Garralda}, N. and {Gavras}, P. and {Gerssen}, J. and {Geyer}, R. and {Gilmore}, G. and {Girona}, S. and {Giuffrida}, G. and {Gomes}, M. and {Gonz{\'a}lez-Marcos}, A. and {Gonz{\'a}lez-N{\'u}{\~n}ez}, J. and {Gonz{\'a}lez-Vidal}, J.~J. and {Granvik}, M. and {Guerrier}, A. and {Guillout}, P. and {Guiraud}, J. and {G{\'u}rpide}, A. and {Guti{\'e}rrez-S{\'a}nchez}, R. and {Guy}, L.~P. and {Haigron}, R. and {Hatzidimitriou}, D. and {Haywood}, M. and {Heiter}, U. and {Helmi}, A. and {Hobbs}, D. and {Hofmann}, W. and {Holl}, B. and {Holland}, G. and {Hunt}, J.~A.~S. and {Hypki}, A. and {Icardi}, V. and {Irwin}, M. and {Jevardat de Fombelle}, G. and {Jofr{\'e}}, P. and {Jonker}, P.~G. and {Jorissen}, A. and {Julbe}, F. and {Karampelas}, A. and {Kochoska}, A. and {Kohley}, R. and {Kolenberg}, K. and {Kontizas}, E. and {Koposov}, S.~E. and {Kordopatis}, G. and {Koubsky}, P. and {Kowalczyk}, A. and {Krone-Martins}, A. and {Kudryashova}, M. and {Kull}, I. and {Bachchan}, R.~K. and {Lacoste-Seris}, F. and {Lanza}, A.~F. and {Lavigne}, J. -B. and {Le Poncin-Lafitte}, C. and {Lebreton}, Y. and {Lebzelter}, T. and {Leccia}, S. and {Leclerc}, N. and {Lecoeur-Taibi}, I. and {Lemaitre}, V. and {Lenhardt}, H. and {Leroux}, F. and {Liao}, S. and {Licata}, E. and {Lindstr{\o}m}, H.~E.~P. and {Lister}, T.~A. and {Livanou}, E. and {Lobel}, A. and {L{\"o}ffler}, W. and {L{\'o}pez}, M. and {Lopez-Lozano}, A. and {Lorenz}, D. and {Loureiro}, T. and {MacDonald}, I. and {Magalh{\~a}es Fernandes}, T. and {Managau}, S. and {Mann}, R.~G. and {Mantelet}, G. and {Marchal}, O. and {Marchant}, J.~M. and {Marconi}, M. and {Marie}, J. and {Marinoni}, S. and {Marrese}, P.~M. and {Marschalk{\'o}}, G. and {Marshall}, D.~J. and {Mart{\'\i}n-Fleitas}, J.~M. and {Martino}, M. and {Mary}, N. and {Matijevi{\v{c}}}, G. and {Mazeh}, T. and {McMillan}, P.~J. and {Messina}, S. and {Mestre}, A. and {Michalik}, D. and {Millar}, N.~R. and {Miranda}, B.~M.~H. and {Molina}, D. and {Molinaro}, R. and {Molinaro}, M. and {Moln{\'a}r}, L. and {Moniez}, M. and {Montegriffo}, P. and {Monteiro}, D. and {Mor}, R. and {Mora}, A. and {Morbidelli}, R. and {Morel}, T. and {Morgenthaler}, S. and {Morley}, T. and {Morris}, D. and {Mulone}, A.~F. and {Muraveva}, T. and {Musella}, I. and {Narbonne}, J. and {Nelemans}, G. and {Nicastro}, L. and {Noval}, L. and {Ord{\'e}novic}, C. and {Ordieres-Mer{\'e}}, J. and {Osborne}, P. and {Pagani}, C. and {Pagano}, I. and {Pailler}, F. and {Palacin}, H. and {Palaversa}, L. and {Parsons}, P. and {Paulsen}, T. and {Pecoraro}, M. and {Pedrosa}, R. and {Pentik{\"a}inen}, H. and {Pereira}, J. and {Pichon}, B. and {Piersimoni}, A.~M. and {Pineau}, F. -X. and {Plachy}, E. and {Plum}, G. and {Poujoulet}, E. and {Pr{\v{s}}a}, A. and {Pulone}, L. and {Ragaini}, S. and {Rago}, S. and {Rambaux}, N. and {Ramos-Lerate}, M. and {Ranalli}, P. and {Rauw}, G. and {Read}, A. and {Regibo}, S. and {Renk}, F. and {Reyl{\'e}}, C. and {Ribeiro}, R.~A. and {Rimoldini}, L. and {Ripepi}, V. and {Riva}, A. and {Rixon}, G. and {Roelens}, M. and {Romero-G{\'o}mez}, M. and {Rowell}, N. and {Royer}, F. and {Rudolph}, A. and {Ruiz-Dern}, L. and {Sadowski}, G. and {Sagrist{\`a} Sell{\'e}s}, T. and {Sahlmann}, J. and {Salgado}, J. and {Salguero}, E. and {Sarasso}, M. and {Savietto}, H. and {Schnorhk}, A. and {Schultheis}, M. and {Sciacca}, E. and {Segol}, M. and {Segovia}, J.~C. and {Segransan}, D. and {Serpell}, E. and {Shih}, I. -C. and {Smareglia}, R. and {Smart}, R.~L. and {Smith}, C. and {Solano}, E. and {Solitro}, F. and {Sordo}, R. and {Soria Nieto}, S. and {Souchay}, J. and {Spagna}, A. and {Spoto}, F. and {Stampa}, U. and {Steele}, I.~A. and {Steidelm{\"u}ller}, H. and {Stephenson}, C.~A. and {Stoev}, H. and {Suess}, F.~F. and {S{\"u}veges}, M. and {Surdej}, J. and {Szabados}, L. and {Szegedi-Elek}, E. and {Tapiador}, D. and {Taris}, F. and {Tauran}, G. and {Taylor}, M.~B. and {Teixeira}, R. and {Terrett}, D. and {Tingley}, B. and {Trager}, S.~C. and {Turon}, C. and {Ulla}, A. and {Utrilla}, E. and {Valentini}, G. and {van Elteren}, A. and {Van Hemelryck}, E. and {van Leeuwen}, M. and {Varadi}, M. and {Vecchiato}, A. and {Veljanoski}, J. and {Via}, T. and {Vicente}, D. and {Vogt}, S. and {Voss}, H. and {Votruba}, V. and {Voutsinas}, S. and {Walmsley}, G. and {Weiler}, M. and {Weingrill}, K. and {Werner}, D. and {Wevers}, T. and {Whitehead}, G. and {Wyrzykowski}, {\L}. and {Yoldas}, A. and {{\v{Z}}erjal}, M. and {Zucker}, S. and {Zurbach}, C. and {Zwitter}, T. and {Alecu}, A. and {Allen}, M. and {Allende Prieto}, C. and {Amorim}, A. and {Anglada-Escud{\'e}}, G. and {Arsenijevic}, V. and {Azaz}, S. and {Balm}, P. and {Beck}, M. and {Bernstein}, H. -H. and {Bigot}, L. and {Bijaoui}, A. and {Blasco}, C. and {Bonfigli}, M. and {Bono}, G. and {Boudreault}, S. and {Bressan}, A. and {Brown}, S. and {Brunet}, P. -M. and {Bunclark}, P. and {Buonanno}, R. and {Butkevich}, A.~G. and {Carret}, C. and {Carrion}, C. and {Chemin}, L. and {Ch{\'e}reau}, F. and {Corcione}, L. and {Darmigny}, E. and {de Boer}, K.~S. and {de Teodoro}, P. and {de Zeeuw}, P.~T. and {Delle Luche}, C. and {Domingues}, C.~D. and {Dubath}, P. and {Fodor}, F. and {Fr{\'e}zouls}, B. and {Fries}, A. and {Fustes}, D. and {Fyfe}, D. and {Gallardo}, E. and {Gallegos}, J. and {Gardiol}, D. and {Gebran}, M. and {Gomboc}, A. and {G{\'o}mez}, A. and {Grux}, E. and {Gueguen}, A. and {Heyrovsky}, A. and {Hoar}, J. and {Iannicola}, G. and {Isasi Parache}, Y. and {Janotto}, A. -M. and {Joliet}, E. and {Jonckheere}, A. and {Keil}, R. and {Kim}, D. -W. and {Klagyivik}, P. and {Klar}, J. and {Knude}, J. and {Kochukhov}, O. and {Kolka}, I. and {Kos}, J. and {Kutka}, A. and {Lainey}, V. and {LeBouquin}, D. and {Liu}, C. and {Loreggia}, D. and {Makarov}, V.~V. and {Marseille}, M.~G. and {Martayan}, C. and {Martinez-Rubi}, O. and {Massart}, B. and {Meynadier}, F. and {Mignot}, S. and {Munari}, U. and {Nguyen}, A. -T. and {Nordlander}, T. and {Ocvirk}, P. and {O'Flaherty}, K.~S. and {Olias Sanz}, A. and {Ortiz}, P. and {Osorio}, J. and {Oszkiewicz}, D. and {Ouzounis}, A. and {Palmer}, M. and {Park}, P. and {Pasquato}, E. and {Peltzer}, C. and {Peralta}, J. and {P{\'e}turaud}, F. and {Pieniluoma}, T. and {Pigozzi}, E. and {Poels}, J. and {Prat}, G. and {Prod'homme}, T. and {Raison}, F. and {Rebordao}, J.~M. and {Risquez}, D. and {Rocca-Volmerange}, B. and {Rosen}, S. and {Ruiz-Fuertes}, M.~I. and {Russo}, F. and {Sembay}, S. and {Serraller Vizcaino}, I. and {Short}, A. and {Siebert}, A. and {Silva}, H. and {Sinachopoulos}, D. and {Slezak}, E. and {Soffel}, M. and {Sosnowska}, D. and {Strai{\v{z}}ys}, V. and {ter Linden}, M. and {Terrell}, D. and {Theil}, S. and {Tiede}, C. and {Troisi}, L. and {Tsalmantza}, P. and {Tur}, D. and {Vaccari}, M. and {Vachier}, F. and {Valles}, P. and {Van Hamme}, W. and {Veltz}, L. and {Virtanen}, J. and {Wallut}, J. -M. and {Wichmann}, R. and {Wilkinson}, M.~I. and {Ziaeepour}, H. and {Zschocke}, S.},
        title = "{The Gaia mission}",
      journal = {\aap},
         year = 2016,
        month = nov,
       volume = {595},
          eid = {A1},
        pages = {A1},
          doi = {10.1051/0004-6361/201629272},
archivePrefix = {arXiv},
       eprint = {1609.04153},
 primaryClass = {astro-ph.IM},
       adsurl = {https://ui.adsabs.harvard.edu/abs/2016A&A...595A...1G}
}

@misc{GaiaDR3,
  doi = {10.48550/ARXIV.2208.00211},
  url = {https://arxiv.org/abs/2208.00211},
  author = {{Gaia Collaboration} and {Vallenari}, A. and {Brown}, A.~G.~A. and {Prusti}, T. and {de Bruijne}, J.~H.~J. and {Arenou}, F. and {Babusiaux}, C. and {Biermann}, M. and {Creevey}, O.~L. and {Ducourant}, C. and {Evans}, D.~W. and {Eyer}, L. and {Guerra}, R. and {Hutton}, A. and {Jordi}, C. and {Klioner}, S.~A. and {Lammers}, U.~L. and {Lindegren}, L. and {Luri}, X. and {Mignard}, F. and {Panem}, C. and {Pourbaix}, D. and {Randich}, S. and {Sartoretti}, P. and {Soubiran}, C. and {Tanga}, P. and {Walton}, N.~A. and {Bailer-Jones}, C.~A.~L. and {Bastian}, U. and {Drimmel}, R. and {Jansen}, F. and {Katz}, D. and {Lattanzi}, M.~G. and {van Leeuwen}, F. and {Bakker}, J. and {Cacciari}, C. and {Casta{\~n}eda}, J. and {De Angeli}, F. and {Fabricius}, C. and {Fouesneau}, M. and {Fr{\'e}mat}, Y. and {Galluccio}, L. and {Guerrier}, A. and {Heiter}, U. and {Masana}, E. and {Messineo}, R. and {Mowlavi}, N. and {Nicolas}, C. and {Nienartowicz}, K. and {Pailler}, F. and {Panuzzo}, P. and {Riclet}, F. and {Roux}, W. and {Seabroke}, G.~M. and {Sordo}, R. and {Th{\'e}venin}, F. and {Gracia-Abril}, G. and {Portell}, J. and {Teyssier}, D. and {Altmann}, M. and {Andrae}, R. and {Audard}, M. and {Bellas-Velidis}, I. and {Benson}, K. and {Berthier}, J. and {Blomme}, R. and {Burgess}, P.~W. and {Busonero}, D. and {Busso}, G. and {C{\'a}novas}, H. and {Carry}, B. and {Cellino}, A. and {Cheek}, N. and {Clementini}, G. and {Damerdji}, Y. and {Davidson}, M. and {de Teodoro}, P. and {Nu{\~n}ez Campos}, M. and {Delchambre}, L. and {Dell'Oro}, A. and {Esquej}, P. and {Fern{\'a}ndez-Hern{\'a}ndez}, J. and {Fraile}, E. and {Garabato}, D. and {Garc{\'\i}a-Lario}, P. and {Gosset}, E. and {Haigron}, R. and {Halbwachs}, J. -L. and {Hambly}, N.~C. and {Harrison}, D.~L. and {Hern{\'a}ndez}, J. and {Hestroffer}, D. and {Hodgkin}, S.~T. and {Holl}, B. and {Jan{\ss}en}, K. and {Jevardat de Fombelle}, G. and {Jordan}, S. and {Krone-Martins}, A. and {Lanzafame}, A.~C. and {L{\"o}ffler}, W. and {Marchal}, O. and {Marrese}, P.~M. and {Moitinho}, A. and {Muinonen}, K. and {Osborne}, P. and {Pancino}, E. and {Pauwels}, T. and {Recio-Blanco}, A. and {Reyl{\'e}}, C. and {Riello}, M. and {Rimoldini}, L. and {Roegiers}, T. and {Rybizki}, J. and {Sarro}, L.~M. and {Siopis}, C. and {Smith}, M. and {Sozzetti}, A. and {Utrilla}, E. and {van Leeuwen}, M. and {Abbas}, U. and {{\'A}brah{\'a}m}, P. and {Abreu Aramburu}, A. and {Aerts}, C. and {Aguado}, J.~J. and {Ajaj}, M. and {Aldea-Montero}, F. and {Altavilla}, G. and {{\'A}lvarez}, M.~A. and {Alves}, J. and {Anders}, F. and {Anderson}, R.~I. and {Anglada Varela}, E. and {Antoja}, T. and {Baines}, D. and {Baker}, S.~G. and {Balaguer-N{\'u}{\~n}ez}, L. and {Balbinot}, E. and {Balog}, Z. and {Barache}, C. and {Barbato}, D. and {Barros}, M. and {Barstow}, M.~A. and {Bartolom{\'e}}, S. and {Bassilana}, J. -L. and {Bauchet}, N. and {Becciani}, U. and {Bellazzini}, M. and {Berihuete}, A. and {Bernet}, M. and {Bertone}, S. and {Bianchi}, L. and {Binnenfeld}, A. and {Blanco-Cuaresma}, S. and {Blazere}, A. and {Boch}, T. and {Bombrun}, A. and {Bossini}, D. and {Bouquillon}, S. and {Bragaglia}, A. and {Bramante}, L. and {Breedt}, E. and {Bressan}, A. and {Brouillet}, N. and {Brugaletta}, E. and {Bucciarelli}, B. and {Burlacu}, A. and {Butkevich}, A.~G. and {Buzzi}, R. and {Caffau}, E. and {Cancelliere}, R. and {Cantat-Gaudin}, T. and {Carballo}, R. and {Carlucci}, T. and {Carnerero}, M.~I. and {Carrasco}, J.~M. and {Casamiquela}, L. and {Castellani}, M. and {Castro-Ginard}, A. and {Chaoul}, L. and {Charlot}, P. and {Chemin}, L. and {Chiaramida}, V. and {Chiavassa}, A. and {Chornay}, N. and {Comoretto}, G. and {Contursi}, G. and {Cooper}, W.~J. and {Cornez}, T. and {Cowell}, S. and {Crifo}, F. and {Cropper}, M. and {Crosta}, M. and {Crowley}, C. and {Dafonte}, C. and {Dapergolas}, A. and {David}, M. and {David}, P. and {de Laverny}, P. and {De Luise}, F. and {De March}, R. and {De Ridder}, J. and {de Souza}, R. and {de Torres}, A. and {del Peloso}, E.~F. and {del Pozo}, E. and {Delbo}, M. and {Delgado}, A. and {Delisle}, J. -B. and {Demouchy}, C. and {Dharmawardena}, T.~E. and {Di Matteo}, P. and {Diakite}, S. and {Diener}, C. and {Distefano}, E. and {Dolding}, C. and {Edvardsson}, B. and {Enke}, H. and {Fabre}, C. and {Fabrizio}, M. and {Faigler}, S. and {Fedorets}, G. and {Fernique}, P. and {Fienga}, A. and {Figueras}, F. and {Fournier}, Y. and {Fouron}, C. and {Fragkoudi}, F. and {Gai}, M. and {Garcia-Gutierrez}, A. and {Garcia-Reinaldos}, M. and {Garc{\'\i}a-Torres}, M. and {Garofalo}, A. and {Gavel}, A. and {Gavras}, P. and {Gerlach}, E. and {Geyer}, R. and {Giacobbe}, P. and {Gilmore}, G. and {Girona}, S. and {Giuffrida}, G. and {Gomel}, R. and {Gomez}, A. and {Gonz{\'a}lez-N{\'u}{\~n}ez}, J. and {Gonz{\'a}lez-Santamar{\'\i}a}, I. and {Gonz{\'a}lez-Vidal}, J.~J. and {Granvik}, M. and {Guillout}, P. and {Guiraud}, J. and {Guti{\'e}rrez-S{\'a}nchez}, R. and {Guy}, L.~P. and {Hatzidimitriou}, D. and {Hauser}, M. and {Haywood}, M. and {Helmer}, A. and {Helmi}, A. and {Sarmiento}, M.~H. and {Hidalgo}, S.~L. and {Hilger}, T. and {H{\l}adczuk}, N. and {Hobbs}, D. and {Holland}, G. and {Huckle}, H.~E. and {Jardine}, K. and {Jasniewicz}, G. and {Jean-Antoine Piccolo}, A. and {Jim{\'e}nez-Arranz}, {\'O}. and {Jorissen}, A. and {Juaristi Campillo}, J. and {Julbe}, F. and {Karbevska}, L. and {Kervella}, P. and {Khanna}, S. and {Kontizas}, M. and {Kordopatis}, G. and {Korn}, A.~J. and {K{\'o}sp{\'a}l}, {\'A}. and {Kostrzewa-Rutkowska}, Z. and {Kruszy{\'n}ska}, K. and {Kun}, M. and {Laizeau}, P. and {Lambert}, S. and {Lanza}, A.~F. and {Lasne}, Y. and {Le Campion}, J. -F. and {Lebreton}, Y. and {Lebzelter}, T. and {Leccia}, S. and {Leclerc}, N. and {Lecoeur-Taibi}, I. and {Liao}, S. and {Licata}, E.~L. and {Lindstr{\o}m}, H.~E.~P. and {Lister}, T.~A. and {Livanou}, E. and {Lobel}, A. and {Lorca}, A. and {Loup}, C. and {Madrero Pardo}, P. and {Magdaleno Romeo}, A. and {Managau}, S. and {Mann}, R.~G. and {Manteiga}, M. and {Marchant}, J.~M. and {Marconi}, M. and {Marcos}, J. and {Marcos Santos}, M.~M.~S. and {Mar{\'\i}n Pina}, D. and {Marinoni}, S. and {Marocco}, F. and {Marshall}, D.~J. and {Martin Polo}, L. and {Mart{\'\i}n-Fleitas}, J.~M. and {Marton}, G. and {Mary}, N. and {Masip}, A. and {Massari}, D. and {Mastrobuono-Battisti}, A. and {Mazeh}, T. and {McMillan}, P.~J. and {Messina}, S. and {Michalik}, D. and {Millar}, N.~R. and {Mints}, A. and {Molina}, D. and {Molinaro}, R. and {Moln{\'a}r}, L. and {Monari}, G. and {Mongui{\'o}}, M. and {Montegriffo}, P. and {Montero}, A. and {Mor}, R. and {Mora}, A. and {Morbidelli}, R. and {Morel}, T. and {Morris}, D. and {Muraveva}, T. and {Murphy}, C.~P. and {Musella}, I. and {Nagy}, Z. and {Noval}, L. and {Oca{\~n}a}, F. and {Ogden}, A. and {Ordenovic}, C. and {Osinde}, J.~O. and {Pagani}, C. and {Pagano}, I. and {Palaversa}, L. and {Palicio}, P.~A. and {Pallas-Quintela}, L. and {Panahi}, A. and {Payne-Wardenaar}, S. and {Pe{\~n}alosa Esteller}, X. and {Penttil{\"a}}, A. and {Pichon}, B. and {Piersimoni}, A.~M. and {Pineau}, F. -X. and {Plachy}, E. and {Plum}, G. and {Poggio}, E. and {Pr{\v{s}}a}, A. and {Pulone}, L. and {Racero}, E. and {Ragaini}, S. and {Rainer}, M. and {Raiteri}, C.~M. and {Rambaux}, N. and {Ramos}, P. and {Ramos-Lerate}, M. and {Re Fiorentin}, P. and {Regibo}, S. and {Richards}, P.~J. and {Rios Diaz}, C. and {Ripepi}, V. and {Riva}, A. and {Rix}, H. -W. and {Rixon}, G. and {Robichon}, N. and {Robin}, A.~C. and {Robin}, C. and {Roelens}, M. and {Rogues}, H.~R.~O. and {Rohrbasser}, L. and {Romero-G{\'o}mez}, M. and {Rowell}, N. and {Royer}, F. and {Ruz Mieres}, D. and {Rybicki}, K.~A. and {Sadowski}, G. and {S{\'a}ez N{\'u}{\~n}ez}, A. and {Sagrist{\`a} Sell{\'e}s}, A. and {Sahlmann}, J. and {Salguero}, E. and {Samaras}, N. and {Sanchez Gimenez}, V. and {Sanna}, N. and {Santove{\~n}a}, R. and {Sarasso}, M. and {Schultheis}, M. and {Sciacca}, E. and {Segol}, M. and {Segovia}, J.~C. and {S{\'e}gransan}, D. and {Semeux}, D. and {Shahaf}, S. and {Siddiqui}, H.~I. and {Siebert}, A. and {Siltala}, L. and {Silvelo}, A. and {Slezak}, E. and {Slezak}, I. and {Smart}, R.~L. and {Snaith}, O.~N. and {Solano}, E. and {Solitro}, F. and {Souami}, D. and {Souchay}, J. and {Spagna}, A. and {Spina}, L. and {Spoto}, F. and {Steele}, I.~A. and {Steidelm{\"u}ller}, H. and {Stephenson}, C.~A. and {S{\"u}veges}, M. and {Surdej}, J. and {Szabados}, L. and {Szegedi-Elek}, E. and {Taris}, F. and {Taylor}, M.~B. and {Teixeira}, R. and {Tolomei}, L. and {Tonello}, N. and {Torra}, F. and {Torra}, J. and {Torralba Elipe}, G. and {Trabucchi}, M. and {Tsounis}, A.~T. and {Turon}, C. and {Ulla}, A. and {Unger}, N. and {Vaillant}, M.~V. and {van Dillen}, E. and {van Reeven}, W. and {Vanel}, O. and {Vecchiato}, A. and {Viala}, Y. and {Vicente}, D. and {Voutsinas}, S. and {Weiler}, M. and {Wevers}, T. and {Wyrzykowski}, {\L}. and {Yoldas}, A. and {Yvard}, P. and {Zhao}, H. and {Zorec}, J. and {Zucker}, S. and {Zwitter}, T.},
  title = {Gaia Data Release 3: Summary of the content and survey properties},
  publisher = {arXiv},
  year = {2022},
  journal = {A\&A}
}

@article{Pamyatnykh1999,
       author = {{Pamyatnykh}, A.~A.},
        title = "{Pulsational Instability Domains in the Upper Main Sequence}",
      journal = {\actaa},
         year = 1999,
        month = jun,
       volume = {49},
        pages = {119-148},
       adsurl = {https://ui.adsabs.harvard.edu/abs/1999AcA....49..119P}
}

@ARTICLE{houdek2008,
       author = {{Houdek}, G.},
        title = "{The effect of convection on pulsational stability}",
      journal = {Communications in Asteroseismology},
         year = 2008,
        month = dec,
       volume = {157},
        pages = {137-143},
          doi = {10.48550/arXiv.0810.5228},
archivePrefix = {arXiv},
       eprint = {0810.5228},
 primaryClass = {astro-ph},
       adsurl = {https://ui.adsabs.harvard.edu/abs/2008CoAst.157..137H}
}

@article{antoci2014,
    doi = {10.1088/0004-637X/796/2/118},
    url = {https://dx.doi.org/10.1088/0004-637X/796/2/118},
    year = {2014},
    publisher = {The American Astronomical Society},
    volume = {796},
    number = {2},
    pages = {118},
    author = {V. Antoci and M. Cunha and G. Houdek and H. Kjeldsen and R. Trampedach and G. Handler and T. Lüftinger and T. Arentoft and S. Murphy},
    title = {THE ROLE OF TURBULENT PRESSURE AS A COHERENT PULSATIONAL DRIVING MECHANISM: THE CASE OF THE δ SCUTI STAR HD 187547},
    journal = {\apj}
}

@article{antoci2019,
       author = {{Antoci}, V. and {Cunha}, M.~S. and {Bowman}, D.~M. and {Murphy}, S.~J. and {Kurtz}, D.~W. and {Bedding}, T.~R. and {Borre}, C.~C. and {Christophe}, S. and {Daszy{\'n}ska-Daszkiewicz}, J. and {Fox-Machado}, L. and {Garc{\'\i}a Hern{\'a}ndez}, A. and {Ghasemi}, H. and {Handberg}, R. and {Hansen}, H. and {Hasanzadeh}, A. and {Houdek}, G. and {Johnston}, C. and {Justesen}, A.~B. and {Kahraman Alicavus}, F. and {Kotysz}, K. and {Latham}, D. and {Matthews}, J.~M. and {M{\o}nster}, J. and {Niemczura}, E. and {Paunzen}, E. and {S{\'a}nchez Arias}, J.~P. and {Pigulski}, A. and {Pepper}, J. and {Richey-Yowell}, T. and {Safari}, H. and {Seager}, S. and {Smalley}, B. and {Shutt}, T. and {S{\'o}dor}, A. and {Su{\'a}rez}, J. -C. and {Tkachenko}, A. and {Wu}, T. and {Zwintz}, K. and {Barcel{\'o} Forteza}, S. and {Brunsden}, E. and {Bogn{\'a}r}, Z. and {Buzasi}, D.~L. and {Chowdhury}, S. and {De Cat}, P. and {Evans}, J.~A. and {Guo}, Z. and {Guzik}, J.~A. and {Jevtic}, N. and {Lampens}, P. and {Lares Martiz}, M. and {Lovekin}, C. and {Li}, G. and {Mirouh}, G.~M. and {Mkrtichian}, D. and {Monteiro}, M.~J.~P.~F.~G. and {Nemec}, J.~M. and {Ouazzani}, R. -M. and {Pascual-Granado}, J. and {Reese}, D.~R. and {Rieutord}, M. and {Rodon}, J.~R. and {Skarka}, M. and {Sowicka}, P. and {Stateva}, I. and {Szab{\'o}}, R. and {Weiss}, W.~W.},
        title = "{The first view of {\ensuremath{\delta}} Scuti and {\ensuremath{\gamma}} Doradus stars with the TESS mission}",
      journal = {MNRAS},
         year = 2019,
       volume = {490},
       number = {3},
        pages = {4040-4059},
          doi = {10.1093/mnras/stz2787},
       adsurl = {https://ui.adsabs.harvard.edu/abs/2019MNRAS.490.4040A}
}

@article{guzik2000,
    doi = {10.1086/312908},
    year = {2000},
    volume = {542},
    number = {1},
    pages = {L57},
    author = {Joyce A. Guzik and Anthony B. Kaye and Paul A. Bradley and Arthur N. Cox and Corinne Neuforge},
    title = {Driving the Gravity-Mode Pulsations in γ Doradus
    Variables},
    journal = {\apj}
}

@article{dupret2004,
	author = {Dupret, M.-A. and {Grigahc\`ene, A.} and {Garrido, R.} and {Gabriel, M.} and {Scuflaire, R.}},
	title = {Theoretical instability strips for ti and radus stars},
	DOI= "10.1051/0004-6361:20031740",
	url= "https://doi.org/10.1051/0004-6361:20031740",
	journal = {A\&A},
	year = 2004,
	volume = 414,
	number = 2,
	pages = "L17-L20",
}

@ARTICLE{Gautam2025,
       author = {{Gautam}, Anuj and {Murphy}, Simon J. and {Bedding}, Timothy R.},
        title = "{Modelling $δ$ Scuti pulsations: A new grid of p, g, and f modes across pre-main-sequence to post-main-sequence evolution}",
      journal = {arXiv e-prints},
         year = 2025,
        month = jul,
          eid = {arXiv:2507.03561},
        pages = {arXiv:2507.03561},
          doi = {10.48550/arXiv.2507.03561},
archivePrefix = {arXiv},
       eprint = {2507.03561},
 primaryClass = {astro-ph.SR},
       adsurl = {https://ui.adsabs.harvard.edu/abs/2025arXiv250703561G}
}

@ARTICLE{murphy2013,
       author = {{Murphy}, Simon J. and {Shibahashi}, Hiromoto and {Kurtz}, Donald W.},
        title = "{Super-Nyquist asteroseismology with the Kepler Space Telescope}",
      journal = {\mnras},
         year = 2013,
        month = apr,
       volume = {430},
       number = {4},
        pages = {2986-2998},
          doi = {10.1093/mnras/stt105},
archivePrefix = {arXiv},
       eprint = {1212.5603},
 primaryClass = {astro-ph.SR},
       adsurl = {https://ui.adsabs.harvard.edu/abs/2013MNRAS.430.2986M}
}

@ARTICLE{pourbaix2004,
       author = {{Pourbaix}, D. and {Tokovinin}, A.~A. and {Batten}, A.~H. and {Fekel}, F.~C. and {Hartkopf}, W.~I. and {Levato}, H. and {Morrell}, N.~I. and {Torres}, G. and {Udry}, S.},
        title = "{S$_{B$^{9}$}$: The ninth catalogue of spectroscopic binary orbits}",
      journal = {\aap},
         year = 2004,
        month = sep,
       volume = {424},
        pages = {727-732},
          doi = {10.1051/0004-6361:20041213},
archivePrefix = {arXiv},
       eprint = {astro-ph/0406573},
 primaryClass = {astro-ph},
       adsurl = {https://ui.adsabs.harvard.edu/abs/2004A&A...424..727P}
}

@article{renson2009,
    author = {Renson, P. and Manfroid, J.},
    year = {2009},
    pages = {961 - 966},
    title = {Catalogue of Ap, HgMn and Am stars},
    volume = {498},
    number = {3},
    journal = {A\&A},
    doi = {https://doi.org/10.1051/0004-6361/200810788}
}

@ARTICLE{barac2022,
       author = {{Barac}, Natascha and {Bedding}, Timothy R. and {Murphy}, Simon J. and {Hey}, Daniel R.},
        title = "{Revisiting bright {\ensuremath{\delta}} Scuti stars and their period-luminosity relation with TESS and Gaia DR3}",
      journal = {\mnras},
         year = 2022,
        month = oct,
       volume = {516},
       number = {2},
        pages = {2080-2094},
          doi = {10.1093/mnras/stac2132},
archivePrefix = {arXiv},
       eprint = {2207.00343},
 primaryClass = {astro-ph.SR},
       adsurl = {https://ui.adsabs.harvard.edu/abs/2022MNRAS.516.2080B}
}

@ARTICLE{bailerjones2021,
       author = {{Bailer-Jones}, C.~A.~L. and {Rybizki}, J. and {Fouesneau}, M. and {Demleitner}, M. and {Andrae}, R.},
        title = "{Estimating Distances from Parallaxes. V. Geometric and Photogeometric Distances to 1.47 Billion Stars in Gaia Early Data Release 3}",
      journal = {\aj},
         year = 2021,
        month = mar,
       volume = {161},
       number = {3},
          eid = {147},
        pages = {147},
          doi = {10.3847/1538-3881/abd806},
archivePrefix = {arXiv},
       eprint = {2012.05220},
 primaryClass = {astro-ph.SR},
       adsurl = {https://ui.adsabs.harvard.edu/abs/2021AJ....161..147B}
}

@ARTICLE{lallement2019,
       author = {{Lallement}, R. and {Babusiaux}, C. and {Vergely}, J.~L. and {Katz}, D. and {Arenou}, F. and {Valette}, B. and {Hottier}, C. and {Capitanio}, L.},
        title = "{Gaia-2MASS 3D maps of Galactic interstellar dust within 3 kpc}",
      journal = {\aap},
         year = 2019,
        month = may,
       volume = {625},
          eid = {A135},
        pages = {A135},
          doi = {10.1051/0004-6361/201834695},
archivePrefix = {arXiv},
       eprint = {1902.04116},
 primaryClass = {astro-ph.GA},
       adsurl = {https://ui.adsabs.harvard.edu/abs/2019A&A...625A.135L}
}

@ARTICLE{jordi2010,
       author = {{Jordi}, C. and {Gebran}, M. and {Carrasco}, J.~M. and {de Bruijne}, J. and {Voss}, H. and {Fabricius}, C. and {Knude}, J. and {Vallenari}, A. and {Kohley}, R. and {Mora}, A.},
        title = "{Gaia broad band photometry}",
      journal = {\aap},
         year = 2010,
        month = nov,
       volume = {523},
          eid = {A48},
        pages = {A48},
          doi = {10.1051/0004-6361/201015441},
archivePrefix = {arXiv},
       eprint = {1008.0815},
 primaryClass = {astro-ph.IM},
       adsurl = {https://ui.adsabs.harvard.edu/abs/2010A&A...523A..48J}
}

@ARTICLE{tian2023,
       author = {{Tian}, Xiao-man and {Wang}, Zhi-hua and {Zhu}, Li-ying and {Yang}, Xiao-Ling},
        title = "{A New Catalog of Am-type Chemically Peculiar Stars Based on LAMOST}",
      journal = {\apjs},
         year = 2023,
        month = may,
       volume = {266},
       number = {1},
          eid = {14},
        pages = {14},
          doi = {10.3847/1538-4365/acc4b5},
       adsurl = {https://ui.adsabs.harvard.edu/abs/2023ApJS..266...14T}
}

@ARTICLE{duerfeldt2024,
       author = {{D{\"u}rfeldt-Pedros}, O. and {Antoci}, V. and {Smalley}, B. and {Murphy}, S. and {Posilek}, N. and {Niemczura}, E.},
        title = "{Variability and stellar pulsation incidence in Am and Fm stars using TESS and Gaia data}",
      journal = {\aap},
         year = 2024,
        month = oct,
       volume = {690},
          eid = {A104},
        pages = {A104},
          doi = {10.1051/0004-6361/202349076},
archivePrefix = {arXiv},
       eprint = {2408.11657},
 primaryClass = {astro-ph.SR},
       adsurl = {https://ui.adsabs.harvard.edu/abs/2024A&A...690A.104D}
}

@ARTICLE{frasca2016,
       author = {{Frasca}, A. and {Molenda-{\.Z}akowicz}, J. and {De Cat}, P. and {Catanzaro}, G. and {Fu}, J.~N. and {Ren}, A.~B. and {Luo}, A.~L. and {Shi}, J.~R. and {Wu}, Y. and {Zhang}, H.~T.},
        title = "{Activity indicators and stellar parameters of the Kepler targets. An application of the ROTFIT pipeline to LAMOST-Kepler stellar spectra}",
      journal = {\aap},
         year = 2016,
        month = oct,
       volume = {594},
          eid = {A39},
        pages = {A39},
          doi = {10.1051/0004-6361/201628337},
archivePrefix = {arXiv},
       eprint = {1606.09149},
 primaryClass = {astro-ph.SR},
       adsurl = {https://ui.adsabs.harvard.edu/abs/2016A&A...594A..39F}
}

@article{breger1993,
       author = {{Breger}, M. and {Stich}, J. and {Garrido}, R. and {Martin}, B. and {Jiang}, S. -Y. and {Li}, Z. -P. and {Hube}, D.~P. and {Ostermann}, W. and {Paparo}, M. and {Scheck}, M.},
        title = "{Nonradial pulsation of the delta Scuti star BU CANCRI in the Praesepe cluster.}",
      journal = {\aap},
         year = 1993,
        month = apr,
       volume = {271},
        pages = {482-486},
       adsurl = {https://ui.adsabs.harvard.edu/abs/1993A&A...271..482B}
}

@article{zwintz2020,
	author = {{Zwintz} and {Neiner, C.} and {Kochukhov, O.} and {Ryabchikova, T.} and {Pigulski, A.} and {M\"ullner, M.} and {Steindl, T.} and {Kuschnig, R.} and {Handler, G.} and {Moffat, A. F. J.} and {Pablo, H.} and {Popowicz, A.} and {Wade, G. A.}},
	title = {Beta Cas: The first Delta Scuti star with a dynamo magnetic field},
	DOI= "10.1051/0004-6361/202038210",
	url= "https://doi.org/10.1051/0004-6361/202038210",
	journal = {\aap},
	year = 2020,
	volume = 643,
	pages = "A110",
}

@article{lovekin2017,
       author = {{Lovekin}, C.~C. and {Guzik}, J.~A.},
        title = "{Convection and Overshoot in Models of {\ensuremath{\gamma}} Doradus and {\ensuremath{\delta}} Scuti Stars}",
      journal = {\apj},
         year = 2017,
        month = nov,
       volume = {849},
       number = {1},
          eid = {38},
        pages = {38},
          doi = {10.3847/1538-4357/aa8e01},
archivePrefix = {arXiv},
       eprint = {1709.06857},
 primaryClass = {astro-ph.SR},
       adsurl = {https://ui.adsabs.harvard.edu/abs/2017ApJ...849...38L}
}

@ARTICLE{murphy2018,
       author = {{Murphy}, Simon J. and {Moe}, Maxwell and {Kurtz}, Donald W. and {Bedding}, Timothy R. and {Shibahashi}, Hiromoto and {Boffin}, Henri M.~J.},
        title = "{Finding binaries from phase modulation of pulsating stars with Kepler: V. Orbital parameters, with eccentricity and mass-ratio distributions of 341 new binaries}",
      journal = {\mnras},
         year = 2018,
        month = mar,
       volume = {474},
       number = {4},
        pages = {4322-4346},
          doi = {10.1093/mnras/stx3049},
archivePrefix = {arXiv},
       eprint = {1712.00022},
 primaryClass = {astro-ph.SR},
       adsurl = {https://ui.adsabs.harvard.edu/abs/2018MNRAS.474.4322M}
}

@ARTICLE{Paxton2011,
       author = {{Paxton}, Bill and {Bildsten}, Lars and {Dotter}, Aaron and {Herwig}, Falk and {Lesaffre}, Pierre and {Timmes}, Frank},
        title = "{Modules for Experiments in Stellar Astrophysics (MESA)}",
      journal = {\apjs},
         year = 2011,
        month = jan,
       volume = {192},
       number = {1},
          eid = {3},
        pages = {3},
          doi = {10.1088/0067-0049/192/1/3},
archivePrefix = {arXiv},
       eprint = {1009.1622},
 primaryClass = {astro-ph.SR},
       adsurl = {https://ui.adsabs.harvard.edu/abs/2011ApJS..192....3P}
}

@ARTICLE{Paxton2013,
       author = {{Paxton}, Bill and {Cantiello}, Matteo and {Arras}, Phil and {Bildsten}, Lars and {Brown}, Edward F. and {Dotter}, Aaron and {Mankovich}, Christopher and {Montgomery}, M.~H. and {Stello}, Dennis and {Timmes}, F.~X. and {Townsend}, Richard},
        title = "{Modules for Experiments in Stellar Astrophysics (MESA): Planets, Oscillations, Rotation, and Massive Stars}",
      journal = {\apjs},
         year = 2013,
        month = sep,
       volume = {208},
       number = {1},
          eid = {4},
        pages = {4},
          doi = {10.1088/0067-0049/208/1/4},
archivePrefix = {arXiv},
       eprint = {1301.0319},
 primaryClass = {astro-ph.SR},
       adsurl = {https://ui.adsabs.harvard.edu/abs/2013ApJS..208....4P}
}

@ARTICLE{Paxton2015,
       author = {{Paxton}, Bill and {Marchant}, Pablo and {Schwab}, Josiah and {Bauer}, Evan B. and {Bildsten}, Lars and {Cantiello}, Matteo and {Dessart}, Luc and {Farmer}, R. and {Hu}, H. and {Langer}, N. and {Townsend}, R.~H.~D. and {Townsley}, Dean M. and {Timmes}, F.~X.},
        title = "{Modules for Experiments in Stellar Astrophysics (MESA): Binaries, Pulsations, and Explosions}",
      journal = {\apjs},
         year = 2015,
        month = sep,
       volume = {220},
       number = {1},
          eid = {15},
        pages = {15},
          doi = {10.1088/0067-0049/220/1/15},
archivePrefix = {arXiv},
       eprint = {1506.03146},
 primaryClass = {astro-ph.SR},
       adsurl = {https://ui.adsabs.harvard.edu/abs/2015ApJS..220...15P}
}

@ARTICLE{Paxton2018,
       author = {{Paxton}, Bill and {Schwab}, Josiah and {Bauer}, Evan B. and {Bildsten}, Lars and {Blinnikov}, Sergei and {Duffell}, Paul and {Farmer}, R. and {Goldberg}, Jared A. and {Marchant}, Pablo and {Sorokina}, Elena and {Thoul}, Anne and {Townsend}, Richard H.~D. and {Timmes}, F.~X.},
        title = "{Modules for Experiments in Stellar Astrophysics (MESA): Convective Boundaries, Element Diffusion, and Massive Star Explosions}",
      journal = {\apjs},
         year = 2018,
        month = feb,
       volume = {234},
       number = {2},
          eid = {34},
        pages = {34},
          doi = {10.3847/1538-4365/aaa5a8},
archivePrefix = {arXiv},
       eprint = {1710.08424},
 primaryClass = {astro-ph.SR},
       adsurl = {https://ui.adsabs.harvard.edu/abs/2018ApJS..234...34P}
}

@ARTICLE{Paxton2019,
       author = {{Paxton}, Bill and {Smolec}, R. and {Schwab}, Josiah and {Gautschy}, A. and {Bildsten}, Lars and {Cantiello}, Matteo and {Dotter}, Aaron and {Farmer}, R. and {Goldberg}, Jared A. and {Jermyn}, Adam S. and {Kanbur}, S.~M. and {Marchant}, Pablo and {Thoul}, Anne and {Townsend}, Richard H.~D. and {Wolf}, William M. and {Zhang}, Michael and {Timmes}, F.~X.},
        title = "{Modules for Experiments in Stellar Astrophysics (MESA): Pulsating Variable Stars, Rotation, Convective Boundaries, and Energy Conservation}",
      journal = {\apjs},
         year = 2019,
        month = jul,
       volume = {243},
       number = {1},
          eid = {10},
        pages = {10},
          doi = {10.3847/1538-4365/ab2241},
archivePrefix = {arXiv},
       eprint = {1903.01426},
 primaryClass = {astro-ph.SR},
       adsurl = {https://ui.adsabs.harvard.edu/abs/2019ApJS..243...10P}
}

@ARTICLE{Mombarg2021,
       author = {{Mombarg}, J.~S.~G. and {Van Reeth}, T. and {Aerts}, C.},
        title = "{Constraining stellar evolution theory with asteroseismology of {\ensuremath{\gamma}} Doradus stars using deep learning. Stellar masses, ages, and core-boundary mixing}",
      journal = {\aap},
         year = 2021,
        month = jun,
       volume = {650},
          eid = {A58},
        pages = {A58},
          doi = {10.1051/0004-6361/202039543},
archivePrefix = {arXiv},
       eprint = {2103.13394},
 primaryClass = {astro-ph.SR},
       adsurl = {https://ui.adsabs.harvard.edu/abs/2021A&A...650A..58M}
}

@ARTICLE{Jermyn2023,
       author = {{Jermyn}, Adam S. and {Bauer}, Evan B. and {Schwab}, Josiah and {Farmer}, R. and {Ball}, Warrick H. and {Bellinger}, Earl P. and {Dotter}, Aaron and {Joyce}, Meridith and {Marchant}, Pablo and {Mombarg}, Joey S.~G. and {Wolf}, William M. and {Sunny Wong}, Tin Long and {Cinquegrana}, Giulia C. and {Farrell}, Eoin and {Smolec}, R. and {Thoul}, Anne and {Cantiello}, Matteo and {Herwig}, Falk and {Toloza}, Odette and {Bildsten}, Lars and {Townsend}, Richard H.~D. and {Timmes}, F.~X.},
        title = "{Modules for Experiments in Stellar Astrophysics (MESA): Time-dependent Convection, Energy Conservation, Automatic Differentiation, and Infrastructure}",
      journal = {\apjs},
         year = 2023,
        month = mar,
       volume = {265},
       number = {1},
          eid = {15},
        pages = {15},
          doi = {10.3847/1538-4365/acae8d},
archivePrefix = {arXiv},
       eprint = {2208.03651},
 primaryClass = {astro-ph.SR},
       adsurl = {https://ui.adsabs.harvard.edu/abs/2023ApJS..265...15J}
}

@ARTICLE{Townsend2013,
       author = {{Townsend}, R.~H.~D. and {Teitler}, S.~A.},
        title = "{GYRE: an open-source stellar oscillation code based on a new Magnus Multiple Shooting scheme}",
      journal = {\mnras},
         year = 2013,
        month = nov,
       volume = {435},
       number = {4},
        pages = {3406-3418},
          doi = {10.1093/mnras/stt1533},
archivePrefix = {arXiv},
       eprint = {1308.2965},
 primaryClass = {astro-ph.SR},
       adsurl = {https://ui.adsabs.harvard.edu/abs/2013MNRAS.435.3406T}
}

@ARTICLE{Townsend2018,
       author = {{Townsend}, R.~H.~D. and {Goldstein}, J. and {Zweibel}, E.~G.},
        title = "{Angular momentum transport by heat-driven g-modes in slowly pulsating B stars}",
      journal = {\mnras},
         year = 2018,
        month = mar,
       volume = {475},
       number = {1},
        pages = {879-893},
          doi = {10.1093/mnras/stx3142},
archivePrefix = {arXiv},
       eprint = {1712.02420},
 primaryClass = {astro-ph.SR},
       adsurl = {https://ui.adsabs.harvard.edu/abs/2018MNRAS.475..879T}
}

@ARTICLE{Garcia2022,
       author = {{Garcia}, S. and {Van Reeth}, T. and {De Ridder}, J. and {Tkachenko}, A. and {IJspeert}, L. and {Aerts}, C.},
        title = "{Detection of period-spacing patterns due to the gravity modes of rotating dwarfs in the TESS southern continuous viewing zone}",
      journal = {\aap},
         year = 2022,
        month = jun,
       volume = {662},
          eid = {A82},
        pages = {A82},
          doi = {10.1051/0004-6361/202141926},
archivePrefix = {arXiv},
       eprint = {2202.10507},
 primaryClass = {astro-ph.SR},
       adsurl = {https://ui.adsabs.harvard.edu/abs/2022A&A...662A..82G}
}

@ARTICLE{Christophe2018,
       author = {{Christophe}, S. and {Ballot}, J. and {Ouazzani}, R. -M. and {Antoci}, V. and {Salmon}, S.~J.~A.~J.},
        title = "{Deciphering the oscillation spectrum of {\ensuremath{\gamma}} Doradus and SPB stars}",
      journal = {\aap},
         year = 2018,
        month = oct,
       volume = {618},
          eid = {A47},
        pages = {A47},
          doi = {10.1051/0004-6361/201832782},
archivePrefix = {arXiv},
       eprint = {1807.03707},
 primaryClass = {astro-ph.SR},
       adsurl = {https://ui.adsabs.harvard.edu/abs/2018A&A...618A..47C}
}

@ARTICLE{GangLi2020,
       author = {{Li}, Gang and {Van Reeth}, Timothy and {Bedding}, Timothy R. and {Murphy}, Simon J. and {Antoci}, Victoria and {Ouazzani}, Rhita-Maria and {Barbara}, Nicholas H.},
        title = "{Gravity-mode period spacings and near-core rotation rates of 611 {\ensuremath{\gamma}} Doradus stars with Kepler}",
      journal = {\mnras},
         year = 2020,
        month = jan,
       volume = {491},
       number = {3},
        pages = {3586-3605},
          doi = {10.1093/mnras/stz2906},
archivePrefix = {arXiv},
       eprint = {1910.06634},
 primaryClass = {astro-ph.SR},
       adsurl = {https://ui.adsabs.harvard.edu/abs/2020MNRAS.491.3586L}
}

@ARTICLE{Iglesias1993,
       author = {{Iglesias}, Carlos A. and {Rogers}, Forrest J.},
        title = "{Radiative Opacities for Carbon- and Oxygen-rich Mixtures}",
      journal = {\apj},
         year = 1993,
        month = aug,
       volume = {412},
        pages = {752},
          doi = {10.1086/172958},
       adsurl = {https://ui.adsabs.harvard.edu/abs/1993ApJ...412..752I}
}

@ARTICLE{GS98,
       author = {{Grevesse}, N. and {Sauval}, A.~J.},
        title = "{Standard Solar Composition}",
      journal = {\ssr},
         year = 1998,
        month = may,
       volume = {85},
        pages = {161-174},
          doi = {10.1023/A:1005161325181},
       adsurl = {https://ui.adsabs.harvard.edu/abs/1998SSRv...85..161G}
}

@ARTICLE{Kurtz2015,
       author = {{Kurtz}, Donald W. and {Shibahashi}, Hiromoto and {Murphy}, Simon J. and {Bedding}, Timothy R. and {Bowman}, Dominic M.},
        title = "{A unifying explanation of complex frequency spectra of {\ensuremath{\gamma}} Dor, SPB and Be stars: combination frequencies and highly non-sinusoidal light curves}",
      journal = {\mnras},
         year = 2015,
        month = jul,
       volume = {450},
       number = {3},
        pages = {3015-3029},
          doi = {10.1093/mnras/stv868},
archivePrefix = {arXiv},
       eprint = {1504.04245},
 primaryClass = {astro-ph.SR},
       adsurl = {https://ui.adsabs.harvard.edu/abs/2015MNRAS.450.3015K}
}

@ARTICLE{2023AJ....165..239P,
       author = {{Pedersen}, May G. and {Bell}, Keaton J.},
        title = "{Contamination in TESS Light Curves: The Case of the Fast Yellow Pulsating Supergiants}",
      journal = {\aj},
         year = 2023,
        month = jun,
       volume = {165},
       number = {6},
          eid = {239},
        pages = {239},
          doi = {10.3847/1538-3881/accc31},
archivePrefix = {arXiv},
       eprint = {2304.05706},
 primaryClass = {astro-ph.SR},
       adsurl = {https://ui.adsabs.harvard.edu/abs/2023AJ....165..239P}
}

@ARTICLE{2023A&A...676A..55L,
       author = {{Labadie-Bartz}, J. and {H{\"u}mmerich}, S. and {Bernhard}, K. and {Paunzen}, E. and {Shultz}, M.~E.},
        title = "{Photometric variability of the LAMOST sample of magnetic chemically peculiar stars as seen by TESS}",
      journal = {\aap},
         year = 2023,
        month = aug,
       volume = {676},
          eid = {A55},
        pages = {A55},
          doi = {10.1051/0004-6361/202346657},
archivePrefix = {arXiv},
       eprint = {2306.12861},
 primaryClass = {astro-ph.SR},
       adsurl = {https://ui.adsabs.harvard.edu/abs/2023A&A...676A..55L}
}

@ARTICLE{2023AJ....165..141H,
       author = {{Higgins}, Michael E. and {Bell}, Keaton J.},
        title = "{Localizing Sources of Variability in Crowded TESS Photometry}",
      journal = {\aj},
         year = 2023,
        month = apr,
       volume = {165},
       number = {4},
          eid = {141},
        pages = {141},
          doi = {10.3847/1538-3881/acb20c},
archivePrefix = {arXiv},
       eprint = {2204.06020},
 primaryClass = {astro-ph.IM},
       adsurl = {https://ui.adsabs.harvard.edu/abs/2023AJ....165..141H}
}

@MISC{2018ascl.soft12013L,
   author = {{Lightkurve Collaboration} and {Cardoso}, J.~V.~d.~M. and
             {Hedges}, C. and {Gully-Santiago}, M. and {Saunders}, N. and
             {Cody}, A.~M. and {Barclay}, T. and {Hall}, O. and
             {Sagear}, S. and {Turtelboom}, E. and {Zhang}, J. and
             {Tzanidakis}, A. and {Mighell}, K. and {Coughlin}, J. and
             {Bell}, K. and {Berta-Thompson}, Z. and {Williams}, P. and
             {Dotson}, J. and {Barentsen}, G.},
    title = "{Lightkurve: Kepler and TESS time series analysis in Python}",
howpublished = {Astrophysics Source Code Library},
     year = 2018,
    month = dec,
archivePrefix = "ascl",
   eprint = {1812.013},
   adsurl = {http://adsabs.harvard.edu/abs/2018ascl.soft12013L},
}

@ARTICLE{Henriksen_2023_rotational_modulation,
       author = {{Henriksen}, Andreea I. and {Antoci}, Victoria and {Saio}, Hideyuki and {Cantiello}, Matteo and {Kjeldsen}, Hans and {Kurtz}, Donald W. and {Murphy}, Simon J. and {Mathur}, Savita and {Garc{\'\i}a}, Rafael A. and {Santos}, {\^A}ngela R.~G.},
        title = "{Rotational modulation in A and F stars: magnetic stellar spots or convective core rotation?}",
      journal = {\mnras},
         year = 2023,
        month = mar,
       volume = {520},
       number = {1},
        pages = {216-232},
          doi = {10.1093/mnras/stad153},
archivePrefix = {arXiv},
       eprint = {2301.04974},
 primaryClass = {astro-ph.SR},
       adsurl = {https://ui.adsabs.harvard.edu/abs/2023MNRAS.520..216H}
}

@ARTICLE{Henriksen_2023_Rossby_Modes,
       author = {{Henriksen}, Andreea I. and {Antoci}, Victoria and {Saio}, Hideyuki and {Grundahl}, Frank and {Kjeldsen}, Hans and {Van Reeth}, Timothy and {Bowman}, Dominic M. and {P{\'a}pics}, P{\'e}ter I. and {De Cat}, Peter and {Kr{\"u}ger}, Joachim and {Andersen}, M. Fredslund and {Pall{\'e}}, P.~L.},
        title = "{Unresolved Rossby and gravity modes in 214 A and F stars showing rotational modulation}",
      journal = {\mnras},
         year = 2023,
        month = sep,
       volume = {524},
       number = {3},
        pages = {4196-4211},
          doi = {10.1093/mnras/stad1971},
archivePrefix = {arXiv},
       eprint = {2306.16766},
 primaryClass = {astro-ph.SR},
       adsurl = {https://ui.adsabs.harvard.edu/abs/2023MNRAS.524.4196H}
}

@ARTICLE{Antoci_2025,
       author = {{Antoci}, V. and {Cantiello}, M. and {Khalack}, V. and {Henriksen}, A. and {Saio}, H. and {White}, T.~R. and {Buchhave}, L.},
        title = "{Magnetic fields or overstable convective modes in HR 7495: Exploring the underlying causes of the spike in the 'hump and spike' features}",
      journal = {\aap},
         year = 2025,
        month = apr,
       volume = {696},
          eid = {A111},
        pages = {A111},
          doi = {10.1051/0004-6361/202450640},
archivePrefix = {arXiv},
       eprint = {2502.11879},
 primaryClass = {astro-ph.SR},
       adsurl = {https://ui.adsabs.harvard.edu/abs/2025A&A...696A.111A}
}

@ARTICLE{Kurtz_1982,
       author = {{Kurtz}, D.~W.},
        title = "{Rapidly oscillating AP stars.}",
      journal = {\mnras},
         year = 1982,
        month = sep,
       volume = {200},
        pages = {807-859},
          doi = {10.1093/mnras/200.3.807},
       adsurl = {https://ui.adsabs.harvard.edu/abs/1982MNRAS.200..807K}
}

@ARTICLE{Kaye_1999,
       author = {{Kaye}, Anthony B. and {Handler}, Gerald and {Krisciunas}, Kevin and {Poretti}, Ennio and {Zerbi}, Filippo M.},
        title = "{Gamma Doradus Stars: Defining a New Class of Pulsating Variables}",
      journal = {\pasp},
         year = 1999,
        month = jul,
       volume = {111},
       number = {761},
        pages = {840-844},
          doi = {10.1086/316399},
archivePrefix = {arXiv},
       eprint = {astro-ph/9905042},
 primaryClass = {astro-ph},
       adsurl = {https://ui.adsabs.harvard.edu/abs/1999PASP..111..840K}
}

@ARTICLE{Balona_1994,
       author = {{Balona}, L.~A. and {Krisciunas}, K. and {Cousins}, A.~W.~J.},
        title = "{Gamma Doradus : evidence for a new class of pulsating star.}",
      journal = {\mnras},
         year = 1994,
        month = oct,
       volume = {270},
        pages = {905-913},
          doi = {10.1093/mnras/270.4.905},
       adsurl = {https://ui.adsabs.harvard.edu/abs/1994MNRAS.270..905B}
}

@ARTICLE{Holdsworth_2024,
       author = {{Holdsworth}, D.~L. and {Cunha}, M.~S. and {Lares-Martiz}, M. and {Kurtz}, D.~W. and {Antoci}, V. and {Barcel{\'o} Forteza}, S. and {De Cat}, P. and {Derekas}, A. and {Kayhan}, C. and {Ozuyar}, D. and {Skarka}, M. and {Hey}, D.~R. and {Shi}, F. and {Bowman}, D.~M. and {Kobzar}, O. and {Ayala G{\'o}mez}, A. and {Bogn{\'a}r}, Zs and {Buzasi}, D.~L. and {Ebadi}, M. and {Fox-Machado}, L. and {Garc{\'\i}a Hern{\'a}ndez}, A. and {Ghasemi}, H. and {Guzik}, J.~A. and {Handberg}, R. and {Handler}, G. and {Hasanzadeh}, A. and {Jayaraman}, R. and {Khalack}, V. and {Kochukhov}, O. and {Lovekin}, C.~C. and {Miko{\l}ajczyk}, P. and {Mkrtichian}, D. and {Murphy}, S.~J. and {Niemczura}, E. and {Olafsson}, B.~G. and {Pascual-Granado}, J. and {Paunzen}, E. and {Posi{\l}ek}, N. and {Ram{\'o}n-Ballesta}, A. and {Safari}, H. and {Samadi-Ghadim}, A. and {Smalley}, B. and {S{\'o}dor}, {\'A}. and {Stateva}, I. and {Su{\'a}rez}, J.~C. and {Szab{\'o}}, R. and {Wu}, T. and {Ziaali}, E. and {Zong}, W. and {Seager}, S.},
        title = "{TESS Cycle 2 observations of roAp stars with 2-min cadence data}",
      journal = {\mnras},
         year = 2024,
        month = feb,
       volume = {527},
       number = {4},
        pages = {9548-9580},
          doi = {10.1093/mnras/stad3800},
archivePrefix = {arXiv},
       eprint = {2312.04199},
 primaryClass = {astro-ph.SR},
       adsurl = {https://ui.adsabs.harvard.edu/abs/2024MNRAS.527.9548H}
}

@ARTICLE{Shibahashi_1985,
       author = {{Shibahashi}, H. and {Saio}, H.},
        title = "{Rapid Oscillations of Ap Stars}",
      journal = {\pasj},
         year = 1985,
        month = aug,
       volume = {37},
       number = {2},
        pages = {245-259},
          doi = {10.1093/pasj/37.2.245},
       adsurl = {https://ui.adsabs.harvard.edu/abs/1985PASJ...37..245S}
}

@ARTICLE{Handler_Shobbrook_2002,
       author = {{Handler}, G. and {Shobbrook}, R.~R.},
        title = "{On the relationship between the {\ensuremath{\delta}} Scuti and {\ensuremath{\gamma}} Doradus pulsators}",
      journal = {\mnras},
         year = 2002,
        month = jun,
       volume = {333},
       number = {2},
        pages = {251-262},
          doi = {10.1046/j.1365-8711.2002.05401.x},
archivePrefix = {arXiv},
       eprint = {astro-ph/0202152},
 primaryClass = {astro-ph},
       adsurl = {https://ui.adsabs.harvard.edu/abs/2002MNRAS.333..251H}
}

@ARTICLE{Grigahcene_2010,
       author = {{Grigahc{\`e}ne}, A. and {Antoci}, V. and {Balona}, L. and {Catanzaro}, G. and {Daszy{\'n}ska-Daszkiewicz}, J. and {Guzik}, J.~A. and {Handler}, G. and {Houdek}, G. and {Kurtz}, D.~W. and {Marconi}, M. and {Monteiro}, M.~J.~P.~F.~G. and {Moya}, A. and {Ripepi}, V. and {Su{\'a}rez}, J.-C. and {Uytterhoeven}, K. and {Borucki}, W.~J. and {Brown}, T.~M. and {Christensen-Dalsgaard}, J. and {Gilliland}, R.~L. and {Jenkins}, J.~M. and {Kjeldsen}, H. and {Koch}, D. and {Bernabei}, S. and {Bradley}, P. and {Breger}, M. and {Di Criscienzo}, M. and {Dupret}, M.-A. and {Garc{\'\i}a}, R.~A. and {Garc{\'\i}a Hern{\'a}ndez}, A. and {Jackiewicz}, J. and {Kaiser}, A. and {Lehmann}, H. and {Mart{\'\i}n-Ruiz}, S. and {Mathias}, P. and {Molenda-{\.Z}akowicz}, J. and {Nemec}, J.~M. and {Nuspl}, J. and {Papar{\'o}}, M. and {Roth}, M. and {Szab{\'o}}, R. and {Suran}, M.~D. and {Ventura}, R.},
        title = "{Hybrid {\ensuremath{\gamma}} Doradus-{\ensuremath{\delta}} Scuti Pulsators: New Insights into the Physics of the Oscillations from Kepler Observations}",
      journal = {\apjl},
         year = 2010,
        month = apr,
       volume = {713},
       number = {2},
        pages = {L192-L197},
          doi = {10.1088/2041-8205/713/2/L192},
archivePrefix = {arXiv},
       eprint = {1001.0747},
 primaryClass = {astro-ph.SR},
       adsurl = {https://ui.adsabs.harvard.edu/abs/2010ApJ...713L.192G}
}

@ARTICLE{Smalley_2014,
       author = {{Smalley}, B. and {Southworth}, J. and {Pintado}, O.~I. and {Gillon}, M. and {Holdsworth}, D.~L. and {Anderson}, D.~R. and {Barros}, S.~C.~C. and {Collier Cameron}, A. and {Delrez}, L. and {Faedi}, F. and {Haswell}, C.~A. and {Hellier}, C. and {Horne}, K. and {Jehin}, E. and {Maxted}, P.~F.~L. and {Norton}, A.~J. and {Pollacco}, D. and {Skillen}, I. and {Smith}, A.~M.~S. and {West}, R.~G. and {Wheatley}, P.~J.},
        title = "{Eclipsing Am binary systems in the SuperWASP survey}",
      journal = {\aap},
         year = 2014,
        month = apr,
       volume = {564},
          eid = {A69},
        pages = {A69},
          doi = {10.1051/0004-6361/201323158},
archivePrefix = {arXiv},
       eprint = {1402.7168},
 primaryClass = {astro-ph.SR},
       adsurl = {https://ui.adsabs.harvard.edu/abs/2014A&A...564A..69S}
}

@INPROCEEDINGS{Crawford_2010,
       author = {{Crawford}, Steven M. and {Still}, Martin and {Schellart}, Pim and {Balona}, Luis and {Buckley}, David A.~H. and {Dugmore}, Garith and {Gulbis}, Amanda A.~S. and {Kniazev}, Alexei and {Kotze}, Marissa and {Loaring}, Nicola and {Nordsieck}, Kenneth H. and {Pickering}, Timothy E. and {Potter}, Stephen and {Romero Colmenero}, Encarni and {Vaisanen}, Petri and {Williams}, Theodore and {Zietsman}, Ewald},
        title = "{PySALT: the SALT science pipeline}",
    booktitle = {Observatory Operations: Strategies, Processes, and Systems III},
         year = 2010,
       editor = {{Silva}, David R. and {Peck}, Alison B. and {Soifer}, B. Thomas},
       series = {Society of Photo-Optical Instrumentation Engineers (SPIE) Conference Series},
       volume = {7737},
        month = jul,
          eid = {773725},
        pages = {773725},
          doi = {10.1117/12.857000},
       adsurl = {https://ui.adsabs.harvard.edu/abs/2010SPIE.7737E..25C}
}

@ARTICLE{Saio_2018,
       author = {{Saio}, Hideyuki and {Kurtz}, Donald W. and {Murphy}, Simon J. and {Antoci}, Victoria L. and {Lee}, Umin},
        title = "{Theory and evidence of global Rossby waves in upper main-sequence stars: r-mode oscillations in many Kepler stars}",
      journal = {\mnras},
         year = 2018,
        month = feb,
       volume = {474},
       number = {2},
        pages = {2774-2786},
          doi = {10.1093/mnras/stx2962},
archivePrefix = {arXiv},
       eprint = {1711.04908},
 primaryClass = {astro-ph.SR},
       adsurl = {https://ui.adsabs.harvard.edu/abs/2018MNRAS.474.2774S}
}

@ARTICLE{Van_Reeth_2015,
       author = {{Van Reeth}, T. and {Tkachenko}, A. and {Aerts}, C. and {P{\'a}pics}, P.~I. and {Triana}, S.~A. and {Zwintz}, K. and {Degroote}, P. and {Debosscher}, J. and {Bloemen}, S. and {Schmid}, V.~S. and {De Smedt}, K. and {Fremat}, Y. and {Fuentes}, A.~S. and {Homan}, W. and {Hrudkova}, M. and {Karjalainen}, R. and {Lombaert}, R. and {Nemeth}, P. and {{\O}stensen}, R. and {Van De Steene}, G. and {Vos}, J. and {Raskin}, G. and {Van Winckel}, H.},
        title = "{Gravity-mode Period Spacings as a Seismic Diagnostic for a Sample of {\ensuremath{\gamma}} Doradus Stars from Kepler Space Photometry and High-resolution Ground-based Spectroscopy}",
      journal = {\apjs},
         year = 2015,
        month = jun,
       volume = {218},
       number = {2},
          eid = {27},
        pages = {27},
          doi = {10.1088/0067-0049/218/2/27},
archivePrefix = {arXiv},
       eprint = {1504.02119},
 primaryClass = {astro-ph.SR},
       adsurl = {https://ui.adsabs.harvard.edu/abs/2015ApJS..218...27V}
}

@ARTICLE{Papics_2012,
       author = {{P{\'a}pics}, P.~I.},
        title = "{The puzzle of combination frequencies found in heat-driven pulsators}",
      journal = {Astronomische Nachrichten},
         year = 2012,
        month = dec,
       volume = {333},
       number = {10},
        pages = {1053},
          doi = {10.1002/asna.201211809},
       adsurl = {https://ui.adsabs.harvard.edu/abs/2012AN....333.1053P}
}

@ARTICLE{Miglio_2008,
       author = {{Miglio}, Andrea and {Montalb{\'a}n}, Josefina and {Noels}, Arlette and {Eggenberger}, Patrick},
        title = "{Probing the properties of convective cores through g modes: high-order g modes in SPB and {\ensuremath{\gamma}} Doradus stars}",
      journal = {\mnras},
         year = 2008,
        month = may,
       volume = {386},
       number = {3},
        pages = {1487-1502},
          doi = {10.1111/j.1365-2966.2008.13112.x},
archivePrefix = {arXiv},
       eprint = {0802.2057},
 primaryClass = {astro-ph},
       adsurl = {https://ui.adsabs.harvard.edu/abs/2008MNRAS.386.1487M}
}

@ARTICLE{Cunha_2019,
       author = {{Cunha}, M.~S. and {Antoci}, V. and {Holdsworth}, D.~L. and {Kurtz}, D.~W. and {Balona}, L.~A. and {Bogn{\'a}r}, Zs and {Bowman}, D.~M. and {Guo}, Z. and {Ko{\l}aczek-Szyma{\'n}ski}, P.~A. and {Lares-Martiz}, M. and {Paunzen}, E. and {Skarka}, M. and {Smalley}, B. and {S{\'o}dor}, {\'A}. and {Kochukhov}, O. and {Pepper}, J. and {Richey-Yowell}, T. and {Ricker}, G.~R. and {Seager}, S. and {Buzasi}, D.~L. and {Fox-Machado}, L. and {Hasanzadeh}, A. and {Niemczura}, E. and {Quitral-Manosalva}, P. and {Monteiro}, M.~J.~P.~F.~G. and {Stateva}, I. and {De Cat}, P. and {Garc{\'\i}a Hern{\'a}ndez}, A. and {Ghasemi}, H. and {Handler}, G. and {Hey}, D. and {Matthews}, J.~M. and {Nemec}, J.~M. and {Pascual-Granado}, J. and {Safari}, H. and {Su{\'a}rez}, J.~C. and {Szab{\'o}}, R. and {Tkachenko}, A. and {Weiss}, W.~W.},
        title = "{Rotation and pulsation in Ap stars: first light results from TESS sectors 1 and 2}",
      journal = {\mnras},
         year = 2019,
        month = aug,
       volume = {487},
       number = {3},
        pages = {3523-3549},
          doi = {10.1093/mnras/stz1332},
archivePrefix = {arXiv},
       eprint = {1906.01111},
 primaryClass = {astro-ph.SR},
       adsurl = {https://ui.adsabs.harvard.edu/abs/2019MNRAS.487.3523C}
}

@ARTICLE{Holdsworth_2021,
       author = {{Holdsworth}, D.~L. and {Cunha}, M.~S. and {Kurtz}, D.~W. and {Antoci}, V. and {Hey}, D.~R. and {Bowman}, D.~M. and {Kobzar}, O. and {Buzasi}, D.~L. and {Kochukhov}, O. and {Niemczura}, E. and {Ozuyar}, D. and {Shi}, F. and {Szab{\'o}}, R. and {Samadi-Ghadim}, A. and {Bogn{\'a}r}, Zs and {Fox-Machado}, L. and {Khalack}, V. and {Lares-Martiz}, M. and {Lovekin}, C.~C. and {Miko{\l}ajczyk}, P. and {Mkrtichian}, D. and {Pascual-Granado}, J. and {Paunzen}, E. and {Richey-Yowell}, T. and {S{\'o}dor}, {\'A}. and {Sikora}, J. and {Yang}, T.~Z. and {Brunsden}, E. and {David-Uraz}, A. and {Derekas}, A. and {Garc{\'\i}a Hern{\'a}ndez}, A. and {Guzik}, J.~A. and {Hatamkhani}, N. and {Handberg}, R. and {Lambert}, T.~S. and {Lampens}, P. and {Murphy}, S.~J. and {Monier}, R. and {Pollard}, K.~R. and {Quitral-Manosalva}, P. and {Ram{\'o}n-Ballesta}, A. and {Smalley}, B. and {Stateva}, I. and {Vanderspek}, R.},
        title = "{TESS cycle 1 observations of roAp stars with 2-min cadence data}",
      journal = {\mnras},
         year = 2021,
        month = sep,
       volume = {506},
       number = {1},
        pages = {1073-1110},
          doi = {10.1093/mnras/stab1578},
archivePrefix = {arXiv},
       eprint = {2105.13274},
 primaryClass = {astro-ph.SR},
       adsurl = {https://ui.adsabs.harvard.edu/abs/2021MNRAS.506.1073H}
}

@ARTICLE{Schmid_2016,
       author = {{Schmid}, V.~S. and {Aerts}, C.},
        title = "{Asteroseismic modelling of the two F-type hybrid pulsators KIC 10080943A and KIC 10080943B}",
      journal = {\aap},
         year = 2016,
        month = aug,
       volume = {592},
          eid = {A116},
        pages = {A116},
          doi = {10.1051/0004-6361/201628617},
archivePrefix = {arXiv},
       eprint = {1605.07958},
 primaryClass = {astro-ph.SR},
       adsurl = {https://ui.adsabs.harvard.edu/abs/2016A&A...592A.116S}
}

@ARTICLE{Rodriguez_2001,
       author = {{Rodr{\'\i}guez}, E. and {Breger}, M.},
        title = "{delta Scuti and related stars: Analysis of the R00 Catalogue}",
      journal = {\aap},
         year = 2001,
        month = jan,
       volume = {366},
        pages = {178-196},
          doi = {10.1051/0004-6361:20000205},
       adsurl = {https://ui.adsabs.harvard.edu/abs/2001A&A...366..178R}
}

@ARTICLE{2014MNRAS.444..102K,
       author = {{Kurtz}, Donald W. and {Saio}, Hideyuki and {Takata}, Masao and {Shibahashi}, Hiromoto and {Murphy}, Simon J. and {Sekii}, Takashi},
        title = "{Asteroseismic measurement of surface-to-core rotation in a main-sequence A star, KIC 11145123}",
      journal = {\mnras},
         year = 2014,
        month = oct,
       volume = {444},
       number = {1},
        pages = {102-116},
          doi = {10.1093/mnras/stu1329},
archivePrefix = {arXiv},
       eprint = {1405.0155},
 primaryClass = {astro-ph.SR},
       adsurl = {https://ui.adsabs.harvard.edu/abs/2014MNRAS.444..102K}
}

@ARTICLE{Ginestet_2003,
       author = {{Ginestet}, N. and {Prieur}, J.-L. and {Carquillat}, J.-M. and {Griffin}, R.~F.},
        title = "{Contribution to the search for binaries among Am stars - IV. HD 100054B and 187258}",
      journal = {\mnras},
         year = 2003,
        month = jun,
       volume = {342},
       number = {1},
        pages = {61-68},
          doi = {10.1046/j.1365-8711.2003.06489.x},
       adsurl = {https://ui.adsabs.harvard.edu/abs/2003MNRAS.342...61G}
}

@ARTICLE{Tanner_1949,
       author = {{Tanner}, R.~W.},
        title = "{The orbit of the spectroscopic binary HD 201032.}",
      journal = {Publications of the David Dunlap Observatory},
         year = 1949,
        month = jan,
       volume = {1},
        pages = {507-508},
       adsurl = {https://ui.adsabs.harvard.edu/abs/1949PDDO....1..507T}
}

@ARTICLE{Torres_2021,
       author = {{Torres}, Guillermo and {Latham}, David W. and {Quinn}, Samuel N.},
        title = "{Long-term Spectroscopic Survey of the Pleiades Cluster: The Binary Population}",
      journal = {\apj},
         year = 2021,
        month = nov,
       volume = {921},
       number = {2},
          eid = {117},
        pages = {117},
          doi = {10.3847/1538-4357/ac1585},
archivePrefix = {arXiv},
       eprint = {2107.10259},
 primaryClass = {astro-ph.SR},
       adsurl = {https://ui.adsabs.harvard.edu/abs/2021ApJ...921..117T}
}

@ARTICLE{Kurtz_2022,
       author = {{Kurtz}, Donald W.},
        title = "{Asteroseismology Across the Hertzsprung-Russell Diagram}",
      journal = {\araa},
         year = 2022,
        month = aug,
       volume = {60},
        pages = {31-71},
          doi = {10.1146/annurev-astro-052920-094232},
archivePrefix = {arXiv},
       eprint = {2201.11629},
 primaryClass = {astro-ph.SR},
       adsurl = {https://ui.adsabs.harvard.edu/abs/2022ARA&A..60...31K}
}

@ARTICLE{Dziembowski_1977,
       author = {{Dziembowski}, W.},
        title = "{Light and radial velocity variations in a nonradially oscillating star.}",
      journal = {\actaa},
         year = 1977,
        month = jan,
       volume = {27},
        pages = {203-211},
       adsurl = {https://ui.adsabs.harvard.edu/abs/1977AcA....27..203D}
}

@ARTICLE{1973ApJ...182..809A,
       author = {{Abt}, H.~A. and {Moyd}, K.~I.},
        title = "{Rotation and shell spectra among A-type dwarfs.}",
      journal = {\apj},
         year = 1973,
        month = jun,
       volume = {182},
        pages = {809},
          doi = {10.1086/152184},
       adsurl = {https://ui.adsabs.harvard.edu/abs/1973ApJ...182..809A}
}

@ARTICLE{1989ApJS...70..623G,
       author = {{Gray}, R.~O. and {Garrison}, R.~F.},
        title = "{The Late A-Type Stars: Refined MK Classification, Confrontation with Stroemgren Photometry, and the Effects of Rotation}",
      journal = {\apjs},
         year = 1989,
        month = jul,
       volume = {70},
        pages = {623},
          doi = {10.1086/191349},
       adsurl = {https://ui.adsabs.harvard.edu/abs/1989ApJS...70..623G}
}

@ARTICLE{2019RAA....19...64Q,
       author = {{Qian}, Sheng-Bang and {Shi}, Xiang-Dong and {Zhu}, Li-Ying and {Li}, Lin-Jia and {Zhang}, Jia and {Zhao}, Er-Gang and {Han}, Zhong-Tao and {Zhou}, Xiao and {Fang}, Xiao-Hui and {Liao}, Wen-Ping},
        title = "{More than two hundred and fifty thousand spectroscopic binary or variable star candidates discovered by LAMOST}",
      journal = {Research in Astronomy and Astrophysics},
         year = 2019,
        month = may,
       volume = {19},
       number = {5},
          eid = {064},
        pages = {064},
          doi = {10.1088/1674-4527/19/5/64},
       adsurl = {https://ui.adsabs.harvard.edu/abs/2019RAA....19...64Q}
}

@ARTICLE{1951PRCO....2...85O,
       author = {{O'Connell}, D.~J.~K.},
        title = "{The so-called periastron effect in close eclipsing binaries ; New variable stars (fifth list)}",
      journal = {Publications of the Riverview College Observatory},
         year = 1951,
        month = aug,
       volume = {2},
       number = {6},
        pages = {85-100},
       adsurl = {https://ui.adsabs.harvard.edu/abs/1951PRCO....2...85O}
}

\appendix

\section{TESS data}

\begin{table*}
\centering
\caption{Available TESS observations for the stars in our sample. 
The table lists all sectors observed in 2-minute cadence and in full-frame images (FFIs).}
\label{app:tesssectors}
\begin{tabular}{llll}
\hline\hline
TIC & HD Name & 2-min cadence sectors & FFI sectors \\
\hline
TIC~7597696     & HD~27079     & 4/5; 31/32                       & 4/5; 31/32 \\
TIC~125736216   & HD~23488     & 71                               & 42/43/44; 70/71 \\
TIC~137003360   & HD~112515    & 15; 22                           & 15; 22; 49; 76 \\
TIC~150183718   & HD~78388     & 47                               & 21; 47 \\
TIC~151769040   & HD~97160     & 10; 36; 37; 63; 90               & same as 2-min \\
TIC~179033962   & HD~27230     & 4/5; 31/32                       & same as 2-min \\
TIC~197647472   & HD~208139    & 1; 28                            & 01; 28; 68; 95 \\
TIC~299779198   & HD~16232     & 1; 12/13; 27/28; 39              & 01; 12/13; 27/28; 39; 66/67/68; 93/94/95 \\
TIC~305679500   & HD~201032    & 15/16/17; 24                     & 15/16/17; 24; 56/57/58; 76/77/78; 83/84/85 \\
TIC~357469812   & HD~107340    & 11/12                            & 11/12; 38/39; 65/66; 93 \\
TIC~391070709   & HD~187258    & 41; 54                           & 14; 41; 54; 81 \\
TIC~394818541   & HD~17784     & 27; 39                           & 12/13; 27; 39; 66/67; 93/94 \\
TIC~450302084   & HD~86167     & 14; 21; 48                       & same as 2-min \\
TIC~452590255   & HD~83094     & 10/11/12; 37/38/39               & 10/11/12; 37/38/39; 63/64/65; 90 \\
TIC~466443867   & HD~2523A     & 42/43                            & 42/43; 57; 70; 84 \\
TIC~1506355332  & HD~158251A   & 25/26; 52/53                     & 25/26; 52/53; 79 \\
\hline
\end{tabular}
\end{table*}

\clearpage

\section{Stellar parameters}
\onecolumn
\begin{landscape}
\begin{longtable}{l c c c c c c c c c c c c c}
\caption{Stellar photometric and Gaia-based parameters, including absolute magnitudes, bolometric magnitudes estimated from $M_V$, effective temperatures, surface gravities, and luminosities. Here, $T_\mathrm{eff}^{\mathrm{single}}$ denotes the effective temperature derived under the assumption of a single-star spectrum. A representative uncertainty of $\pm110\,\mathrm{K}$ is adopted for the effective temperatures. See Section~\ref{sec:gaiadata} for details.}\\
\hline\hline
Name & TESS mag & $\varpi$ & $\sigma_\varpi$ & $M_V$ & $M_V^\mathrm{low}$ & $M_V^\mathrm{upp}$ & $M_\mathrm{bol}^{M_V}$ & $T_\mathrm{eff}^{\mathrm{single}}$ & $T_\mathrm{eff}^{\mathrm{gspphot}}$ & $\log g$ & $L/L_\odot$ & $L_\mathrm{low}$ & $L_\mathrm{upp}$ \\
 &  & [mas] & [mas] & [mag] & [mag] & [mag] & [mag] & [K] & [K] & [dex] &  &  &  \\
\hline
\endfirsthead
\caption{continued.} \\
\hline\hline
Name & TESS mag & $\varpi$ & $\sigma_\varpi$ & $M_V$ & $M_V^\mathrm{low}$ & $M_V^\mathrm{upp}$ & $M_\mathrm{bol}^{M_V}$ & $T_\mathrm{eff}^{\mathrm{single}}$ & $T_\mathrm{eff}^{\mathrm{gspphot}}$ & $\log g$ & $L/L_\odot$ & $L_\mathrm{low}$ & $L_\mathrm{upp}$ \\
\hline
\endhead
\endfoot
\hline
\endlastfoot
KIC 2310586 & 13.06 & 0.62 & 0.12 & 1.04 & 0.93 & 1.16 & 0.96 & 7163 & 7136 & 3.8 & 9.4 & 7.2 & 15.5 \\
KIC 5038228 & 11.05 & 1.37 & 0.01 & 2.07 & 2.05 & 2.09 & 1.99 & 7117 & - & 3.8 & 13.4 & 14.0 & 12.7 \\
KIC 5390069 & 14.72 & 0.33 & 0.02 & 1.56 & 1.41 & 1.66 & 1.45 & 6892 & 7593 & 4.2 & 10.5 & 9.5 & 11.6 \\
KIC 5443410 & 13.14 & 0.62 & 0.01 & 2.17 & 2.10 & 2.30 & 2.08 & 7210 & 7193 & 4.1 & 9.5 & 9.2 & 9.7 \\
KIC 5459805 & 13.10 & 0.49 & 0.01 & 1.72 & 1.65 & 1.77 & 1.64 & 6923 & - & 4.3 & 19.0 & 20.2 & 17.7 \\
KIC 5725443 & 13.13 & 0.60 & 0.01 & 1.27 & 1.17 & 1.32 & 1.18 & 6917 & 7137 & 3.9 & 12.6 & 12.2 & 13.2 \\
KIC 6595315 & 13.61 & 0.56 & 0.01 & 2.14 & 1.92 & 2.21 & 2.05 & 7601 & 7223 & 4.1 & 7.8 & 7.5 & 8.1 \\
KIC 6875337 & 12.31 & 0.38 & 0.07 & 2.02 & 1.95 & 2.08 & 1.94 & 6441 & 6913 & 4.0 & 7.6 & 7.3 & 7.9 \\
KIC 6937123 & 11.21 & 1.30 & 0.01 & 1.58 & 1.55 & 1.61 & 1.49 & 7380 & 7182 & 3.8 & 12.4 & 12.1 & 12.6 \\
KIC 6951231 & 13.07 & 0.65 & 0.01 & 2.14 & 2.06 & 2.24 & 2.05 & 7283 & 7270 & 4.2 & 9.3 & 9.0 & 9.6 \\
KIC 7045685 & 12.54 & 0.67 & 0.01 & 1.86 & 1.82 & 1.90 & 1.77 & 6821 & - & 3.8 & 12.8 & 13.7 & 11.9 \\
KIC 7352776 & 10.78 & 2.11 & 0.01 & 2.18 & 2.17 & 2.20 & 2.07 & 7401 & 7606 & 4.2 & 7.5 & 7.5 & 7.6 \\
KIC 7430757 & 12.31 & 0.65 & 0.01 & 1.37 & 1.24 & 1.44 & 1.28 & 7100 & 6961 & 3.8 & 16.9 & 16.4 & 17.3 \\
KIC 7900367 & 10.92 & 1.53 & 0.01 & 1.61 & 1.59 & 1.63 & 1.52 & 7400 & 7289 & 4.0 & 12.1 & 11.9 & 12.3 \\
KIC 7973199 & 12.60 & 0.81 & 0.03 & 1.51 & 1.48 & 1.55 & 1.42 & 6591 & 7364 & 4.1 & 10.7 & 10.0 & 11.2 \\
KIC 8177748 & 11.25 & 1.41 & 0.01 & 1.83 & 1.75 & 1.85 & 1.74 & 7419 & 7199 & 3.9 & 10.7 & 10.5 & 10.9 \\
KIC 8299332 & 13.08 & 0.73 & 0.02 & 1.73 & 1.67 & 1.77 & 1.62 & 7136 & 7447 & 4.0 & 8.2 & 7.9 & 8.7 \\
KIC 8460993 & 10.84 & 1.83 & 0.01 & 2.15 & 2.12 & 2.19 & 2.06 & 7019 & 6950 & 4.0 & 8.7 & 8.5 & 8.8 \\
KIC 9347095 & 10.39 & 1.48 & 0.01 & 1.33 & 1.31 & 1.38 & 1.25 & 7421 & 6945 & 3.7 & 19.7 & 19.4 & 20.1 \\
KIC 9875566 & 11.27 & 0.81 & 0.01 & 1.23 & 1.19 & 1.26 & 1.14 & 7093 & - & 3.5 & 32.8 & 30.7 & 35.0 \\
KIC 10014548 & 10.33 & 2.03 & 0.01 & 2.20 & 2.19 & 2.21 & 2.10 & 7470 & - & 4.0 & 12.0 & 13.4 & 11.0 \\
KIC 10154966 & 11.76 & 1.36 & 0.01 & 2.20 & 2.18 & 2.22 & 2.10 & 7401 & - & 4.3 & 6.8 & 6.6 & 6.4 \\
KIC 11822789 & 12.67 & 0.84 & 0.01 & 1.96 & 1.94 & 1.99 & 1.86 & 7401 & 7475 & 4.1 & 8.6 & 8.4 & 8.8 \\
TIC 7597696 & 7.79 & 6.45 & 0.02 & 2.12 & 2.12 & 2.13 & 2.04 & 7147 & 6997 & 3.9 & 11.0 & 10.9 & 11.1 \\
TIC 125736216 & 8.38 & 7.38 & 0.16 & 2.32 & 2.27 & 2.37 & 2.23 & 7400 & 7313 & 4.1 & 6.0 & 5.7 & 6.2 \\
TIC 137003360 & 8.23 & 6.38 & 0.02 & 2.55 & 2.54 & 2.55 & 2.46 & 7243 & 7113 & 4.1 & 7.3 & 7.2 & 7.3 \\
TIC 150183718 & 7.24 & 6.93 & 0.04 & 1.71 & 1.66 & 1.75 & 1.63 & 7214 & 6875 & 3.8 & 16.1 & 15.9 & 16.4 \\
TIC 151769040 & 8.21 & 5.90 & 0.02 & 2.26 & 2.22 & 2.27 & 2.18 & 7206 & 7110 & 4.0 & 9.2 & 9.1 & 9.2 \\
TIC 179033962 & 7.87 & 6.43 & 0.03 & 2.03 & 2.02 & 2.04 & 1.94 & 7257 & 7124 & 3.9 & 10.4 & 10.3 & 10.5 \\
TIC 197647472 & 7.29 & 7.88 & 0.12 & 1.93 & 1.88 & 1.98 & 1.84 & 7824 & 7039 & 3.9 & 11.8 & 11.4 & 12.1 \\
TIC 299779198 & 8.74 & 5.14 & 0.01 & 2.44 & 2.43 & 2.44 & 2.35 & 7521 & 7220 & 4.1 & 7.6 & 7.5 & 7.6 \\
TIC 305679500 & 6.98 & 8.00 & 0.02 & 1.74 & 1.69 & 1.75 & 1.66 & 7219 & 6868 & 3.8 & 14.6 & 14.4 & 15.0 \\
TIC 357469812 & 7.47 & 5.33 & 0.04 & -0.29 & -0.32 & -0.27 & -0.72 & 7401 & 9957 & 4.1 & 57.1 & 56.0 & 58.1 \\
TIC 391070709 & 7.26 & 7.52 & 0.03 & 1.95 & 1.93 & 1.97 & 1.87 & 6842 & 6873 & 3.8 & 13.0 & 12.8 & 13.1 \\
TIC 394818541 & 8.80 & 4.44 & 0.01 & 1.65 & 1.65 & 1.66 & 1.57 & 7590 & 7070 & 3.9 & 11.4 & 11.3 & 11.4 \\
TIC 450302084 & 8.11 & 7.20 & 0.02 & 2.69 & 2.67 & 2.72 & 2.60 & 7601 & 7213 & 4.1 & 6.6 & 6.6 & 6.7 \\
TIC 452590255 & 8.87 & 5.39 & 0.01 & 2.48 & 2.47 & 2.49 & 2.38 & 7600 & 7348 & 4.2 & 6.7 & 6.6 & 6.7 \\
TIC 466443867 & 7.77 & 8.02 & 0.11 & 2.34 & 2.29 & 2.40 & 2.26 & 7976 & 7027 & 4.0 & 7.3 & 7.1 & 7.4 \\
TIC 1506355332 & 7.00 & 12.10 & 0.02 & 2.70 & 2.70 & 2.71 & 2.61 & 7976 & 7231 & 4.1 & 6.6 & 6.6 & 6.6 

\label{tab:params}
\end{longtable}

\end{landscape}
\twocolumn

\section{Fundamental radial modes}
\onecolumn

\begin{landscape}
\begin{longtable}{l c c c c c c c c c c c l}
\caption{
Summary of fundamental radial mode properties and agglomerated region (AR) classification.  
Here, $p_1$ denotes the frequency of the fundamental radial mode. $f^\mathrm{min}$ and $f^\mathrm{max}$ denote the lower and upper frequencies of the AR, respectively.
The column “AR $\geq p_1$” indicates the relative location of the radial mode with respect to the agglomerated power excess:  
a value of 0 means $p_1$ is higher than the AR,  
1 means $p_1$ is lower than the AR, and  
0.5 indicates either conflicting results from different methods or that $p_1$ overlaps with the AR.  
The column “Type” refers to the morphological pulsation type:  
GDOR = $\gamma$~Doradus, HYB = hybrid (both p- and g modes), ROT = rotational modulation, BIN = binary signal.
} \\
\hline\hline
Name & $p_1^\mathrm{PL}$ & $p_1^\mathrm{lower}$ & $p_1^\mathrm{upper}$ & $p_1^\mathrm{Q}$-$T_\mathrm{eff}$(GSP) & $p_1^\mathrm{Q}$(single star) & $p_1^\mathrm{model}$ & $\Pi_0^\mathrm{this\ work}$ & $\Pi_0^\mathrm{GangLi+}$ & $f^\mathrm{min}$ & $f^\mathrm{max}$ & AR$\geq p_1$ & Type \\
 & [d$^{-1}$] & [d$^{-1}$] & [d$^{-1}$] & [d$^{-1}$] & [d$^{-1}$] & [d$^{-1}$] & [s] & [s] & [d$^{-1}$] & [d$^{-1}$] &  &  \\
\hline
\endfirsthead
\endhead
\endfoot
\hline
\endlastfoot
KIC 2310586 & 6.48 & 7.09 & 5.93 & 7.29 & 7.32 & 7.346 & 3300 & --- & 10.3 & 17.46 & 1 & GDOR \\
KIC 5038228 & 14.22 & 14.46 & 13.98 & 9.80 & 9.98 & 10.064 & 4082 & 4152 & 4.75 & 8.7 & 0 & HYB \\
KIC 5390069 & 9.61 & 10.40 & 8.61 & 13.45 & 12.21 & --- & --- & --- & 7 & 9 & 0 & GDOR \\
KIC 5443410 & 15.32 & 16.91 & 14.55 & 13.47 & 13.50 & 13.367 & 3984 & --- & 4 & 9.6 & 0 & HYB \\
KIC 5459805 & 10.87 & 11.33 & 10.32 & 14.60 & 14.47 & 8.297 & 4505 & 4649 & 3.98 & 7.8 & 0 & HYB \\
KIC 5725443 & 7.69 & 7.99 & 7.12 & 9.31 & 9.02 & --- & --- & --- & 12.49 & 18 & 1 & GDOR \\
KIC 6595315 & 15.03 & 15.86 & 12.67 & 13.24 & 13.94 & 7.950 & 2967 & --- & 9.9 & 14.8 & 0.5 & GDOR \\
KIC 6875337 & 13.71 & 14.31 & 12.97 & 12.05 & 11.23 & --- & --- & --- & 10.72 & 14 & 0.5 & GDOR \\
KIC 6937123 & 9.76 & 9.98 & 9.54 & 8.70 & 8.94 & 7.285 & 3378 & 3486 & 2.31 & 5 & 0 & HYB \\
KIC 6951231 & 15.01 & 16.20 & 14.15 & 15.03 & 15.06 & --- & --- & --- & 8.29 & 14.53 & 0 & GDOR \\
KIC 7045685 & 12.06 & 12.46 & 11.72 & 8.26 & 8.27 & --- & --- & --- & 3.35 & 9.12 & 0 & HYB \\
KIC 7352776 & 15.52 & 15.68 & 15.35 & 16.84 & 16.39 & 14.373 & 3676 & --- & 13.85 & 19.7 & 0.5 & HYB \\
KIC 7430757 & 8.31 & 8.78 & 7.53 & 8.01 & 8.17 & --- & --- & --- & 7 & 11.01 & 0.5 & ROT \\
KIC 7900367 & 9.99 & 10.13 & 9.89 & 11.18 & 11.35 & --- & --- & --- & 7.28 & 9.58 & 0 & HYB \\
KIC 7973199 & 9.29 & 9.54 & 9.05 & 11.58 & 10.37 & --- & --- & --- & 5.49 & 12.38 & 0 & HYB \\
KIC 8177748 & 11.80 & 12.02 & 11.14 & 10.75 & 11.08 & --- & --- & --- & 3 & 6.7 & 0 & HYB \\
KIC 8299332 & 10.93 & 11.27 & 10.50 & 10.98 & 10.53 & --- & --- & --- & 5.2 & 9.3 & 0 & HYB \\
KIC 8460993 & 15.08 & 15.57 & 14.81 & 11.57 & 11.69 & --- & --- & --- & 3.8 & 10 & 0 & HYB \\
KIC 9347095 & 8.09 & 8.41 & 7.96 & 7.02 & 7.50 & --- & --- & --- & 4.12 & 8.36 & 0 & HYB \\
KIC 9875566 & 7.45 & 7.65 & 7.25 & 5.76 & 5.68 & 7.79753 & 4762 & 5034 & 4.05 & 7.79 & 0.5 & HYB \\
KIC 10014548 & 15.71 & 15.88 & 15.53 & 12.38 & 12.54 & --- & --- & --- & 7.82 & 11.5 & 0 & GDOR \\
KIC 10154966 & 15.69 & 15.93 & 15.47 & 17.35 & 17.44 & --- & --- & --- & 5.63 & 12.38 & 0 & HYB \\
KIC 11822789 & 13.11 & 13.35 & 12.83 & 13.28 & 13.15 & 16.75181 & 3774 & --- & 5.56 & 10.95 & 0 & HYB \\
TIC 7597696 & 14.82 & 14.93 & 14.72 & 10.36 & 10.58 & --- & --- & --- & 5.78 & 12.6 & 0 & GDOR \\
TIC 125736216 & 17.25 & 17.89 & 16.60 & 14.75 & 14.93 & --- & --- & --- & 8.67 & 14.6 & 0 & HYB \\
TIC 137003360 & 20.52 & 20.59 & 20.44 & 15.72 & 16.01 & --- & --- & --- & 6.17 & 15.69 & 0 & HYB \\
TIC 150183718 & 10.81 & 11.12 & 10.40 & 8.48 & 8.90 & --- & --- & --- & 2 & 5 & 0 & HYB(?) \\
TIC 151769040 & 16.46 & 16.62 & 16.00 & 12.58 & 12.75 & --- & --- & --- & 11.68 & 17 & 0.5 & HYB \\
TIC 179033962 & 13.79 & 13.93 & 13.66 & 10.95 & 11.15 & --- & --- & --- & 10 & 15 & 0.5 & GDOR \\
TIC 197647472 & 12.74 & 13.32 & 12.25 & 9.96 & 11.07 & --- & --- & --- & 5 & 11.1 & 0 & HYB \\
TIC 299779198 & 18.81 & 18.90 & 18.71 & 14.77 & 15.39 & --- & --- & --- & 8.6 & 15.6 & 0 & GDOR \\
TIC 305679500 & 11.02 & 11.12 & 10.63 & 8.44 & 8.88 & --- & --- & --- & 3.85 & 7.96 & 0 & HYB(?)/ROT(?)/BIN(?) \\
TIC 357469812 & 2.34 & 2.37 & 2.29 & 9.82 & 7.30 & --- & --- & --- & 9.18 & 12.9 & 1 & HYB \\
TIC 391070709 & 12.98 & 13.21 & 12.81 & 9.46 & 9.42 & --- & --- & --- & 2.67 & 6.24 & 0 & HYB(?)/ROT(?)/BIN(?) \\
TIC 394818541 & 10.33 & 10.40 & 10.28 & 9.78 & 10.50 & --- & --- & --- & 7.9 & 15.5 & 1 & GDOR \\
TIC 450302084 & 22.77 & 23.32 & 22.58 & 16.60 & 17.50 & --- & --- & --- & 7.92 & 13.52 & 0 & HYB \\
TIC 452590255 & 19.45 & 19.57 & 19.29 & 16.98 & 17.56 & --- & --- & --- & 18.5 & 24 & 1 & HYB(?)/ROT(?)/BIN(?) \\
TIC 466443867 & 17.47 & 18.35 & 16.83 & 13.03 & 14.79 & --- & --- & --- & 4.85 & 8.53 & 0 & HYB \\
TIC 1506355332 & 23.10 & 23.17 & 23.03 & 15.92 & 17.57 & --- & --- & --- & 3.4 & 8.9 & 0 & HYB
\label{tab:fundamental}
\end{longtable}
\end{landscape}
\twocolumn

\section{Neighbours and blending} \label{sec:blending_analysis}

All nearby contaminates within 5 magnitudes (in Gaia $G$ band) for our targets are given in Table~\ref{tab:neighbours}. 

For the TESS sample, a pixel-level analysis of the TESS Full Frame Images was done as in \citet{2023A&A...676A..55L}. Except for TIC 137003360, all variations can confidently be attributed to the target star via a visual inspection of the frequency content of the pixels within a 30$\times$30 grid centered on the target star. For TIC 137003360, there is a close neighbour (SBS 81; F6V) 13 arcsec away (about half of a TESS pixel) that is one magnitude fainter and thus requires a more careful analysis. The Python package \textit{TESS-Localize}\footnote{\url{https://github.com/Higgins00/TESS-Localize}} was designed to identify the source of periodically variable stars to sub-pixel precision \citep{2023AJ....165..141H}, and was therefore employed to determine if the signals in the light curve of TIC 137003360 could be due to its close neighbour. This analysis found that the probability that the signals (both the low-frequency g modes and the higher-frequency agglomerated hump) arise in the neighbour SBS 81 is only $\sim$10$^{-5}$, and a probability of 0.99995 that the signals come from TIC 137003360. There is therefore no source confusion for the TESS sample -- all signals originate on-target. 

For the 11 Kepler targets in Table~\ref{tab:neighbours}, the target pixel files were inspected in a similar way as for the TESS sample using the lightkurve package\footnote{\url{https://lightkurve.github.io/lightkurve/}} \citep{2018ascl.soft12013L}. The position of any nearby Gaia sources was identified, and custom aperture masks were used to extract light curves for these neighbouring stars (avoiding flux from the brighter target star as much as possible) and for the target star (avoiding flux from contaminating neighbours as much as possible). The amplitude spectra for these light curves were then compared. In all cases, the amplitudes of the signals of interest (including at low, intermediate, and high frequencies) were highest for the light curve generated from the small aperture centered on the target star, and lower for the light curves generated for neighbouring sources (which in general were heavily contaminated by flux from the target star). We thus conclude that all of the signals identified in the Kepler sample are intrinsic to the target star, and any effects of blending are merely a modest suppression of the observed amplitudes. 

%\textbf{TESS Localize}\footnote{\url{https://github.com/Higgins00/TESS-Localize}} \citep{2023AJ....165..141H} shows a 100\% probability that the low-frequency g mode signals originate in the target star, and only a 1e-5 probability that they arise from the neighbour SBS 81. The same analysis for the higher-frequency signals belonging to the \textbf{wild oscillating hump} also indicate only a 7.005438e-06 probability that the source is the neighbour. All relevant signals for TIC 137003360 are therefore on-target. 

\begin{table*}
\caption{neighbouring stars from Gaia.}
\label{tab:neighbours}
\centering
\begin{tabular}{llll}
\hline\hline
Name & $\Delta_{mag}$ & Distance (\arcsec) & Gaia ID \\
\hline
KIC 2310586 & 4.10 & 6.1 & Gaia DR2 2051663752270306688 \\
\hline
KIC 5038228 & 4.82 & 16.5 & Gaia DR2 2073578122832626944 \\
\hline
KIC 5390069 & 4.65 & 4.4 & Gaia DR2 2073612963601047936 \\
            & 4.00 & 5.4 & Gaia DR2 2073612963608767616 \\
            & 4.19 & 8.1 & Gaia DR2 2073612959294685184 \\
            & 4.63 & 10.1 & Gaia DR2 2073612963608769920 \\
            & 2.53 & 12.3 & Gaia DR2 2073612993654446592 \\
            & 4.04 & 15.7 & Gaia DR2 2073612997960806016 \\
            & 4.45 & 15.9 & Gaia DR2 2073612959294687360 \\
            & 3.13 & 16.0 & Gaia DR2 2073612963608766208 \\
            & 0.61 & 17.8 & Gaia DR2 2073612993654414336 \\
\hline
KIC 5443410 & 4.30 & 9.7 & Gaia DR2 2101199740392605184 \\
\hline
KIC 5459805 & 3.84 & 12.9 & Gaia DR2 2076605658144756096 \\
            & 4.11 & 14.4 & Gaia DR2 2076605662446045184 \\
            & 4.65 & 14.6 & Gaia DR2 2076605658144758528 \\
\hline
KIC 5725443 & 0.01 & 7.4 & Gaia DR2 2076766191145701888 \\
\hline
KIC 6875337 & 4.93 & 10.3 & Gaia DR2 2077877552578232960 \\
\hline
KIC 6937123 & 4.68 & 9.3 & Gaia DR2 2102507476331736576 \\
\hline
KIC 6951231 & 4.52 & 1.6 & Gaia DR2 2125738473396386816 \\
\hline
KIC 7045685 & 3.64 & 13.3 & Gaia DR2 2077902016709267584 \\
\hline
KIC 7900367 & 4.80 & 11.2 & Gaia DR2 2078093267317792000 \\
\hline
TIC 137003360 & 1.00 & 12.7 & Gaia DR2 1530736690872982400 \\
\hline
TIC 151769040 & 3.63 & 114.1 & Gaia DR2 5389359391479027840 \\
\hline
TIC 391070709 & 3.13 & 91.2 & Gaia DR2 1824557877545994496 \\
\hline
TIC 394818541 & 4.31 & 116.3 & Gaia DR2 4612981556659440256 \\
\hline
TIC 450302084 & 4.28 & 5.1 & Gaia DR2 743556798553137792 \\
\hline
TIC 452590255 & 4.60 & 31.1 & Gaia DR2 5217976697291592064 \\
              & 4.32 & 68.1 & Gaia DR2 5217977075248712832 \\
              & 1.20 & 88.5 & Gaia DR2 5217976727353813760 \\
\hline
TIC 466443867 & 4.66 & 4.1 & Gaia DR2 2754555834813394560 \\
\hline
TIC 1506355332 & 2.98 & 2.3 & Gaia DR2 4544054890700058880 \\
               & 4.10 & 48.8 & Gaia DR2 4544057811279981568 \\
\hline
\end{tabular}
\tablefoot{neighbouring stars from Gaia within 5 magnitudes of the target star, and within an 18 arcsecond radius for the Kepler sample, and within a 120 arcsecond radius for the TESS sample.
}
\end{table*}

\section{MESA inlist}\label{sec:MESAinlist}
The typical MESA inlist for the main sequence evolution of an intermediate-mass star is shown below. \\

\noindent\&\texttt{star\_job} \par
   \texttt{load\_model\_filename = 'PMS.mod'}\par
   \texttt{save\_model\_filename = 'MS.mod'}\par
   \texttt{show\_log\_description\_at\_start = .true.}\par
   \texttt{load\_saved\_model = .true.}\par
   \texttt{history\_columns\_file = 'history\_columns.list'}\par
   \texttt{profile\_columns\_file = 'profile\_columns.list'}\par
   \texttt{pgstar\_flag = .false.}\par
   \texttt{change\_net = .true.}\par
   \texttt{change\_initial\_net = .true.}\par
   \texttt{new\_net\_name = 'pp\_cno\_extras\_o18\_ne22.net'}\par
   \texttt{save\_model\_when\_terminate = .true.} \\
/ \\

\noindent\&\texttt{kap}\par
   \texttt{use\_Type2\_opacities = .true.}\par
   \texttt{Zbase = } \\
/ \\

\noindent\&\texttt{controls} \par
   \texttt{max\_allowed\_nz = 60000}\par
   \texttt{mesh\_delta\_coeff = 0.4}\par
   \texttt{max\_dq = 1d-3}\par
   \texttt{varcontrol\_target = 1d-3} \par
   \texttt{timestep\_dt\_factor = 0.9} \par
   \texttt{energy\_eqn\_option = 'dedt'}\par
   \texttt{use\_gold2\_tolerances = .true.} \par
   \texttt{set\_min\_D\_mix = .true.}\par
   \texttt{min\_D\_mix = 10.0}\par
   \texttt{MLT\_option = 'Cox'}\par
   \texttt{mixing\_length\_alpha = 1.8}\par
   \texttt{use\_Ledoux\_criterion = .true.}\par
   \texttt{alpha\_semiconvection = 0d0}\par
   \texttt{thermohaline\_coeff = 0d0}\par
   \texttt{predictive\_mix(1) = .true.}\par
   \texttt{predictive\_zone\_type(1) = 'burn\_H'}\par
   \texttt{predictive\_zone\_loc(1) = 'core'}\par
   \texttt{predictive\_bdy\_loc(1) = 'top'}\par
   \texttt{overshoot\_scheme(1) = 'exponential'}\par
   \texttt{overshoot\_zone\_type(1) = 'burn\_H'}\par
   \texttt{overshoot\_zone\_loc(1) = 'core'}\par
   \texttt{overshoot\_bdy\_loc(1) = 'top'}\par
   \texttt{overshoot\_f(1) = 0.017}\par
   \texttt{overshoot\_f0(1) = 0.002}\par
   \texttt{overshoot\_D\_min = 1d-2}\par
   \texttt{overshoot\_brunt\_B\_max = 0d0}\par
   \texttt{num\_cells\_for\_smooth\_brunt\_B = 0}\par
   \texttt{num\_cells\_for\_smooth\_gradL\_composition\_term=0}\par
   \texttt{remove\_mixing\_glitches = .false.}\par
   \texttt{write\_pulse\_data\_with\_profile = .true.}\par
   \texttt{pulse\_data\_format = 'GYRE'}\par
   \texttt{add\_atmosphere\_to\_pulse\_data = .false.}\par
   \texttt{keep\_surface\_point\_for\_pulse\_data = .true.}\par
   \texttt{add\_double\_points\_to\_pulse\_data = .true.}\par
   \texttt{threshold\_grad\_mu\_for\_double\_point = 5d0}\par
   \texttt{interpolate\_rho\_for\_pulse\_data = .true.}\par
   \texttt{log\_center\_density\_upper\_limit = 3.25} \par
   \texttt{max\_num\_profile\_models = 500}\par
   \texttt{photo\_interval = 500} \\
/

\section{GYRE inlist}\label{sec:GYRE_inlist}
\&\texttt{model} \par
   \texttt{model\_type = 'EVOL'} \par
   \texttt{file = 'profile9.data.GYRE'} \par
   \texttt{file\_format = 'MESA'} \\
/ 

\noindent \&\texttt{constants} \\
/

\noindent\&\texttt{mode} \par

   \texttt{l = 0} \par
   \texttt{m = 0} \par
   \texttt{n\_pg\_min = 0} \par
   \texttt{n\_pg\_max = 2} \\
/

\noindent \&\texttt{osc} \par
   \texttt{outer\_bound = 'VACUUM'} \\
/

\noindent \&\texttt{rot} \par
   \texttt{Omega\_rot\_source = 'UNIFORM'} \par
   \texttt{Omega\_rot = 0.0} \par
   \texttt{Omega\_rot\_units = 'CYC\_PER\_DAY'} \par
   \texttt{coriolis\_method = 'TAR'} \par
   \texttt{rossby = .FALSE.} \\
/

\noindent \&\texttt{num} \par
   \texttt{diff\_scheme = 'COLLOC\_GL4'} \par
   \texttt{n\_iter\_max = 30} \\
/

\noindent \&\texttt{scan} \par
   \texttt{grid\_type = 'LINEAR'} \par
   \texttt{grid\_frame = 'INERTIAL'} \par
   \texttt{freq\_units = 'CYC\_PER\_DAY'} \par
   \texttt{freq\_min = 1} \par
   \texttt{freq\_max = 40} \par
   \texttt{n\_freq = 1000} \\
/

\noindent \&\texttt{grid} \par
   \texttt{w\_osc = 10} \par
   \texttt{w\_ctr = 10} \par
   \texttt{w\_exp = 2} \\
/

\noindent \&\texttt{ad\_output} \par
   \texttt{freq\_units = 'CYC\_PER\_DAY'} \par
   \texttt{summary\_file = 'profile9\_l0\_m0-freqs.dat'} \par
   \texttt{summary\_file\_format = 'TXT'} \par
   \texttt{summary\_item\_list = 'l, m, n\_pg, n\_p, n\_g, freq, omega, E\_norm'} \par
   \texttt{detail\_template = 'profile9\_l\%l\%m\%n.txt'} \par
   \texttt{detail\_file\_format = 'TXT'} \par
   \texttt{detail\_item\_list = 'Gamma\_1, V\_2, c\_1, x, xi\_r, xi\_h, Delta\_p, Delta\_g, H, W\_eps, dW\_dx, dE\_dx, nabla\_ad, eta, rho'} \\
/

\noindent \&\texttt{nad\_output} \\
/

\section{Amplitude-phase variability Monte-Carlo simulations} \label{sec:montecarlo}
To assess whether the observed amplitude and phase variability could be reproduced solely by beating between closely spaced frequencies, we generated 1000 synthetic light curves of the form $x(t)=\sum_i A_i \sin(2\pi \nu_i t + \phi_i)$, sampled at the observed \textit{Kepler} timestamps so that the time sampling and window function of the data were preserved. The synthetic frequency content was organised into ridge-like structures motivated by the observed \'echelle diagrams. For each ridge, the frequencies were defined as $\nu_k=\nu_{\rm center}+k\,\Delta\nu+\alpha k^2$, with $\Delta\nu=\nu_{\rm rot}(1+\epsilon)$, where $\epsilon$ sets the ridge tilt in \'echelle space and $\alpha$ its curvature. In each realization, $\epsilon$ and $\alpha$ were drawn once and applied to all ridges, so that all ridges in a given realization shared the same global geometry.

Each realization contained one anchor ridge centred on the analysed frequency and 3--5 additional ridges placed relative to the anchor directly in \'echelle space through a signed horizontal modulo-frequency offset with magnitude drawn uniformly from $[0,\nu_{\rm rot}]$ and a vertical offset of $N\nu_{\rm rot}$, where $N$ was drawn from 2 to 17. Each ridge contained 5, 7, 9, or 11 components. The ridge amplitudes followed a Gaussian envelope in the component index $k$, with an additional multiplicative scatter applied to the individual component amplitudes, while the phases were drawn randomly from a uniform distribution. We adopted $\epsilon\in[-0.1,0.0]$, $\alpha\in[-0.008,0.0]$, $\sigma_k\in[1.0,2.5]$, central amplitudes in the range 60--160\,ppm, and component-amplitude scatter in the range 0.75--1.25. No additional white noise was included. The rotation frequency was fixed to $\nu_{\rm rot}=0.1459\,{\rm d}^{-1}$, the analysed frequency ($\nu_{\rm center}$) to $\nu=12.3\,{\rm d}^{-1}$ (similar to KIC~6875337; see Sect.\ref{sec:kic6875337}), and the light curves were divided into 10 rotation-phase bins.

For each realization, the light curve was phase-folded with the rotation period and, in each phase bin, a sinusoid at fixed frequency was fit by linear least squares to recover amplitude and phase as functions of rotation phase. We then considered the anchor ridge and the first two contaminating ridge centres and evaluated all three pairwise combinations. For each pair we computed the amplitude correlation $r_A$, the phase correlation $r_\phi$, and an effective phase-scatter metric $S_{\phi}$, defined as the smaller of ${\rm std}(\phi_i-\phi_j)$ and ${\rm std}(\phi_i+\phi_j)$, such that correlated and anticorrelated phase behaviour were treated symmetrically. A pair was classified as coherent if $|r_A|>0.7$, $|r_\phi|>0.7$, and $S_{\phi}<40^\circ$. A triplet was considered coherent only if all three pairs satisfied these criteria simultaneously. This extends the usual close-frequency beating tests by requiring mutually consistent behaviour across three independent pairwise relations.

\begin{figure}[h]
\centering
\includegraphics[width=\hsize]{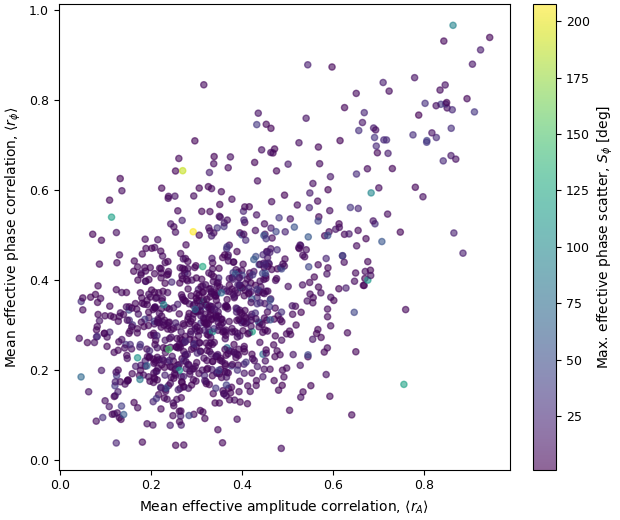}
\caption{Monte Carlo triplet-coherence test. Each point corresponds to one synthetic realization and is plotted as a function of the mean effective amplitude correlation and mean effective phase correlation of the three pairwise combinations among the tested ridge centres. The colour scale shows the maximum effective phase scatter, $S_{\phi}$, among the three pairs. Realizations that simultaneously exhibit strong amplitude coherence, strong phase coherence, and low phase scatter are rare, showing that such triplet behaviour is difficult to reproduce through beating alone.}
\label{fig:montecarlo}
\end{figure}

The outcome of the Monte Carlo simulations is summarized in Fig.~\ref{fig:montecarlo}, which shows the mean effective amplitude correlation and mean effective phase correlation of the tested triplets, with colour indicating the maximum value of $S_{\phi}$ among the three pairwise combinations. Most realizations cluster at low to moderate correlation values, and only a small fraction simultaneously reach high amplitude and phase coherence while maintaining low phase scatter. This indicates that, although beating between closely spaced frequencies can readily produce apparent amplitude and phase variability in individual peaks, reproducing mutually consistent behaviour across three ridge centres is substantially less likely ($\lesssim 1\%$.).

\end{document}